\documentclass[12pt]{article} 
\pdfoutput=1
\usepackage[utf8]{inputenc}%
\usepackage{graphicx}
\usepackage[english]{babel}
\usepackage{amsmath}	% Les symboles les plus fréquents.
\usepackage{amssymb}	% Des symboles.

\usepackage{hyperref}

\def\be{\begin{equation}}
\def\ee{\end{equation}}

\newcommand{\ket}[4]{\vert #1, #2 \rangle \otimes \vert #3, #4 \rangle}
\newcommand\ZZ{\mathbb{Z}}
\newcommand{\eqdef}{\ensuremath{\mathrel{\stackrel{\mathrm{def}}{=}}}}

\def\beq{\begin{equation}}
\def\eeq{\end{equation}}

\newcommand\NN{\mathbb{N}}
\newcommand\QQ{\mathbb{Q}}

\newcommand{\RR}{\mathbb{R}}

\newcommand\beqn{\begin{eqnarray}}
\newcommand\eeqn{\end{eqnarray}}
\newcommand\beqa{\begin{eqnarray}}
\newcommand\eeqa{\end{eqnarray}}
\def\eq#1{(\ref{#1})}
\def \keti{\rangle\!\rangle}
\def \ket{\rangle}

\def \bra{\langle}
\def \ie{{\it i.e.}~}
\def \eg{{\it e.g.}~}
\def \cf{{\it c.f.}~}
\def\p{{}^\prime}
\def\wt#1{\widetilde{#1}}

\newcommand{\OOO}{\mathcal O}

\newcommand{\vect}[2]{\left(\begin{array}{c} #1 \\
    #2\end{array}\right)}
\newcommand{\mat}[4]{\left(\begin{array}{cc} #1 & #2 \\
    #3 & #4 \end{array}\right)}

\newcommand{\wh}{\widehat}
\begin{document}
%%%%%%%%%%%%%%%%%%%%%

\vspace*{-2cm}
\thispagestyle{empty}
\begin{flushright}
\vskip -2cm
RUNHETC-2012-06\\
LPTENS-2012-21\\
LMU-ASC 33/12
%hep-th/yymmnnn
\end{flushright}
\vspace*{1.5cm}
 
\begin{center}
{\Large 
{\bf A worldsheet extension of  $O(d, d\vert \ZZ)$ }}
\vspace{1.5cm}

{\large C.~Bachas${}^{\, \sharp}$}% 
 ,   
\hspace*{0.1cm} {\large I.~Brunner${}^{\, \flat, \, c}$}%
\hspace*{0.1cm} {\large and} \hspace*{0.1cm} {\large D.~Roggenkamp${}^{\, \natural}$}%
\vspace*{0.8cm}

${}^\sharp$ Laboratoire de Physique Th\'eorique de l'Ecole 
Normale Sup\'erieure 
\footnote{Unit\'e mixte de recherche (UMR 8549)
du CNRS  et de l'ENS, associ\'ee \`a l'Universit\'e  Pierre et Marie Curie et aux 
f\'ed\'erations de recherche 
FR684  et FR2687.}
\\
 24 rue Lhomond, 75231 Paris cedex, France\\
\vspace*{0.4cm}

${}^\flat$ 
Arnold Sommerfeld Center, Ludwig Maximilians Universit\"at \\
Theresienstra\ss{}e  37, 80333 M\"unchen, Germany \\
\vspace*{0.4cm}

${}^c$  
Excellence Cluster Universe, Technische Universit\"at   
M\"unchen \\
Boltzmannstra\ss{}e  2, 85748 Garching, Germany \\
 
 \vspace*{0.4cm}

${}^\natural$
Department of Physics and Astronomy, Rutgers University \\
Piscataway, NJ 08855-0849, USA \\ 
 
\vspace*{1.7cm}

{\bf Abstract}
\end{center}

We study superconformal interfaces between  
${\mathcal N}=(1,1)$ supersymmetric sigma models on tori, which preserve a $\widehat u(1)^{2d}$ current algebra. Their fusion is non-singular and, using
parallel transport on CFT deformation space, it can be reduced to fusion of defect lines in a single torus model. 
We show that  the latter is  described by a semi-group extension of $O(d,d|\QQ)$, and 
 that (on the level of Ramond charges) fusion of interfaces agrees with composition of associated geometric integral transformations. 
 This generalizes the well-known fact that T-duality can be geometrically represented by Fourier-Mukai transformations. 
 
Interestingly, we find that 
  the topological interfaces between torus models  form the same semi-group upon fusion.  
We argue that 
this semi-group of orbifold equivalences
 can be regarded as   the  $\alpha^\prime$   deformation of  the continuous $O(d,d)$ symmetry of classical supergravity.

\newpage

\section{Introduction and summary of results}

String theory compactified on a $d$-dimensional torus  is invariant under the group  $O(d, d\vert \ZZ)$ of
 T-duality transformations  \cite{Giveon:1994fu}.
This is the subgroup of  U-dualities realized   
as automorphisms of the worldsheet  sigma model. 
It is, however,  also a subgroup of the much larger continuous  group  
 $O(d, d\vert \mathbb{R})$,  which is the group of  symmetries  of the  classical   low-energy supergravity theory. 
This  larger continuous symmetry is broken by quantum effects,  in particular by the fact that
  the string momentum  and winding vectors are quantized.

 \smallskip 

In this paper we show that a certain  relic  of $O(d, d\vert \mathbb{R})$   does survive   as  
 a symmetry of a subset of  observables, at leading order in the string-loop expansion but to all 
  orders in  $\alpha\p$.  These  ``quasi-symmetries''  are implemented on the string worldsheet by topological
  interfaces (also referred to as defect lines).
  Topological interfaces have played a role  in various contexts in recent years, see  for example 
  \cite{Petkova:2000ip,bbdo, Graham:2003nc,Frohlich:2004ef, Bachas:2004sy,Frohlich:2006ch,Fuchs:2007tx,
arXiv:0712.0076,Brunner:2008fa,Runkel:2008gr,Bachas:2008jd,Sarkissian:2009aa,Carqueville:2009ev,Bachas:2009mc,Drukker:2010jp,Davydov:2010rm,Fuchs:2010hk,Kapustin:2010zc,Suszek:2011hg}.

We are interested in topological interfaces between $d$-dimensional torus models which preserve a $\widehat u(1)^{2d}$
current algebra. It turns out that they are associated to elements  $\hat\Lambda\in O(d,d|\QQ)$,
the group of $O(d,d)$-matrices with rational entries. 
   Their action on perturbative string states transforms
   an integer momentum and winding vector $\hat\gamma \in \ZZ^{d,d}$\  to\  $\hat\Lambda \hat\gamma$ whenever  this is consistent with
   charge quantization, \ie whenever    $\hat\Lambda \hat\gamma$ is also in  $\ZZ^{d,d}$;  
   otherwise it projects the string state to zero.  
   We will argue that the transformation  also rescales
      the effective string-coupling constant by
   \beq\label{coupling1}
    \lambda_{\rm eff} \mapsto \lambda_{\rm eff}\, \sqrt{{\rm ind} (\hat\Lambda)} \ .   
   \eeq
Here, ${\rm ind} (\hat\Lambda)$ denotes the index of the 
sublattice of charges that survives the projection, \ie the smallest positive  integer $K$ such that
  $K  \hat\Lambda$ has only  integer entries.  Clearly these  transformations can  only be inverted if $K=1$, 
in which case they are the familiar  T-dualities of string theory. The   
transformations for general $K$ do not  form a group  but rather a semi-group. It turns out to be a semi-group  extension of $O(d, d\vert \mathbb{Q})$. 
  \smallskip
  
  Topological interfaces for the free boson compactified on a circle, \ie for $d=1$ have been analyzed in \cite{Fuchs:2007tx,arXiv:0712.0076}. 
We extend this analysis  to torus models of arbitrary dimension $d\geq 1$, and also  to theories with  ${\cal N} = (1,1)$  worldsheet supersymmetry. 
  Following  \cite{arXiv:0712.0076} we actually  compute the
  composition, or ``fusion'' of the more general superconformal but not necessarily topological interfaces. These do not separately commute with left and right moving superconformal algebras of the bulk SCFTs as is the case for topological ones, but only with the diagonal 
  subalgebra.\footnote{Such   conformal interfaces arise as generic fixed points of  renormalization-group  flows, 
   see for instance   \cite{Oshikawa:1996dj,bbdo,Quella:2002ct, Bachas:2004sy,
   Quella2,Brunner:2007qu,Brunner:2010xm,Gaiotto:2012wh} and references therein.}
  In the purely bosonic CFT  this requires the introduction of a regulator   and  the subtraction of a divergent Casimir energy. 
  For interfaces preserving a mutually compatible supersymmetry, on the other hand, 
   the divergent Casimir energy  cancels  between bosons and fermions and there is no need
 for an infinite  subtraction.\footnote{This was shown to be also the case for  ${\cal N} = (2,2)$ supersymmetric  interfaces 
between   Landau-Ginzburg models
in \cite{Brunner:2007qu}.  For the free  theories considered in this paper  ${\cal N} = (1,1)$  supersymmetry is sufficient
to remove the singularities of fusion \cite{bbdo}. }
%  The finite part of this energy ensures that  the fusion of  non-topological interfaces
%  does not violate  Cardy's  consistency condition  \cite{UCSB-TH-89-06}. 
 The finite part of this energy   
  contributes to the $g$ factor of the fusion product,  just as expected from
    Cardy's  consistency condition  \cite{UCSB-TH-89-06}. 
 \vskip 0.7mm

 %HERE
 
 Non-topological interfaces can be used to parallel transform the torus CFT along   moduli space. 
  This makes it  possible to pull-back all interfaces to defects in a fixed, reference CFT, and to  associate to them
 a universal defect algebra.  The calculation of  this algebra  of non-topological defects  is the main technical result in  the present 
 paper. 
 
       On a different note, conformal interfaces and defects can be realized as quantum junctions and quantum  impurities in $(1+1)$-dimensional systems
        (for an introduction see \cite{Saleur:1998hq, Saleur:2000gp}).
    Our results on the  fusion of such  defects 
    could thus find more direct applications in the study of the infrared  properties
 of condensed-matter or statistical-mechanical systems. 
 A by-product of our results is,  for instance,  the calculation of the  fusion of conformal defects    in the critical two-dimensional  Ising model.

\vskip 1mm 
 
%HERE 
 
Conformal interfaces on the superstring worldsheet have been constructed  recently in  
   \cite{Satoh:2011an}. There, the Green-Schwarz formulation was used instead of the NSR formulation employed in this paper, 
   and space-time instead of worldsheet supersymmetry was imposed. 
   It was furthermore argued that the  requirement of space-time supersymmetry forces the  interface  either to be  topological
   or  to be  a (totally-reflecting) tensor product of supersymmetric D-branes. 
    Since the $O(d, d\vert \mathbb{Q})$ quasi-symmetries
   are implemented on the NSR worldsheet by  topological interfaces, 
   it should  be possible to rederive our results  in the Green-Schwarz formulation adopted in 
     \cite{Satoh:2011an} as well. However, we will not pursue  this approach   here.

\smallskip     
     The effective  action for the  moduli  and  the associated $u(1)^{2d}$ abelian gauge fields of toroidally-compactified string theory 
   reads  \cite{Maharana:1992my} 
      \beq
  S =  M^2_{\rm Planck}  \int d^{10-d}x \sqrt{-g} \, \left[ {1\over 8}  {\rm Tr}(   \partial_\mu M^{-1}  \partial^\mu M)   
   - {1\over 4}   (F_{\mu\nu})^T  (M^{-1}) F^{\mu\nu}
   \right] \ , 
   \eeq  
where 
 \beq\label{M3}
 M = \left( \begin{matrix}  G^{-1} &  -G^{-1} B \\  B G^{-1} &  G - B G^{-1} B \end{matrix} \right)  
 \eeq
    is a symmetric $O(d,d)$ matrix  that obeys $M\hat \eta M =\hat \eta$,  with $\hat\eta= ( \begin{smallmatrix} 0& {\bf 1}\\ {\bf 1} & 0 \end{smallmatrix})$. 
    Here $G$ is the  metric of the torus in the string frame, 
     $B$ the NS 2-form field and  $F_{\mu\nu}$ a $2d$-vector of gauge field strengths; 
     $M_{\rm Planck}$ is  the Planck scale of the
   effective (super)gravity.     This action is invariant under the global   $O(d,d)$ transformations $F_{\mu\nu} \mapsto  \hat\Lambda F_{\mu\nu}$ 
  and   $M \mapsto \hat\Lambda M\hat\Lambda^{T}$ with $\hat\Lambda^T \hat\eta \hat\Lambda = \hat\eta$. 
  Charge quantization restricts  $\hat\Lambda$  to  the T-duality subgroup $O(d,d\vert\ZZ)$.
 
    The topological interfaces constructed in this paper
  are associated to elements $\hat\Lambda$ of the larger group $O(d,d\vert\mathbb{Q})$, 
  but  they project out sublattices of charges whenever  $\hat\Lambda\notin O(d,d|\ZZ)$.
  \smallskip
  
   The matrix $M$ can be expressed in terms of an auxiliary  ``vielbein'' field 
   \beq\label{vielbeins}
   M =  2\, U^T U\  \leftrightarrow   M^{-1} =  2\,  (U\hat\eta)^T (U\hat\eta)\ . 
   \eeq 
  Using this vielbein one can define a vector of ``physical'' charges $\gamma =  U \hat\eta \hat\gamma$, associated to a vector of integer charges $\hat\gamma$.
The physical-charge vectors $\gamma$ take values in an even self-dual lattice
  $\Gamma^{d,d}$ of left and right momenta, with metric $\eta = {\rm diag} ( {\bf 1}  , -{\bf 1} )$. 
    A general (super)conformal interface
     transforms  $\gamma$ to $\Lambda\gamma$ with
  $\Lambda\in O(d,d)$. It is topological if $\Lambda\in O(d)\times O(d)$.  
    Physical properties,  such as the mass of a fundamental string,  
  only depend on $\gamma$  modulo arbitrary  $O(d)\times O(d)$ rotations.  
    \smallskip
  
    One of the most interesting aspects of our analysis is the way in which 
     the semi-group of topological interfaces 
      acts on  D-branes and on  their Ramond charges. 
 It turns out that just as the masses of fundamental string states, also the D-brane  masses stay  invariant.
  The vectors of  integer Ramond charges, 
  on the other  hand,  transform according to the spinor representation:
  \beq\label{5ext}
  \hat\gamma_D\  \to\     \hat S \,  \hat\gamma_D \ , \qquad 
  {\rm with}\ \ \ \hat S :=  \sqrt{ {\rm ind}(\hat\Lambda)} \, S(\hat\Lambda)\ \in \ GL(2^d\vert \ZZ) \ . 
  \eeq
 Here $S$ is the spinor representation of   $O(d,d\vert\mathbb{Q})$, while
the square root of the index in  the above  expression can be interpreted as the rescaling \eqref{coupling1}
of the effective string coupling.  Interestingly, the latter ensures that 
   $\hat S$ acts as  an endomorphism on the space of integer-component spinors.\footnote{The T-duality 
   group  $O(d,d\vert\mathbb{Z})$ is usually defined as the stabilizer
   of the lattice of
    fundamental-string charges, which transform in the vector representation of the continuous group. 
    That the same discrete group  also stabilizes the lattice of spinor charges is a  subtle mathematical  fact, see 
   for instance  \cite{Mizoguchi:1999fu,Matsu}. The transformation  \eqref{5ext}  is the generalization of this statement to  the  semi-group  extension of     
    $O(d,d\vert\mathbb{Q})$. }
This should be contrasted to $\hat\Lambda$  whose action was restricted to a sublattice of  the lattice
of  integer-component vectors. 
    
   The transformations \eq{5ext} also have a nice geometric meaning. Namely, we show that the action of all superconformal $\widehat u(1)^{2d}$ preserving interfaces on the space of Ramond ground states descends from the action of geometric integral transformations on D-branes. If invertible, such transformations are known as Fourier-Mukai transformations, and it is indeed well known that T-dualities can be realized by Fourier-Mukai transformations 
\cite{Andreas:2004uf, Bruzzo}. 
     
    Although a topological interface  with index $K\not=  1$ cannot be inverted,  its fusion 
 with its parity-transform always yields a sum of invertible defects.  The authors of 
 \cite{Frohlich:2006ch} have argued very generally that interfaces with the above property 
  separate   CFTs that are related by orbifold constructions, and in particular preserve the sphere correlation
 functions of invariant untwisted states. Our results provide a concrete application of these ideas to the torus theories.
 The interfaces associated to elements of $O(d,d\vert\mathbb{Q})$ and $O(d,d\vert\mathbb{Z})$ are, in the language of \cite{Frohlich:2006ch}, 
  examples respectively of ``duality defects'' and the subclass of ``group-like defects''. 
 \smallskip
 
 Let us stress that  $O(d,d\vert\mathbb{Q})$ is  {\it not}  an exact symmetry  of string theory 
     but an orbifold equivalence, \ie   a  classical invariance  of a subset of  observables. It   does,  however, survive  $\alpha^\prime$ corrections. 
   It remains to be seen whether  this ``quasi-symmetry'' has any  profound meaning, or  whether it  is related to  other fascinating glimpses
   on  the arithmetic properties  of string theory (see \eg \cite{Cartier:2007zz}   and references therein).  
  
%         Quite independently, conformal interfaces can be realized as quantum junctions and quantum  impurities in $(1+1)$-dimensional systems
%         (for an introduction see \cite{Saleur:1998hq, Saleur:2000gp}).
%     Our results on the  fusion of $\widehat u(1)^{2d}$ invariant defects 
%     could thus find more direct applications in the study of the infrared  properties
%  of such systems.

\smallskip

  The rest of the paper gives  the technical details behind the  claims made in this introduction. 
  We begin in Section 2 with the construction of  interfaces between bosonic circle theories that preserve  $\widehat u(1)^2$
  symmetry. We present both the explicit interface operators,  and the corresponding boundary states of the
  two-boson theory that is obtained by folding the worldsheet along  the interface. 
  This material is already contained in \cite{bbdo, arXiv:0712.0076}. But we formulate it in a way that easily
  generalizes to higher target-space dimensions.
 
  In Section 3 we extend the construction of Section 2  to superconformal interfaces between ${\cal N} = (1,1)$ supersymmetric   $c=3/2$  circle theories. We emphasize the GSO projection, and in particular establish a precise correspondence of superconformal interfaces in the GSO projected theory and Cardy 
  defects in the Ising model   \cite{Petkova:2000ip}. 

   In Section 4 we derive the  fusion of the $\widehat u(1)^2$-preserving superconformal interfaces between the $c=3/2$ circle theories. 
We show that fusion is non-singular for interfaces preserving the same supersymmetry, even if none of these interfaces is topological. We also explain how any interface can be 
 parallel-transported to a defect in a given reference bulk theory, and compute the monoid of  
 superconformal  defects.  This monoid  turns out to be a semi-group extension of $O(1,1\vert \mathbb{Q})$, tensored for the GSO projected theory  with the fusion algebra of the Ising model. 
 We furthermore show that parallel transport provides a one-to-one correspondence of $\widehat u(1)^2$-preserving superconformal defects in circle theories and the $\widehat u(1)^2$-preserving topological interfaces starting in any given circle theory. This correspondence is compatible with fusion, so that the category 
 of $\widehat u(1)^2$-preserving topological interfaces between circle theories can be completely described in terms of the monoid of  $\widehat u(1)^2$-preserving superconformal defects. 
General  conformal defects of the Ising model have been 
studied in  \cite{Oshikawa:1996dj, Quella2}. 
A  by-product of our analysis is  the fusion algebra of  these Ising  defects. 

  In Section 5 we explain the relation between the defect monoid and the  $O(1,1\vert \mathbb{Q})$  quasi-symmetries of the supergravity action. In particular, we describe their action on perturbative string states on the one hand and D-brane charges on the other. 

Section 6 contains the generalization to   target space dimension $d>1$. We construct the $\widehat u(1)^{2d}$-preserving superconformal interfaces between $d$-dimensional torus models, and calculate their fusion. As in the case of $d=1$, also for arbitrary $d$, parallel transport reduces the fusion structure to the monoid of defects in a fixed
reference  torus model. We determine this monoid  to be the extension \eq{extension} of $O(d,d|\QQ)$ by the semi-group of
maximal rank sublattices of $\ZZ^{d,d}$ (where multiplication is given by intersection).  
In addition we also calculate the  fusion of these defects with $\widehat u(1)^{2d}$-preserving superconformal boundary conditions. 
We tried to keep this section to some extent self contained, so as to make it readable independently of the detailed discussion of the $d=1$ case in Sections 2--4. It can therefore also serve as an overview of our analysis of interfaces. 
  
\smallskip  

In Section 7 we relate the action of the superconformal interfaces to geometric integral transformations. More precisely, we show that the interfaces act on Ramond ground states in the same way  that corresonding geometric integral transformations act on D-brane charges. Even though we did not attempt to prove it, we believe that this is in fact true on the level of the full D-brane category, and
that the interfaces fuse as the respective integral transformations compose.  

Finally, in Section 8 we establish  the one-to-one correspondence between conformal defect lines and topological interfaces in torus models. This extends the relation between the defect monoid on one hand,  and $O(d,d|\QQ)$
quasi-symmetries of the effective supergravity action after
compactification   on a torus of  arbitrary dimension  $d\geq 1$.

In  Appendix A we collect some conventions, and 
in Appendix B we prove an identity relating indices of certain sublattices which is needed for the calculation of the fusion of interfaces.

 %%%%%%%%%%%%%%%%%%%%%%%%%%%%%%%%%%%%%%
%%%%%%%%%%%%%%%%%%%%%%%%%%%%%%%%%%%%%%

\section{Free-boson interfaces preserving $\widehat u(1)^2$}\label{secBos}

We begin with a  review of  interfaces between 
two $c=1$ conformal field theories of free bosons   compactified on a circles. 
We limit ourselves to interfaces preserving two $\widehat u(1)$ Kac-Moody symmetries. 
 These interfaces were constructed and discussed
in  references  \cite{bbdo,arXiv:0712.0076}. Here, we give a description
that will easily generalize to higher target-space dimensions. 

%%%%%%%%%%%%%%%%%%%%%%%%%%%%%%%%%%%%

\subsection{Interface operators versus boundary states }
\label{secbosonic}
   
  As explained in  the above references, there are two different ways to
  think about interfaces:  as operators  mapping  the states of CFT2 on the circle to those of CFT1;  or as 
  boundary conditions in  the tensor-product theory CFT1$\otimes$CFT2$^*$, where CFT2$^*$ is the parity transform of CFT2.
  These two approaches are technically equivalent, but it will be  useful in the sequel  to keep them both at hand.

    In this section CFT1 and CFT2 are theories of a free massless bosonic field  $\phi$,  compactified on circles of radii  $R_1$ and $R_2$ respectively.  Our conventions for $\phi$  are detailed in Appendix \ref{conventions}.

    In the first approach,  conformal invariance is equivalent to the statement  that the interface operator 
 $I_{1,2}:{\mathcal H}_2\rightarrow{\mathcal H}_1$ between the Hilbert spaces of the two CFTs commutes with the 
  Virasoro algebra  $\{L_n  - \tilde L_{-n}, \,  n \in \mathbb{Z} \}$. 
Since the Virasoro generators are quadratic in the $\widehat u(1)$ currents,  the gluing conditions for the latter must be of the form
\beq\label{gluingcondition}
 \vect{ a^1_n }{ - \tilde  a^1_{-n} } I_{1,2} =  I_{1,2}\,  \Lambda\vect{ a^2_n }{- \tilde a^2_{-n} }    \qquad
 {\rm for}\qquad  \Lambda\in O(1,1)\,.  
\eeq
Here $ a^1$ and $a^2$ are the modes of the left-moving $\widehat u(1)$ currents of CFT1 and CFT2 respectively, while 
  $\tilde a^1$ and  $\tilde a^2$ are the modes of the right-moving currents.
  The matrix $\Lambda$ obeys  $\Lambda^T \eta \Lambda = \eta$ with
  $\eta = {\rm diag} (1 , -1)$. 
  
%CORRECTION  
  
We stress that 
\eqref{gluingcondition} does not describe {\it all}  possible conformal gluing conditions of CFT1 with CFT2.  
First we have assumed that  two   affine $\widehat u(1)$  symmetries are preserved. Furthermore, 
taking an invertible gluing matrix $\Lambda$   discards the possibility that the interface factorize into separate
boundary conditions for the  currents of CFT1 and CFT2.  In theories with $d>1$ bosons this
assumption  eliminates interfaces at which some of the currents of CFT2 (and also of CFT1) are fully reflected.
Such non-generic interfaces can be analyzed separately, when needed.  

%END 

% We stress that 
% \eqref{gluingcondition} does not describe {\it all}  possible conformal gluing conditions, 
% only  those  preserving two   affine $\widehat u(1)$  symmetries. 

 \vskip 1mm

 To convert  interfaces to boundary states one  reflects   CFT2 to CFT2$^*$,
so that both conformal theories are now defined on the half-cylinder $\tau \geq 0$.  This exchanges 
the left- and  right-moving modes
\beq\label{bosmodefolding}
\vect{ a^2_n }{  \tilde  a^2_{n} }  \mapsto \vect{-\tilde  a^2_{-n} }{ -   a^2_{-n} } \ . 
\eeq
The gluing conditions then become   conformal boundary conditions 
 for the  tensor-product theory
 CFT1$\otimes$CFT2$^*$. 
 This is a two-boson  theory whose target space is an orthogonal torus. 
 The folding operation converts the interface into a
  boundary state that satisfies 
  the gluing conditions\footnote{Throughout this article, we use  double kets to distinguish boundary states from normal CFT states (created by local operators)
  which are denoted by a single ket.}
  \beq\label{gluingcondition1}
\left[  \vect{ a^1_n }{ - \tilde  a^1_{-n} } 
 +\Lambda\vect{\tilde a_{-n}^2}{-a_n^2} \right]   \vert  I_{1,2} \keti=0\,.
\eeq
 One  can put these conditions in the equivalent  but more standard form\footnote{In reference  
 \cite{arXiv:0712.0076}  the symbol $S$ was used in place of  the
orthogonal matrix $\OOO$. Here we prefer to save this
symbol for the spinor representation of $O(1,1)$.}
\beq\label{bosbdgluing}
\left[\vect{a_n^1}{a_n^2}+\OOO\vect{\tilde a_{-n}^1}{\tilde a_{-n}^2}\right] \vert I_{1,2} \keti=0\,,
\eeq
where $\OOO$ is the orthogonal  matrix 
 \beq\label{SintermsofO}
\OOO(\Lambda) =\mat{ \Lambda_{12} \Lambda_{22}^{-1}}
{\Lambda_{11}-\Lambda_{12}\Lambda_{22}^{-1}\Lambda_{21}} {\Lambda_{22}^{-1}} {-\Lambda_{22}^{-1} \Lambda_{21}} \, . 
\eeq
The  inverse to relation \eqref{SintermsofO}  is
\beq\label{OintermsofS}
\Lambda(\OOO)  =\mat{\OOO_{12}-\OOO_{11}\OOO_{21}^{-1}\OOO_{22}}{\OOO_{11}\OOO_{21}^{-1}}{-\OOO_{21}^{-1}\OOO_{22}}{\OOO_{21}^{-1}}\,.
\eeq
 
  Anticipating the generalization to higher target-space dimension $d$, we have written equations \eqref{SintermsofO} and \eqref{OintermsofS}  
so that they hold  for current modes that are $d$-dimensional vectors. It is nevertheless instructive to make the mapping  between 
$O(2)$ and $O(1,1)$ matrices more explicit. One notes  that $O(2)$ has two disconnected components, while 
the number of disconnected components in $O(1,1)$  is four. These are related as follows: 
\beq
\OOO = \mat{{\rm cos}(2 \vartheta)} {{\rm sin}(2 \vartheta)} {-{\rm sin}(2 \vartheta)} {{\rm cos}(2 \vartheta)} \leftrightarrow 
\Lambda =  \pm \mat{{\rm cosh}\alpha}{{\rm sinh}\alpha}{{\rm sinh}\alpha}{{\rm cosh}\alpha}\mat{1}{0}{0}{-1}\ , 
\nonumber
\eeq
\beq\label{218}
\OOO = \mat{{\rm cos}(2 \vartheta)} {{\rm sin}(2 \vartheta)} {{\rm sin}(2 \vartheta)} {-{\rm cos}(2 \vartheta)}\leftrightarrow 
\Lambda =  \pm \mat{{\rm cosh}\alpha}{{\rm sinh}\alpha}{{\rm sinh}\alpha}{{\rm cosh}\alpha} \ , 
\eeq
where  the rotation angle ${2 \vartheta}\in (-\pi, \pi]$ is related to the rapidity $\alpha\in (-\infty, \infty)$
as follows:
\beq\label{219}
{\rm tanh}\alpha = {\rm cos}(2 \vartheta)\ , 
\eeq
and the sign $\pm$ corresponds, respectively, to the ranges $ \vartheta>0$ or $ \vartheta<0$. 
Crossing the singular value  $ \vartheta = 0$ amounts to jumping
among the two  disconnected components of $O(1,1)$  related by 
the reflection  $-{\bf 1}$. 
Note that the identity gluing condition for an interface corresponds to a permutation gluing condition
  for the associated boundary condition, which glues the left (right) $\widehat u(1)$ current of CFT1 to the right (left) current of CFT2$^*$. 
  
 \vskip 1mm

 Let us give a name to
  the sign that distinguishes the two components of  the orthogonal group, 
  \beq\label{epsilon}
  {\rm det}\Lambda = - {\rm det} \OOO \ \eqdef \  \varepsilon\ .
  \eeq
   As shown in \cite{arXiv:0712.0076}, 
 when $\varepsilon = +1$ the  interface corresponds to  a D1-brane in the folded theory
      subtending
  an angle $ \vartheta$  to the $\phi^1$  axis.\footnote{
  This   is the reason 
 for including the factor of $2$ in the definition of the rotation angle.
 } 
    For fixed compactification  radii 
    $R_i$  this angle cannot vary continuously, but is subject to  the rationality  condition  
 \beq\label{quantization}
 {\rm tan}  \vartheta = {k_2R_2\over k_1R_1} \qquad {\rm if}\ \ \varepsilon = +1 \ .
\eeq
Here, $k_1, k_2$  are arbitrary  integers -- the winding numbers of the associated D1-brane,
which we take to be coprime in the following. 
For
 $\varepsilon = -1$ the folded  interface corresponds to a 
D2/D0 bound state, and the rationality condition
  reads  
\beq\label{quantizationprime}
 {\rm tan}  \vartheta = {2k_2 R_1 R_2\over  k_1} \qquad {\rm if}\ \ \varepsilon = -1 \ .    
\eeq
In this case,  the integers $(k_1, k_2)$  are respectively the number of 
D2 -branes  and the gauge flux threading through them. The latter is forced to
be integer by Dirac's quantization condition.

  We also quote here the explicit form of the bosonic boundary states 
   from reference \cite{arXiv:0712.0076}: 
 \begin{equation}\label{k1k2}
  \vert \OOO,\varphi\rangle \hskip-0.6mm \rangle_{\rm bos}  \ =\   
  \prod_{n=1}^\infty   e^{\, {1\over n} \OOO_{ij} a_{-n}^i \widetilde a_{-n}^j}  \ 
\vert  \OOO,\varphi \rangle_{\rm bos} \ , 
\end{equation}
where  the ground states for $\varepsilon=1$ and $\varepsilon=-1$ are respectively given by  
\begin{eqnarray}
&& \vert  \OOO,\varphi \rangle_{\rm bos} =   \sqrt{
 k_1 k_2  \over {\rm sin}\, (2 \vartheta)} 
   \sum_{N,M =-\infty}^\infty  
e^{i{N} \varphi_1 + iM \varphi_2}  \vert k_2N,  k_1N,        k_1M, 
  - k_2M  \rangle  \ ,  \ {\rm and}\,     \nonumber
 \\
&&\vert  \OOO , \varphi \rangle_{\rm bos} =    \sqrt{
 k_1 k_2  \over {\rm sin}\,  (2 \vartheta)} 
   \sum_{N,M =-\infty}^\infty  
e^{i{N} \varphi_1 + iM \varphi_2}  \vert k_1M,   - k_1N,     k_2N , 
  k_2M  \rangle  \,.
  \label{k1k2G}
\end{eqnarray}
Here, $\vert N_1, N_2, M_1 , M_2 \rangle$ denotes
 the highest-weight state with integer momenta $(N_1, N_2)$ and winding
numbers 
$(M_1 , M_2)$ in the two torus directions, while  $\varphi$ parametrizes  angle moduli of the boundary state (position and Wilson lines of the corresponding D-brane).

 \vskip 1mm

The $g$-factor is the coefficient of the $N=M=0$ ground state. Another important parameter   is  the
  reflection coefficient  ${\cal R}$, defined quite generally
   in reference  \cite{Quella2}. For the bosonic interfaces at hand, these two parameters
   are given by  \cite{arXiv:0712.0076, bbdo}
\beq\label{d=1g}
g_{\rm bos}  = \sqrt{{k_1k_2\over {\rm sin}(2 \vartheta)}}\ , \qquad  {\cal R} = {\rm cos}^2 (2 \vartheta)\ . 
\eeq
Note that while ${\cal R}$ varies continuously with the angle $ \vartheta$, 
the $g$-factor depends non-trivially on its arithmetic properties. In string theory the $g$-factor is the
(normalized) mass of the D-brane,  viewed as a point particle in the non-compact spacetime. 
This (for $\varepsilon = +1$)   depends on the length --  not only on the orientation angle of the D1-brane.
The quantization condition
\eqref{quantization} ensures that this length, and hence the  interface entropy,  is finite.

\vskip 1mm

Using the behavior \eq{bosmodefolding} of the modes under folding, the boundary states are easily unfolded to interface operators. 
The mode contributions can be formally expressed as products of exponentials $I_{1,2}^{n,{\rm bos}}$.  For $n>0$ 
\beq\label{opsnonzeromodesb}
I_{1,2}^{n,{\rm bos}}=
    {\rm exp}\hskip -1mm \left(  {1\over n} ( a^1_{-n} \OOO_{11}  \tilde a^1_{-n}  
    -  a^1_{-n}\OOO_{12} a^2_n -\tilde a^1_{-n} \OOO_{21}^t \tilde a^2_n +  a^2_{n} \OOO_{22}^t \tilde a^2_{n}) 
    \right) \ ,  
\eeq
while the zero-mode contributions are given by
\beqn
I_{1,2}^{0,{\rm bos}}&=&
     \sqrt{ k_1 k_2 \over {\rm sin} (2 \vartheta) } 
  \hskip -0.8mm  \sum_{N,M =-\infty}^\infty  
e^{i{N} \varphi_1 + iM \varphi_2}  \vert k_2N, k_1M\, 
\rangle \langle 
k_1N,  k_2M  \vert  \ , \  {\rm and}\nonumber\\
I_{1,2}^{0,{\rm bos}} &=&
  \sqrt{k_1 k_2 \over {\rm sin} (2 \vartheta) } 
      \sum_{N,M =-\infty}^\infty  
 e^{i{N} \varphi_1 + iM \varphi_2}   \vert k_1M, k_2N\, 
\rangle \langle 
k_1N,  k_2M  \vert  \ \label{bos0int}
\eeqn
for $\varepsilon=\det\Lambda=+1$ and $-1$, respectively.
Using a slightly abusive notation  we may express the complete interface operator as
\beq\label{bosinterface}
I_{1,2}^{\rm bos}=\prod_{n\geq 0} I_{1,2}^{n,{\rm bos}}\, , 
\eeq
with the implicit understanding that  the positive-frequency modes of CFT1 act on the left and those of CFT2 on the right 
of the map $I_{1,2}^{0,{\rm bos}}$. This latter map implements the zero-mode gluing conditions
on the ground states of the two $\widehat u(1)$ Kac-Moody algebras.

%%%%%%%%%%%%%%%%%%%%%%%%%%%%%%%%%%%%%

\subsection{Quantization and sublattices}\label{sect22}

 The quantization conditions 
 \eqref{quantization} and \eqref{quantizationprime} cannot  be generalized as such
   to higher  target-space dimensions.  To put them in a more convenient form,   
    note  that in addition to the  $O(1,1)$ matrix $\Lambda$ which enters in the gluing of the $\widehat u(1)$ currents,
    the interface is  characterized by  the choice of the bulk radii,  $R_1$ of  CFT1 and $R_2$ of CFT2.  
 More explicitly, the corresponding charge  lattices can be written as (here $j=1,2$)
   \beq\label{phystointeger}
    \Gamma_j =  \left\{  
      \left(
 \begin{matrix}    \mbox{\fontsize{9}{11}\selectfont $N/2R_j$}   +   \mbox{\fontsize{9}{11}\selectfont $MR_j$}   \\  
  \mbox{\fontsize{9}{11}\selectfont $-N/2R_j$}  +
  \mbox{\fontsize{9}{11}\selectfont $MR_j$} 
 \end{matrix} \right) 
   \Bigl\vert  N, M \in \mathbb{Z} \right\}  \ =\  U_j   \,  \ZZ^{1,1}\ ,  
      \eeq
where the matrices
 \beq\label{Uviel}
 U_j = \left(
 \begin{matrix}    \mbox{\fontsize{9}{11}\selectfont $1/2R_j$}  &   \mbox{\fontsize{9}{11}\selectfont $R_j$}   \\  
  \mbox{\fontsize{9}{11}\selectfont $-1/2R_j$}  &
  \mbox{\fontsize{9}{11}\selectfont $R_j$} 
 \end{matrix} \right) 
 \eeq
are the ``vielbeins'' introduced in \eqref{vielbeins} and $\ZZ^{1,1}$  is the lattice of integer momenta and windings. 
The transformation  \eqref{phystointeger}  corresponds precisely to the change of basis from the physical 
left and right $u(1)$ charges\footnote{Note that in
our conventions $\Gamma_j$ is  the lattice of charges 
$(j_0, -\tilde j_0)$.} to integer momentum and winding, which has been mentioned in the introduction.
\smallskip

Note that  states of CFT2  with physical charge vector $\gamma\in \Gamma_2$  are
mapped to  states of  CFT1 with physical charge vector $\Lambda \gamma$. If $\Lambda\gamma \in \Gamma_1$
then  $\vert \Lambda \gamma\rangle\langle \gamma\vert$ does indeed 
 contribute  to the zero-mode operator $I_{1,2}^{0,{\rm bos}}$. Otherwise, 
 all CFT2 states in the $\widehat u(1)^2$ module with highest-weight  vector $\vert\gamma\rangle$  are mapped 
to zero by $I_{1,2}$. 
The CFT2 charge vectors that contribute to the zero-mode sum   lie therefore in the intersection sublattice
of physical charges
\beq
\Gamma_{1,2}^\Lambda\  := \  \{\gamma\in \Gamma_2 \vert \Lambda\gamma \in \Gamma_1 \} \, = \,
\Gamma_2 \cap \Lambda^{-1}\Gamma_1 \, =\, 
\Gamma_2 \cap \Lambda^{-1}U_1  U_2^{-1}  \Gamma_2 \, . 
\eeq
This   is mapped by $\Lambda$ to the sublattice of CFT1 charge vectors
\beq
\Gamma_{2,1}^{\Lambda^{-1}} \  := \  \{\gamma\in \Gamma_1 \vert \Lambda^{-1}\gamma \in \Gamma_2 \} \, = \,
\Gamma_1 \cap \Lambda \Gamma_2 \, =\, 
\Gamma_1 \cap \Lambda U_2  U_1^{-1} \Gamma_1 \, , 
\eeq
where $\Gamma_1 = U_1  U_2^{-1}   \Gamma_2$. 
 The quantization conditions  \eqref{quantization},  \eqref{quantizationprime} 
 ensure that $\Gamma_{1,2}^\Lambda$ is a maximal-rank  sublattice of $\Gamma_2$ (or
equivalently that $\Gamma_{2,1}^{\Lambda^{-1}}$ is a maximal-rank  sublattice of $\Gamma_1$). 
Gluing matrices obeying this maximal-rank   condition  will be referred to  as ``admissible''  gluing matrices.

\smallskip

This  condition is more transparent in the canonical basis of integer
winding  and momentum. The gluing of these integer-charge vectors is implemented by  
$\hat \Lambda := U_1^{-1} \Lambda U_2$.\footnote{Strictly-speaking, the matrix $\hat\Lambda$ defined
in  the introduction is $\hat\eta U_1^{-1} \Lambda U_2\hat\eta$. Henceforth, we will absorb the $\hat\eta$ 
by redefining  the vector of integer charges.}
This  is a $O(1,1)$ matrix that leaves invariant
 the (off-diagonal)  metric   $\hat\eta = \left( \begin{smallmatrix} 
0&1\\ 1&0 
\end{smallmatrix} \right)$ 
   on $\ZZ^{1,1}$.
It can be read off  easily from the zero-mode maps  \eqref{bos0int} with the result: 
  \beq\label{rationalmatrices}
\hat \Lambda =    \left( \begin{matrix}   k_2/k_1 & 0 \\ 0 & k_1/k_2 \end{matrix} \right)  \  \   \quad {\rm or }\quad\ \ 
\hat \Lambda =   \left( \begin{matrix}   0  & 1  \\ 1 & 0 \end{matrix}  \right)
  \left( \begin{matrix}   k_2/k_1 & 0 \\ 0 & k_1/k_2 \end{matrix} \right) 
 \ 
  \eeq
 for $\varepsilon = +1$ or $\varepsilon =-1$, respectively. In this canonical basis   the admissible gluing
conditions  are, therefore,  in one-to-one correspondence with elements of $O(1,1\vert\mathbb{Q})$, the group of
$O(1,1)$ matrices with rational entries. This form of the quantization condition will generalize easily
to higher  target-space dimension. 
\smallskip

For general $k_1,k_2$, the transformations \eqref{rationalmatrices} do not map all integer vectors
to integer vectors. Only the sublattice 
 \beq
U_2^{-1} \Gamma_{1,2}^\Lambda =  \ZZ^{1,1} \cap  \hat\Lambda^{-1}  \ZZ^{1,1} =  k_1 \ZZ \oplus k_2\ZZ\ 
\eeq
is mapped back to $\ZZ^{1,1}$, more precisely to the sublattice 
\beq
U_1^{-1}\Gamma_{2,1}^{\Lambda^{-1}} =  \ZZ^{1,1}\cap  \hat\Lambda   \ZZ^{1,1} =  k_2 \ZZ \oplus k_1\ZZ\ \
{\rm or}\ \    k_1 \ZZ \oplus k_2\ZZ
\eeq
for $\varepsilon = +1$ and $\varepsilon =-1$, respectively. 
 The index
    \beq
  {\rm ind}(\hat\Lambda) := {\rm ind}(\ZZ^{1,1}\cap\hat\Lambda^{-1}\ZZ^{1,1}\subset\ZZ^{1,1})  = \vert k_1 k_2 \vert 
  \eeq 
of   this intertwiner  sublattice in the charge lattice $\ZZ^{1,1}$ will play  a key role in what follows. 
 It is convenient to define the projector 
\beq
\Pi_{\hat\Lambda} \vert   \hat\gamma \rangle  :=   \begin{cases}   \vert   \hat\gamma \rangle  &\ \ {\rm  if} \ \   \  {\hat\Lambda}  \hat\gamma \in \ZZ^{1,1}  \, ,  \\ 0 & 
\ \ {\rm otherwise}\, \end{cases}
\eeq
on sectors with charges in this sublattice.
Using these definitions  and the identities  $  |\Lambda_{22}| = {\rm cosh}\alpha = { \vert \sin(2 \vartheta)\vert }^{-1}
$, see  \eqref{218} and \eqref{219},  
 we can put  the  ground state maps \eqref{bos0int} in the more elegant form
\beq\label{sec2last}
I_{1,2}^{0,{\rm bos}}\ =\ \sqrt{ {\rm ind}({\hat\Lambda}) \,   \vert \Lambda_{22} \vert  }\
\sum_{   \hat\gamma\in\ZZ^{1,1}}
e^{2\pi i\varphi(  \hat\gamma)} |\hat \Lambda  \hat\gamma\ket\bra  \hat\gamma| \, \Pi_{\hat\Lambda}  \ , 
\eeq
where  $\varphi$ is some linear form on  $\ZZ^{1,1}$. 
This expression easily generalizes to higher dimensions.
\smallskip

We conclude this section with the following  remark:  
the   interfaces discussed here can be  uniquely specified
by the data $(\hat\Lambda , \varphi , U_1 , U_2)$, where $\hat\Lambda \in O(1,1\vert \mathbb{Q})$ while  $U_j\in O(1,1|\RR)$
determine the bulk radii.
Interestingly,  in  the expression \eqref{sec2last} for the zero-mode sum  only  $ \Lambda_{22}$
depends on these bulk  radii.  Furthermore, as
 explained in reference \cite{arXiv:0712.0076}, 
to any choice of the discrete data $\hat\Lambda$ 
and of $R_2$  there corresponds an  $R_1$,
\beq\label{topo9}
R_1  = f_{\hat\Lambda}(R_2) := \begin{cases} \  \left\vert {k_2\over k_1}\right\vert R_2 & \ \ {\rm if}\ \varepsilon = +1 \\
\left\vert {k_1\over k_2}\right\vert {1\over 2 R_2} & \ \ {\rm if}\ \varepsilon = -1  \ , \end{cases}
\eeq
   such  that $ \vert \Lambda_{22} \vert \hskip -0.3mm = \hskip -0.3mm  \vert {\rm sin} (2 \vartheta) \vert  \hskip -0.3mm =  \hskip -0.3mm 1$  and the $g$-factor   is 
 minimized.  Indeed from   \eqref{Uviel},  \eqref{rationalmatrices} and \eqref{topo9}
 one can compute    
 $\Lambda = U_1 \hat\Lambda U_2^{-1}  = {\rm diag} (\pm 1, \pm1)$,  so that the gluing matrix for the $u(1)^2$ currents 
 is a $O(1)\times O(1)$ matrix. This means that these interfaces commute with both, the left and right Virasoro algebra, and are therefore topological. For a given $\hat\Lambda$, they exist for any $R_2$, and the corresponding interface operators do not 
 exhibit an explicit $R_2$ dependence. 

A more detailed discussion of this point in the context of torus models of arbitrary target space dimension $d$ can be found in Section \ref{sec8}.

 %%%%%%%%%%%%%%%%%%%%%%%%%%%%%%%%

%%%%%%%%%%%%%%%%%%%%%%%%%%%%%%%%%%%%%%
%%%%%%%%%%%%%%%%%%%%%%%%%%%%%%%%%%%%%%

\section{${\cal N}=1$ supersymmetry}\label{sec3/2}

 We will now extend the discussion of the previous section to the 
 ${\cal N} = (1,1)$ supersymmetric CFT, consisting of a free boson $\phi$ 
 and a free  Majorana fermion with left and right components $\psi$ and $\tilde\psi$.  
Interfaces preserving ${\cal N}=1$ supersymmetry have been constructed  
  in reference \cite{bbdo}. 
  Here we complete this construction in the GSO projected theory,  where the interface
  operators can have a non-trivial Ramond sector.

%%%%%%%%%%%%%%%%%%%%%%%%%%%%%%%%%%%%%% 
 
\subsection{Superconformal $\widehat u(1)$ invariant  boundary states} \label{DNstates}

As a warm up  we will first consider the superconformal boundary states of the
$c= 3/2$ theory. We limit ourselves to  states  preserving a $\widehat u(1)$ symmetry -- for a more
general discussion see references  
\cite{Gaberdiel:2001zq, Cappelli:2002wq}.
Besides the Virasoro generators $\{L_n  - \tilde L_{-n}, \,  n \in \mathbb{Z} \}$, 
 these 
states are  annihilated by the combinations $\{G_r - i \eta_{\rm S}  \tilde G_{-r}, \, \forall r\}$
 of modes of the left and right supersymmetry currents. The choice of gluing condition $\eta_{\rm S}=\pm 1$ 
 specifies which of the two possible supersymmetries is preserved. 
 Notice the factor of $i$ in these combinations;  it
 ensures that  the  supersymmetry generators anticommute into the   Virasoro generators that  annihilate the boundary state.  
   
\smallskip   
     
States  preserving a $\widehat u(1)$ symmetry 
 are annihilated by the combinations $\{a_n-\varepsilon\widetilde{a}_{-n},\, n\in\ZZ\}$ of modes of the left and right $\widehat u(1)$ currents. 
The choice of the sign $\varepsilon=1$ or $\varepsilon=-1$ distinguishes between Dirichlet and Neumann boundary 
conditions.\footnote{This is consistent with the notation of the previous subsections
 since $-\varepsilon$ can be considered  as a one-dimensional orthogonal gluing matrix.}
 In combination with superconformal invariance these gluing conditions force
separate gluing conditions on the fermionic fields. Namely,
 the fermionic modes $\{\psi_r -   i \epsilon\,   \tilde \psi_{-r},\,\forall r\}$ with $\epsilon\equiv\varepsilon\eta_{\rm S}$ also have to annihilate the boundary state.
Having to satisfy gluing conditions for bosons and fermions independently, the boundary states factorize into tensor products of bosonic and fermionic boundary states,
\beq
\vert {\cal B} \keti_{\rm full}  = \vert {\cal B}  \keti_{\rm bos} \otimes \vert {\cal B} \keti_{\rm ferm}\ . 
\eeq 
 
\smallskip 
 The Dirichlet and Neumann  boundary states  for the boson  are well-known 
 (see for example \cite{Di Vecchia:1997pr, Matthias} and references therein) but
 we repeat them here for the reader's convenience:  
  \begin{equation} {\rm D:}\quad
\vert+,\varphi\keti_{\rm bos}  \ =\   
\prod_{n=1}^\infty   {\rm exp} \left( {1\over n} a_{-n}\tilde a_{-n}\right) \ 
%\Bigl(  {1\over\sqrt{2R}} \sum_{N=-\infty}^\infty  e^{-i{N\over R} \varphi}  \vert N, 0 \rangle \Bigr)\ ,  \nonumber
\Bigl(  {1\over\sqrt{2R}} \sum_{N=-\infty}^\infty  e^{-i{N } \varphi}  \vert N, 0 \rangle \Bigr)\ ,  \nonumber
\end{equation}
\begin{equation}\label{nstate}{\rm N:}\quad
\vert-,\varphi\keti_{\rm bos}
    \ =\   
\prod_{n=1}^\infty   {\rm exp}\left( -{1\over n}  a_{-n}\tilde a_{-n}\right) \ 
%\Bigl(  \sqrt{R} \sum_{M=-\infty}^\infty  e^{-2 iM R \varphi}  \vert 0, M  \rangle \Bigr)\ , 
\Bigl(  \sqrt{R} \sum_{M=-\infty}^\infty  e^{- iM   \varphi}  \vert 0, M  \rangle \Bigr)\ , 
\end{equation}
where $\vert N,M\rangle$ is the normalized ground state  
 in a given momentum and winding sector, and the angle $\varphi$ corresponds, in string-theoretic language, 
 to the position of a  D-particle on the circle 
or the Wilson line of a winding D-string.
  The $g$-factors of the above boundary states,  $\sqrt{R}$ or  $\sqrt{1/2R}$,  will be important for our discussion later on.  

\smallskip

The fermionic boundary states are linear combinations of 
  \begin{eqnarray}\label{fermionicstates}
  \vert {\rm NS}, \epsilon   \rangle\hskip -0.7mm \rangle\  =     \prod_{r \in  {\mathbb{N}}-
{1\over 2}} e^{i \epsilon  \psi_{-r} \tilde \psi_{-r}} \vert 0 \rangle_{\rm NS} 
  \ ,  \qquad 
  \vert {\rm R} , \epsilon  \rangle\hskip -0.7mm \rangle\  =    2^{1\over 4} \prod_{r \in  {\mathbb{N} } }
 e^{ i \epsilon \psi_{-r} \tilde \psi_{-r}} \vert \epsilon   \rangle_{\rm R}\ ,
\label{6}
\end{eqnarray}
where 
$\mathbb{N}$ denotes the set of positive integers. 
Our conventions for the fermion field are given in 
 Appendix \ref{conventions}. 
 The  normalized Ramond  ground states $\vert \epsilon  \rangle_{\rm R}$
  form a  representation of the algebra of fermionic zero modes,\footnote{Note that the factor
 $i$ in the boundary conditions  is not compatible  with the Majorana  property of the spinor field, which implies
 that $\psi_0$ and $\tilde\psi_0$ can be chosen real.  It is however compatible with the Majorana condition in Euclidean time,
$\psi_r^* = i \tilde \psi_r$.   }
\begin{equation}\label{rralgebra}
\psi_0 \vert \pm  \rangle_{\rm R} =  {1\over \sqrt{2}} e^{\pm i\pi/4} \vert \mp \rangle_{\rm R} \  ,  \qquad
\tilde\psi_0 \vert \pm \rangle_{\rm R} =  {1\over \sqrt{2}} e^{\mp i\pi/4} \vert \mp \rangle_{\rm R} \  .
\end{equation}
The
cylinder partition functions associated with the above boundary states can be computed using standard techniques. Setting
 $H=L_0+\wt L_0$  for the Hamiltonian and $q=e^{-\tau}$ (with $\tau$ real)  one finds: 
\begin{eqnarray}
&&   \langle \hskip-0.6mm \langle {\rm NS}, \epsilon  \vert\, 
  q^H \, \vert{\rm NS } , \epsilon  \,  \rangle \hskip-0.6mm \rangle \,
  =\, q^{-{1\over 24}}  \prod_{r\in  \mathbb{N} - 1/2} 
(1 +   q^{2r} ) \, = \, \left\vert  {\theta_3\over \eta}  \right\vert^{1/2} \ , \nonumber
\\
 &&  \langle \hskip-0.6mm \langle {\rm NS}, \epsilon \vert\, 
  q^H \,  \vert {\rm NS } , - \epsilon  \,  \rangle \hskip-0.6mm \rangle \,
  =\, q^{-{1\over 24}}  \prod_{r\in  \mathbb{N} - 1/2} 
(1 -  q^{2r} ) \, = \, \left\vert  {\theta_4\over \eta}  \right\vert^{1/2}\ , \nonumber
\\
 &&  \langle \hskip-0.6mm \langle {\rm R}, \epsilon \vert\, 
  q^H \,  \vert {\rm R } , \epsilon  \,  \rangle \hskip-0.6mm \rangle \,
  =\,    \sqrt{2}\, q^{{1\over 12}}  \prod_{r\in  \mathbb{N} } 
(1 +  \,  q^{2r} )\, = \,    \left\vert  {\theta_2\over  \eta}  \right\vert^{1/2}\ .
\label{annulus1}
\end{eqnarray}
 Here $\eta$ and $\theta_a$ denote the familiar Dedekind-eta  and Jacobi-theta functions. The partition function between Ramond contributions of opposite $\epsilon$ vanishes. 
\vskip 1mm

   The boundary states of the unprojected fermion theory are the states $\vert {\rm NS } , \pm  \,  \rangle \hskip-0.6mm \rangle$. 
     We are interested in the boundary states of the GSO projected theory, which can be thought of as an orbifold 
     by the  $\ZZ_2$ group
generated by the operator $(-1)^{F+\wt F}$. 
Here $F$ and $\wt F$ denote left and right fermion numbers respectively.  Since $\vert {\rm NS } , \pm  \,  \rangle \hskip-0.6mm \rangle$
are invariant under the orbifold group, they  must be resolved by additional contributions from the twisted sectors -- the Ramond sector in the case at hand.
 This gives 
\begin{equation}\label{fullferm}
\vert\epsilon \rangle \hskip-0.6mm \rangle_{\rm ferm}={1\over\sqrt{2}}\left( 
\vert{\rm NS},\epsilon\rangle \hskip-0.6mm \rangle\pm |{\rm R},\epsilon \rangle \hskip-0.6mm \rangle
\right)\ , 
\end{equation}
 with the normalization ${|\ZZ_2|}^{-1/2}= {1/\sqrt{2}}$\  chosen as usual 
  so that the identity appears in the direct (open-string) channel with multiplicity one. 
 To obtain the boundary states in the orbifold theory, 
 one only needs to project on the invariant subsectors, which is done by taking appropriately normalized orbits 
 under the action of the orbifold group.
 
  \vskip 1mm

Since $(-1)^{F+\wt F}$ anti-commutes with all the fermionic modes $\psi_r$ and $\wt \psi_r$, its action 
 is completely determined by its action on the ground states $|0\ket_{\rm NS}$ and $|\epsilon\ket_R$. On the NS ground state it acts trivially, but there are two consistent choices on the twisted, \ie the Ramond ground states:
\beq\label{-1fdef}
(-1)^{F+\wt F}=\left\{\begin{array}{cc} -2i\psi_0\wt{\psi}_0\, & {\rm 0A}\\ 2i\psi_0\wt{\psi}_0\,&{\rm 0B}\end{array}\,.\right.
\eeq
By reference to string theory,  we  call the 
two choices ``type 0A'' and  ``type 0B''.  They are related by the $\mathbb{Z}_2$ duality
 that exchanges the spin with the disorder operator of the Ising model, which is  the orbifold CFT.

 The construction of the projected boundary states in orbifold theories has been discussed in \cite{Billo:2000yb}. One  simply
sums the images  under the action of the orbifold group $G$, and normalizes the result by $(|{\rm Stab}_G| / |G|)^{{1\over 2}}$, 
where the stabilizer ${\rm Stab}_G$ is the subgroup of $G$ which leaves the original unprojected boundary state invariant \footnote{Note that the resolution of the boundary states with non-trivial stabilizer has been taken care of in the intermdiate step (\ref{-1fdef})}. 
  It can be seen that in addition to 
 $|{\rm NS},\epsilon\keti$ also  $|{\rm R},-\keti$ 
is  invariant under the $\ZZ_2$ action 
  in the 0A  orbifold, while $|{\rm R},+\keti$ is invariant in the 0B orbifold. 
 On the other hand $(-1)^{F+\widetilde F}$ multiplies  $|{\rm R},+\keti$ (respectively $|{\rm R},-\keti$) by $-1$.
Thus, applying the orbifold construction to 
 the boundary states \eq{fullferm} 
 yields the boundary states 
 \beqn\label{ferm0A}
  \vert  {\rm  {charged}}, \pm  \rangle\hskip -0.7mm \rangle_{\rm ferm}^{\rm 0A}  &=&   {1\over \sqrt{2}} \left( \vert {\rm NS}, -  \rangle\hskip -0.7mm \rangle\  \pm
\vert {\rm R} , -   \rangle\hskip -0.7mm \rangle \right),   \\
\vert {\rm  { neutral } } \rangle\hskip -0.7mm \rangle_{\rm ferm}^{\rm 0A} 
&=&     \vert {\rm NS}, +  \rangle\hskip -0.7mm \rangle\ \nonumber 
\eeqn
for the 0A orbifold,  and 
\beqn\label{ferm0B}
 \vert  {\rm  {charged}}, \pm  \rangle\hskip -0.7mm \rangle_{\rm ferm}^{\rm 0B}  &=&   {1\over \sqrt{2}}  \left(\vert {\rm NS}, + \rangle\hskip -0.7mm \rangle\  \pm
\vert {\rm R} , +   \rangle\hskip -0.7mm \rangle \right),   \\
\vert {\rm  { neutral } } \rangle\hskip -0.7mm \rangle_{\rm ferm}^{\rm 0B} 
&=&     \vert {\rm NS}, - \rangle\hskip -0.7mm \rangle\ \nonumber
\eeqn
for the 0B orbifold. By reference to string theory, we call a boundary condition charged  if it has a non-vanishing R-charge, 
 \ie if it couples to the Ramond ground states.

\vskip 1mm

  Another way of stating this result is that  the  fermion-parity projection eliminates $\vert + \rangle_{\rm R}$  in the type-0A theory, and 
 $\vert -  \rangle_{\rm R}$ in the type-0B theory. The projection also removes
 the Ishibashi states built on these Ramond ground states, leaving three independent boundary states in each theory.
 Cardy's condition  \cite{UCSB-TH-89-06} fixes the precise linear combinations.

\vskip 1mm
 
Indeed, the GSO-orbifold of the free fermionic theory is nothing but the Ising model, 
 a well-known rational CFT with three primary fields of conformal weights $h = \tilde h = 0, {1/ 2}, {1/16}$. Boundary states in this theory can be obtained by means of Cardy's construction, which expresses them in terms of the associated Ishibashi states as 
\cite{UCSB-TH-89-06} 
\begin{eqnarray}\label{CardyIsing}
&& { {{\rm spin\,up}}}:   \qquad  \ \ \ \  \vert 0 \rangle\hskip -0.7mm \rangle_{\rm C} \  =  {1\over \sqrt{2}} \, \vert 0 \rangle\hskip -0.7mm \rangle_{\rm Ish} + 
{1\over \sqrt{2}}\, \vert {1\over 2}  \rangle\hskip -0.7mm \rangle_{\rm Ish} +
{1\over  2^{1/4} }  \vert {1\over 16}  \rangle\hskip -0.7mm \rangle_{\rm Ish}\ ,  \nonumber
 \\
&& {\rm spin\, down}:   \qquad    \vert {1\over 2}  \rangle\hskip -0.7mm \rangle_{\rm C}\  =
 {1\over \sqrt{2}} \, \vert 0 \rangle\hskip -0.7mm \rangle_{\rm Ish} + 
{1\over \sqrt{2}}\, \vert {1\over 2}  \rangle\hskip -0.7mm \rangle_{\rm Ish} -
{1\over  2^{1/4} }  \vert {1\over 16}  \rangle\hskip -0.7mm \rangle_{\rm Ish}\ ,   \nonumber
 \\  
&&  { {{\rm spin\,free}}}:    \qquad  \ \  \vert {1\over 16}  \rangle\hskip -0.7mm \rangle_{\rm C}\  = 
 \vert 0 \rangle\hskip -0.7mm \rangle_{\rm Ish} 
 - \vert {1\over 2} \rangle\hskip -0.7mm \rangle_{\rm Ish} \  .  
 \end{eqnarray}
 The boundary conditions of the Ising spin are indicated on the left.

\vskip 1mm

One can easily identify the  states in (\ref{fermionicstates}) with the Ising Ishibashi  states by comparing the cylinder partition functions. The result  is 
\begin{equation}
\vert {\rm NS}, \pm  \rangle\hskip -0.7mm \rangle = \vert 0 \rangle\hskip -0.7mm \rangle_{\rm Ish} \mp \vert {1\over 2}  \rangle\hskip -0.7mm \rangle_{\rm Ish}
 \qquad
{\rm and} \qquad
 \vert {\rm R} , -  \rangle\hskip -0.7mm \rangle\  = {2^{-{1\over 4}}}\, \vert {1\over 16}  \rangle\hskip -0.7mm \rangle_{\rm Ish}\ . 
\end{equation}
Thus, the boundary states constructed above are related with the Ising boundary states by
\beqn\label{Fermionstates}
&&|{\rm charged},+\keti_{\rm ferm}^{\rm 0A}=|0\keti_{\rm C}\, , \quad{\rm spin\,up}\,,\nonumber\\
&&|{\rm charged},-\keti_{\rm ferm}^{\rm 0A}=|{1\over 2}\keti_{\rm C}\,,\quad{\rm spin\, down}\,,\nonumber\\
&&|{\rm neutral}\keti_{\rm ferm}^{\rm 0A}=|{1\over 16}\keti_{\rm C}\,,\quad{\rm spin\, free}\,.
\eeqn
The  charged  states correspond to the fixed-spin boundary conditions of the Ising model; they have
  non-vanishing one-point functions with the Ramond ground state.  
  The neutral boundary state,  on the other hand, corresponds to the free-spin
  boundary condition of the Ising model;  its one-point function with the Ramond vacuum vanishes.

\vskip 1 mm 

Let us now go back to  the $c={3\over 2}$  theory and put together the bosonic and fermionic states. 
In the unprojected theory this gives 
\beq\label{full}
|\varepsilon,\varphi,\eta_{\rm S}\keti_{\rm full}=|\varepsilon,\varphi\keti_{\rm bos}\otimes |{\rm NS}, \varepsilon\eta_{\rm S}\keti  
\eeq
where $|\varepsilon,\varphi\keti_{\rm bos}$  is one of the states \eqref{nstate}.  After GSO projection, on the other hand, on finds
for instance
in the type 0A model
\begin{eqnarray}\label{full0A}
|\varepsilon,\varphi,\eta_{\rm S}\keti_{\rm full} &=&  |\varepsilon,\varphi\keti_{\rm bos}\otimes 
\sqrt{ 
{  
 | {\rm Stab}_G |   \over  |G| } 
  }
 \sum_{\rm \ZZ_2\ orbit} |\varepsilon\eta_{\rm S}\keti_{\rm ferm}\\
&=&    |\varepsilon,\varphi\keti_{\rm bos}\otimes 
\vert h \keti_{\rm C}\ , 
\end{eqnarray}
where $|\varepsilon\eta_{\rm S}\keti_{\rm ferm}$ was defined in \eqref{fullferm}
and the orbit sum gives  one of the three  Cardy states   of the Ising model, as just explained. 
\smallskip

The supersymmetry   preserved by  boundary states in the GSO projected theories is summarized in table \ref{default}. 
As shown there,  a charged Neumann  and a neutral Dirichlet  state  preserve
the $\eta_S=+1$ supersymmetry in the type 0A model. The second 
supersymmetry, $\eta_S=-1$,   is preserved by a neutral Dirichlet  and a charged Neumann  state. 
\vskip 1 mm
 
\begin{table}[htdp]
\begin{center}
\begin{tabular}{c||c|c}
\, & {\rm Dirichlet }  &  {\rm Neumann} \\
\hline\hline  
 {\rm charged} &  - &    + \\
 \hline 
  {\rm neutral} & +  &  - 
\end{tabular}
\end{center}
\caption{\small{The value of $\eta_{\rm S}$ determining which superconformal symmetry is preserved by   
 boundary states of the
$c = {3\over 2}$ type-0A model. The boundary states  are   tensor products
of  a Dirchlet or Neumann boundary state for the boson 
with a fermion  state in  (\ref{ferm0A})  or \eqref{Fermionstates}.  
Charged   states are doubly-degenerate. In  the type-0B theory    the sign  
of $\eta_{\rm S}$ has to be  reversed. }}
\label{default}
\end{table}
 
Let us  recapitulate all  the   signs that entered  the construction of  boundary states. 
 The  gluing condition of the $\widehat u(1)$ current  is determined by $\varepsilon$,  and the unbroken supersymmetry by  $\eta_{\rm S}$.
 Together these fix the gluing condition $\epsilon = \varepsilon \eta_{\rm S}$ of the fermionic field.
 If the Ishibashi state implementing this gluing condition in the Ramond sector survives the GSO  projection, the boundary state is charged -- \ie it has non-vanishing overlap with the Ramond ground state.
  If it does not  the (superconformal) boundary state is neutral. 
\smallskip 
 
 We close this subsection with two remarks. First  
  by analogy with the 
  $g$-factor, which is the projection of a boundary state on the NS ground state, one can define
  the Ramond charge(s) as the projection onto Ramond ground state(s).  In the case at hand,
  these two quantities are related in a way reminiscent of a BPS condition for supersymmetric D-branes.  
  There is however no space-time supersymmetry
in the present context; the relation is accidental as will become clear later.

The second remark concerns  the cylinder partition function. 
% It can be checked that  for any two  boundary states preserving the same superymmetry, \ie with the same $\eta_{\rm S}$, this partition function
% is finite in the limit $\tau\to 0$. The singular behavior in the bosonic sectors is exactly cancelled by terms  from the fermionic sector. 
As  is well known, 
 for any two  boundary states preserving the same superymmetry, \ie with the same $\eta_{\rm S}$, this partition function
is finite in the limit $\tau\to 0$. The singular behavior in the bosonic sectors is exactly cancelled by the contribution  of the fermions, 
as follows from the absence of tachyons in the open-string channel.  
The generalization of this fact to superconformal interfaces will be important  in the discussion of  fusion. 
  
 %%%%%%%%%%%%%%%%%%%%%%%%%%%%%%%%%%%%%% 

\subsection{Supersymmetric $\widehat u(1)^2$ invariant interfaces }\label{secfermionic}

   Similarly to boundary conditions, 
  also superconformal interfaces between two  ${\mathcal N}=(1,1)$ circle theories which preserve a $\widehat u(1)^2$ current algebra
   factorize into separate interfaces between the bosonic and the fermionic parts of the theories.
The bosonic interfaces have been discussed in Section \ref{secBos}. Here we will  construct the fermionic interfaces. 
Again, several signs enter
the discussion which require  particular care.

\smallskip 
 The  most general 
  intertwining of the  superconformal generators depends on three signs, which can be organized conveniently as follows \cite{bbdo}:
  \beq\label{intertwi}
  ( G_r ^1- i \eta_{\rm S}^1\, \tilde G_{-r}^1 ) I_{1,2}   =   \eta  I_{1,2}    ( G_r^2 - i \eta_{\rm S}^2\, \tilde G_{-r}^2 )\ . 
  \eeq
Here  $\eta_{\rm S}^1 , \eta_{\rm S}^2 = \pm 1$ define the unbroken supersymmetries of the  bulk theories,
while the overall sign $\eta=\pm 1$ accounts for  automorphisms  of the ${\cal N}=1$ algebra. Given a defect operator $I_{1,2}$ implementing 
the gluing condition for a given $\eta$, the defect operators $(-1)^{F_1+\wt F_1}I_{1,2}$ and $I_{1,2}(-1)^{F_2+\wt F_2}$ satisfy gluing conditions for the opposite $\eta$. They can be regarded as fusion products of the defect $I_{1,2}$
with the topological defects associated to $(-1)^{F_i+\wt F_i}$. 

For any given interface  
  the values of $\eta_{\rm S}^{1}$ and $\eta_{\rm S}^{2}$ are fixed,  whereas in order to implement the GSO projection  both signs of $\eta$  
 have to be taken into account. 
\smallskip

 Equation \eqref{intertwi},  together with the gluing conditions
  \eqref{gluingcondition} for the bosonic modes,    imply the gluing conditions 
    \beq\label{gluingconditionF}
 \vect{ \psi^1_r }{ -  i \eta_{\rm S}^1\, \tilde  \psi^1_{-r} } I_{12}  = 
  I_{12} \,  \eta  \Lambda\vect{ \psi^2_r }{ -  i \eta_{\rm S}^2\, \tilde  \psi^2_{-r} } 
    \eeq
for the fermions. Here $\Lambda$ is the same $O(1,1)$ matrix as for the bosons.  
To lighten the notation  we  absorb the various signs in  a  Lorentz
matrix  for the fermion fields, 
\beq\label{lambdaprime}
\Lambda_{\rm F} =  \eta \mat{1}{0}{0}{\eta_{\rm S}^1} \Lambda \mat{1}{0}{0}{\eta_{\rm S}^2} \ ,  
\eeq
in terms of which the gluing conditions take the simpler form
   \beq\label{gluingconditionFnn}
 \vect{ \psi^1_r }{ \ -  i  \, \tilde  \psi^1_{-r} } I_{12} = 
  I_{12}  \,   \Lambda_{\rm F}  \vect{ \psi^2_r }{ \ -  i \, \tilde  \psi^2_{-r} } 
  \, .  
\eeq

\vskip 1mm

  Folding CFT2 as in Section \ref{secBos}  amounts to applying the time-reversal 
transformation  $(\psi, \tilde \psi ) \to (\psi^*, \tilde \psi^* ) i\gamma^0$,  where the right-hand side is evaluated at 
time $-\tau$.  
Spelled out in terms of the modes this reads\footnote{We have fixed the arbitrary phase 
of the transformation so as to leave invariant
the Wick-rotated Majorana condition $\psi_r^* = i \tilde \psi_r$. }
\beq\label{spell}
\vect{ \psi^2_r }{  \tilde  \psi^2_{r} }  \to \vect{-i \tilde  \psi^2_{-r} }{ i  \psi^2_{-r} } \ . 
\eeq
Notice that this operation exchanges the type-0A with the type-0B models, \cf \eq{-1fdef}. 
The commutation relations \eqref{gluingconditionF}  turn into the boundary gluing 
  conditions 
\beq\label{fermbdgluing2}
\left[\vect{ \psi^1_{r}}{\psi^2_r }  +   i \OOO_{\rm F} \vect{  \tilde  \psi^1_{-r} }{  \tilde  \psi^2_{-r} }\right] 
\vert I_{12}^{\, (\eta)} \keti=0\, , 
\eeq
where  the orthogonal matrix $\OOO_{\rm F}$ is related to $\Lambda_{\rm F}$   as  in equation  \eqref{SintermsofO}.  
Notice for future reference that flipping the sign of $\Lambda_F$  
changes   the sign of the off-diagonal blocs of $\OOO_{\rm F}$, that is
it conjugates this latter matrix  with the matrix diag$(+1 , -1)$.

\vskip 1mm

The  general solution to  \eqref{fermbdgluing2}
is a linear combination of boundary  states   in the NS and  the R sectors: 
\beq\label{NSd1}
  \vert {\rm NS}, \OOO_{\rm F}   \rangle\hskip -0.7mm \rangle\  =     \prod_{r \in  {\mathbb{N}}-
{1\over 2}} e^{- i (\OOO_{\rm F})_{ij}  \psi^i_{-r} \tilde \psi^j_{-r}} \vert 0 \rangle_{\rm NS} 
  \ ,   
\eeq
\beq\label{RRd1}
  \vert {\rm R} ,  \OOO_{\rm F}   \rangle\hskip -0.7mm \rangle\  =    \prod_{r \in  {\mathbb{N} } }
 \sqrt{2}\,  e^{ -i (\OOO_{\rm F})_{ij}  \psi^i_{-r} \tilde \psi^j_{-r}} \vert   \OOO_{\rm F} \rangle_{\rm R}\ ,
 \eeq
where  $\vert   \OOO_{\rm F}  \rangle_{\rm R}$  is a  normalized
 Ramond ground state, 
  which  depends on $\OOO_{\rm F}$ 
  in a way that we will  specify. 
 
 Note   that mixed-sector interfaces, with CFT1  in the NS sector and CFT2   in the R sector or  vice versa,  
are only compatible with supersymmetry if the two sides in equation 
\eqref{intertwi} vanish separately. 
Such interfaces are totally-reflecting,  and we will not consider them here.

\vskip 1mm 

The Ramond ground states in  the folded theory represent the algebra of the zero modes
$\psi^j_0$  and  $ -i \tilde \psi^j_0$. This  is the Clifford algebra of $\RR^{2,2}$, so 
 these states transform as  a four-component $O(2,2)$ spinor.  
    The gluing conditions \eqref{fermbdgluing2} for the zero modes yield  two linear constraints, which
therefore determine uniquely  the  ground state  $\vert   \OOO_{\rm F} \rangle_{\rm R}$. 
  We can construct this state more explicitly starting with  the identity matrix,  $\OOO_{\rm F} = {\bf 1}$. 
  The  conditions  \eqref{fermbdgluing2} in this case  imply that 
  $ \vert   {\bf 1}  \rangle_{\rm R}$ is the (normalized) pure-spinor state:  
  \beq
  \gamma^{j=1,2}_+  \vert   {\bf 1}  \rangle_{\rm R} = 0 \ ,
   \qquad {\rm where}\ \ \gamma^j_\pm \eqdef  {1\over\sqrt{2}}\left( \psi_0^j \pm  i \tilde\psi_0^j\right)\ . 
  \eeq
Using the same notation as in   \eqref{rralgebra} we can  write
$ 
\vert  {\bf 1} \rangle_{\rm R} =  \vert  + +   \rangle_{\rm R}\, , 
$
where the two chiralities refer to the decomposition $O(2,2) \supset O(1,1) \times O(1,1)$. 
  The general Ramond ground state is obtained   by a spinor rotation:  
     \beq\label{RRgorundstate}
   \vert   \OOO_{\rm F} \rangle_{\rm R}\ =   S( \OOO_{\rm F} ) \vert  {\bf 1}  \rangle_{\rm R}\  , 
   \eeq
  where  $S(\OOO)$  denotes the spinor representation of   $\OOO$ 
 considered as an element of  the $O(2)$ subgroup of $O(2,2)$ 
 which  only acts on  the left  part of the spinor.\footnote{Because this subgroup is compact, 
 $ \vert   \OOO_{\rm F} \rangle_{\rm R}$ is  also a normalized state.}
    That  \eqref{RRgorundstate}  indeed enforces 
 the required  gluing conditions on  the zero modes 
 follows from the  identity:
     \beq
   \OOO^j_{\ l}\,  S(\OOO)  \psi_0^l \, S(\OOO)^{-1} =  \psi_0^j\ ,  
   \eeq
where we use the fact that $\sqrt{2} \psi_0^l$ obey the Clifford algebra of $\mathbb{R}^2$,  and are thus
represented by the  gamma matrices of  $O(2)$. 
   \vskip 1mm

   We can give an even more explicit  form of  the state \eqref{RRgorundstate}  
 by  first expressing   $S( \OOO_{\rm F})$ in terms of the  $O(2)$ generator  $i\psi_0^1 \psi_0^2$, 
then  using the fact that  $\gamma^j_+$ annihilates  $\vert  {\bf 1}  \rangle_{\rm R}$ .   
For instance, if $ \OOO_{\rm F}$ is a pure rotation by an angle $2 \vartheta$ this  
operation gives
$$
   \vert  \OOO_{\rm F}   \rangle_{\rm R} =  ( {\rm cos}\,  \vartheta \, {\bf 1}  +  2 \, {\rm sin}\,  \vartheta \, \psi_0^1 \psi_0^2 \, 
  ) \vert   + +    \rangle_{\rm R}  
  $$
  \beq\label{alternativeRR}
 =  {\rm cos}\,   \vartheta\, \vert   + +   \rangle_{\rm R} + {\rm sin}\,   \vartheta\, \vert   - -   \rangle_{\rm R}  
\ =\   {\rm cos}\,  \vartheta\  e^{ {\rm tan}\,   \vartheta\,   \gamma^1_- \gamma_-^2} \, \vert   
+ +    \rangle_{\rm R}\  .  
\eeq
\vskip 1mm \noindent
In case  $\OOO_{\rm F}$ is not continuously-connected to the identity, 
we  decompose it as 
a   rotation by an angle $2 \vartheta$ times 
a reflection
(of say direction 2). Using the reflection in spinor space,
  this gives
\beq\label{notcont}
   \vert  \OOO_{\rm F}  \rangle_{\rm R}\  = \ {\rm cos}\,  \vartheta\  e^{ {\rm tan}\,   \vartheta\, 
     \gamma^1_- \gamma_+^2} \, \vert   + -    \rangle_{\rm R}\ . 
\eeq

\smallskip

 One can obtain these formulae in a different way, which easily generalizes
 to higher dimensions, 
  by formulating the gluing conditions  \eq{fermbdgluing2} of the zero modes in terms of the $\gamma_\pm$:
        \beq\label{fermbdgluing5}
\left[ \vect{\gamma^1_{+}}{\gamma^2_{+}}  +   {\cal F}
 \vect{\gamma^1_{-}}{\gamma^2_{-}} \right]  \vert \OOO_{\rm F} \rangle_{\rm R}   
=0\, . 
\eeq
Here, ${\cal F}$ is the antisymmetric matrix defined by
\beq\label{FintermsofO}
\OOO_{\rm F} =  ({\bf 1}  +  {\cal F})^{-1} ({\bf 1} - {\cal F}) 
 \ \Longleftrightarrow\  {\cal F} =  ({\bf 1} - \OOO_{\rm F}) ({\bf 1}+\OOO_{\rm F})^{-1}\ . 
\eeq
 The normalized solution of equations   \eqref{fermbdgluing5} then reads
 \beq\label{newRR}
 \vert \OOO_{\rm F}  \rangle_{\rm R}\,  =  \,  [{\rm det}(1- {\cal F} )]^{-{1\over 2}} 
    \, {\rm exp}\left( - {1\over 2} {\cal F}_{jl}\,   \gamma_-^l \gamma_-^j \right)  \vert {\bf 1}  \rangle_{\rm R} \, . 
 \eeq 
This expression is again only valid  when  $\OOO_{\rm F}$ is in the identity compoment of $O(2)$.
If ${\rm det} \OOO_{\rm F} = -1$, one of its eigenvalues   is $-1$ and the denominator
in the right-hand-side of  \eqref{FintermsofO} is zero. In this case, we 
write  $\OOO_{\rm F}$ as
 a continuous rotation times a reflection. The effect of the latter is to  
 replace $\vert {\bf 1}  \rangle_{\rm R}$ by a pure spinor of opposite  $O(2,2)$ chirality. 
\vskip 1mm

Like their bosonic counterparts, also the
 fermionic boundary states \eqref{NSd1} and \eqref{RRd1}
  can be unfolded to defect operators using the behavior \eq{spell} of the fermionic modes under folding. 
 The result can be formally expressed as products $\prod_{r>0} I_{1,2}^{r,{\rm ferm}} 
  I_{1,2}^{0,{\rm ferm}}$ of exponentials, where
\beq\label{opsnonzeromodesf}
I_{1,2}^{r,{\rm ferm}} =
    {\rm exp}\hskip -1mm \left(  -i  \psi^1_{-r}\OOO_{11}  \tilde \psi^1_{-r} 
   +  \psi^1_{-r} \OOO_{12}\psi^2_r +  \tilde \psi^1_{-r}\OOO_{21}^t \tilde \psi^2_r + i   \psi^2_{r}\OOO_{22}^t  \tilde \psi^2_{r}  
    \right)   \ 
\eeq
with modes of CFT1 and CFT2 acting respectively on the left and right of maps on the fermionic ground states. 
The matrix $\OOO$ in this expression is the one pertaining to the fermions, $\OOO_{\rm F}$, but we have dropped the subscript
$F$ to uncharge the notation.  
Since the NS ground state is unique, the corresponding map is trivial: 
 \beq\label{zeroNStrivial}
I_{1,2}^{0,{\rm NS}}=|0\ket^1_{\rm NS}\,{}_{\rm NS}^{\phantom{,,}2}\bra0|\,. 
\eeq

The story is less trivial  in the Ramond sector where the zero-mode map 
 can be written as
  \beq\label{R0int}
 I_{1,2}^{0,{\rm R}}=\sqrt{   \vert \sin(2 \vartheta) \vert }\  \imath_{1,2}^{\rm R}\, S(\Lambda_{\rm F})\,.  
 \eeq
 Here 
 $S(\Lambda_{\rm F} )$ is the spinor representation of the $O(1,1)$ matrix $\Lambda_{\rm F}$,  and $\imath_{1,2}^{\rm R}$
 is the isomorphism between Ramond ground states of CFT2 and CFT1, 
  \beq
\imath_{1,2}^{\rm R}=|+\ket^{\!1}_{\rm R}\,{}^{2\!}_{\rm R}\!\bra+|\,+\,|-\ket^{\!1}_{\rm R}\,{}^{2\!}_{\rm R}\!\bra -| \ . 
\eeq 
 That  \eqref{R0int} is,  up to normalization,  the correct map
follows  directly from the gluing conditions \eqref{gluingconditionFnn}  for the zero modes, and from
the $O(1,1)$  invariance  of the   gamma matrices. 
To fix the normalization, one can unfold for instance the ground state  \eqref{notcont},  
which corresponds to a gluing matrix $\Lambda_{\rm F}$ of unit determinant. 
Using the fact that $|\pm \ket^{\!2}_{\rm R}$  unfolds  
 to ${}^{2\!}_{\rm R}\!\bra\mp| $, as dictated by the  unfolding \eqref{spell} for the zero modes, one finds
  \beq\label{68}
\vert \OOO_{\rm F}  \rangle_{R} \ \mapsto \  {\rm cos}  \vartheta |+\ket^{\!1}_{\rm R}\,{}^{2\!}_{\rm R}\!\bra+|\,+
\,  {\rm sin}  \vartheta |-\ket^{\!1}_{\rm R}\,{}^{2\!}_{\rm R}\!\bra -|
=
 \sqrt{\vert {\rm sin} (2 \vartheta) \vert \over 2} \,  \imath_{1,2}^{\rm R} \, S( \Lambda_{\rm F})\ .  
\eeq
The second step follows from the fact that ${\rm det} S( \Lambda_{\rm F}) = \pm 1$  for $ \vartheta \in [0, \pm \pi/2 ]$.
 Indeed, as was explained in Section \ref{secBos}, the matrix
  $ \Lambda_{\rm F}$ corresponding to a rotation angle 
  $ \vartheta \in [0, \pm \pi/2 ]$ has the property that $\pm  \Lambda_{\rm F}$ is
   continuously connected to the identity. 
 Thus   ${\rm det} S(\pm \Lambda_{\rm F}) = 1$, and since $S(- {\bf 1}) = \left( \begin{smallmatrix}  1 & 0 \\ 0 & -1 \end{smallmatrix}\right)$
 we deduce that  ${\rm det} S( \Lambda_{\rm F}) = \pm 1$ as claimed. 
 Multiplying by an extra $\sqrt{2}$ from  \eqref{RRd1},   gives the normalization of the zero-mode map  in
 \eqref{R0int}.

%%%%%%%%%%%%%%%%%%%%%%%%%%%%%%%%%%%
%%%%%%%%%%%%%%%%%%%%%%%%%%%%%%%%%%%

\subsection{Fermion-parity projections}
\label{projsection}

Let us take stock of the results of the previous subsection. For any choice of the bosonic gluing matrix $\Lambda$,
or of its orthogonal counterpart $\OOO$, and for any choice of the supersymmetry signs $\eta, \eta_S^j$, which enter
in the gluing condition \eqref{intertwi},  we have constructed the fermionic boundary states 
$ |{\rm NS},\OOO_{\rm F}\keti$ and $|{\rm R},\OOO_{\rm F}\keti$  that implement these gluing conditions in
the Neveu-Schwarz and  Ramond sectors. Unfolding yields the corresponding  interface operators
\beq\label{ferminterfacemodes}
I_{1,2}^{\rm NS}=\prod_{r\in\NN-{1\over 2}}I_{1,2}^{r,{\rm ferm}}I_{1,2}^{0,{\rm NS}}\,,\quad{\rm and}\quad
I_{1,2}^{\rm R}=\prod_{r\in\NN}I_{1,2}^{r,{\rm ferm}}I_{1,2}^{0,{\rm R}}\,.
\eeq
In the unprojected theory there is only a NS sector, so   the complete interface operators read 
\beq\label{completeinterface}
I_{1,2}^{\rm full} (\Lambda,\varphi, \eta_{\rm S}^i, \eta)=I_{1,2}^{\rm bos}(\Lambda,\varphi)\otimes
I_{1,2}^{\rm NS}(\Lambda_{\rm F})\ . 
\eeq
We will now  implement the fermion-parity or GSO projections, which  add a twisted (Ramond)
sector to the interface operators. 
\smallskip

This is similar to the discussion of the projection of boundary states   in Section \ref{DNstates}. The only difference is
that now we have to project in both CFT1 and CFT2 separately. Thus, we have to take a $\ZZ_2\times\ZZ_2$ orbifold, and we have
 four possible projections given by the choice of 0A or 0B orbifolds in each of the two CFTs.  We distinguish  these possibilities pairwise
by defining the new sign 
\beq
\zeta = \begin{cases} + 1 & \ \ {\rm if\  CFT1\ and\ CFT2\ are\ of\ same\ GSO\ type},  \\ -1 & \  \  {\rm if\  CFT1\ and\ CFT2\ are\ of\ opposite\ type}. \end{cases}
\eeq
 In the following discussion we will perform the projection on the boundary states in the folded picture. 
For this it is important to recall  that under folding of CFT2 0A and 0B models are interchanged.

\smallskip
We will perform the orbifold in two steps, first by projecting with respect to the diagonal $\ZZ_2$ generated by $(-1)^{F+\wt{F}}:=(-1)^{F_1+\wt{F}_1+F_2+\wt{F}_2}$ and then by projecting with respect to the remaining $\ZZ_2$ generated by $(-1)^{F_1+\wt{F_1}}$. 

\smallskip

The operator $(-1)^{F+\wt{F}}$  leaves the NS state  invariant. Hence, as in Section  \ref{DNstates} we resolve
it by the addition of the twisted,  \ie Ramond-Ramond sector: 
 \beq\label{fullfermbd}
|\OOO_{\rm F},\pm\keti_{\rm ferm}={1\over\sqrt{2}}\left(|{\rm NS},\OOO_{\rm F}\keti\pm|{\rm R},\OOO_{\rm F}\keti\right)\,.
\eeq
Next, we have to implement the GSO projection. Since $(-1)^{F+\wt{F}}$ commutes with the exponentials in
\eqref{NSd1} and \eqref{RRd1}, its action on the boundary state is determined by the action on the respective ground states.
 Using \eq{-1fdef} one finds
$$
(-1)^{F + \tilde F}  \vert {\rm NS}, \OOO_{\rm F}   \rangle\hskip -0.7mm \rangle\  =  
\vert {\rm NS}, \OOO_{\rm F}   \rangle\hskip -0.7mm \rangle\    
$$
\beq 
{\rm and}\qquad 
(-1)^{F + \tilde F}  \vert {\rm R}, \OOO_{\rm F}   \rangle\hskip -0.7mm \rangle\ =
-\zeta\,{\rm det} \OOO_{\rm F}   \vert {\rm R}, \OOO_{\rm F}   \rangle\hskip -0.7mm \rangle\  , 
\eeq
\vskip 1mm\noindent  
 where in the Ramond case, up to the factor $-\zeta$ which comes from the choice of orbifold and the folding, $(-1)^{F + \tilde F}$ is  the chirality of the ground state spinor
which equals the determinant $\det(\OOO_{\rm F})$. 
\smallskip

We thus see that the  Ramond contribution to a boundary state survives the $(-1)^{F+\wt{F}}$ projection if
$\det(\OOO_{\rm F})=-\det(\Lambda_{\rm F})=-\eta_{\rm S}^1\eta_{\rm S}^2\det(\Lambda)=-\zeta$, or equivalently if
\beq\label{epsiloncond}
 \varepsilon \, =\,  \eta_{\rm S}^1\, \eta_{\rm S}^2\,  \zeta\ . 
\eeq
When this  condition is  satisfied the interface  has a non-trivial  R component -- we say that it
      is ``charged''.  
     Otherwise   the interface is ``neutral'', i.e.  it projects out all the  Ramond  states.
         
 \vskip 1mm
 
 The situation is summarized in Table \ref{default1}. 
 For any choice of    theories on either side,     and for any choice of the preserved 
 superconformal algebras,  there exists both a   (doubly-degenerate) charged
 interface with $\varepsilon =  {\rm det} \Lambda$ obeying the condition \eqref{epsiloncond},  
 and a neutral interface that violates  this condition. 
  We have assumed in the table that CFT1 and CFT2 are
    of the same type, so that $\zeta = +1$. 
 Thus $\eta_{\rm S}^1\eta_{\rm S}^2 $ equals $ \varepsilon$ in the charged case, and $-\varepsilon$ in the neutral one.
 For theories of opposite type the signs are reversed.
 
\vskip 1mm

\begin{table}[htdp]
\begin{center}
\begin{tabular}{c||c|c}
\, & {\rm D1}  &  {\rm D2/D0} \\
\hline\hline  
 {\rm charged} &  + &   - \\
 \hline 
  {\rm neutral} & -  &  +
\end{tabular}
\end{center}
\caption{\small{The value of $\eta_{\rm S}^1\eta_{\rm S}^2$ that determines
which superconformal algebras are preserved by an
 interface between two theories of the same type  (both type-0A or both type-0B).  The geometric
interpretation of the folded boundary condition depends only on $\varepsilon$,  as discussed in 
the previous subsection. }}
\label{default1}
\end{table}

\vskip 1mm

 The resulting boundary states in the projected theory arise by  taking the appropriately normalized orbits  
of  \eq{fullfermbd} under the orbifold group, \cf the discussion in Section \ref{DNstates}. 
This yields
    \beq
   \vert  \OOO_{\rm F};  {\rm charged}, \pm \rangle\hskip -0.7mm \rangle_{\rm ferm}
   =    {1\over \sqrt{2}}  \left(\vert {\rm NS}, \OOO_{\rm F}   \rangle\hskip -0.7mm \rangle\  \pm 
 \vert {\rm R} , \OOO_{\rm F}   \rangle\hskip -0.7mm \rangle\right)    
 \ \  \ \ {\rm if}\ \  {\rm det} \OOO_{\rm F}  =  - \zeta , \nonumber
   \eeq
\beq\label{ferm43}
{\rm and} \ \ \   \vert  \OOO_{\rm F}; {\rm neutral} \rangle\hskip -0.7mm \rangle_{\rm ferm}
   =     \vert {\rm NS}, \OOO_{\rm F}   \rangle\hskip -0.7mm \rangle
    \ \  \ \ {\rm if}\ \   \,  {\rm det} \OOO_{\rm F}  =   \zeta\, .
\eeq
 When combined with 
 bosonic boundary states, the above states  correspond to GSO projected superconformal boundary conditions
 in $c=3$ SCFTs.  The sign $\zeta$ determines whether these c=3 theories are of type 0A or type 0B. 
 However, such states do not unfold to proper interfaces among local theories, because the operator  
 $(-1)^{F+\wt{F}}$ is a non-local operator after unfolding. 
  In order to obtain proper interfaces between separately GSO projected theories one has to perform the remaining non-diagonal $\ZZ_2$ orbifold,
   generated for instance by $(-1)^{F_1+\wt F_1}$.
\smallskip

  This second orbifold operation is  simple if  we exclude perfectly-reflecting defects, \ie those
  for which  $\OOO$ is a diagonal matrix. Namely, the orbifold acts freely on the boundary states: 
\beq\label{signsamb}
(-)^{F_1 + \tilde F_1} \vert {\rm NS \ or\  R},  \OOO(\Lambda_F) \rangle\hskip -0.7mm \rangle  =
\vert {\rm NS\  or \ R}, \OOO(-\Lambda_F) \rangle\hskip -0.7mm \rangle \ ,  
\eeq
as follows  from the definitions  \eqref{NSd1} and \eqref{RRd1} of these states,\footnote{Actually, there
 is an overall  sign   in the R sector which determines whether CFT1 is type 0A or 0B.
 Since   $S(\OOO)$ is only defined up to a sign for
 given $\OOO$, we can always absorb the above overall sign by defining the Ramond states such
  that the relation  \eqref{signsamb}  holds.} 
 and the fact that 
 $\OOO(\Lambda_F) =\OOO(-\Lambda_F)$
  only if $\OOO$ is diagonal (\cf  equations \eqref{SintermsofO} and \eqref{OintermsofS}). 
Furthermore,   twisted sectors of this second orbifold would correspond to having CFT1  in the NS (R) and CFT2  in the R (NS) sector.
As mentioned already in Section \ref{secfermionic}, such sectors are only possible for perfectly-reflecting defects, which we do not consider here. 
 Thus,  the second orbifold construction  simply gives
 \beq
 |\OOO ;   {\rm any}  \keti_{\rm ferm}^{\rm proj}  =
{1\over\sqrt{2}}\left(|\OOO(\Lambda_F);   {\rm any}   \keti_{\rm ferm}+|\OOO(-\Lambda_F);   {\rm any}   \keti_{\rm ferm}\right)\, , 
\eeq
where ``any'' denotes  the three possibilities  in \eqref{ferm43}. Note that to avoid cumbersome notation, we do  not
indicate  here  the dependence on $\eta_{\rm S}^i$,  even though these signs determine whether the interface is neutral or charged. Charged and neutral interfaces have different $g$-factors, for the charged ones one obtains $g_{{\rm charged} \pm}=1$, whereas $g_{{\rm neutral}}=\sqrt{2}$. 
\smallskip
\smallskip

  Let us now collect our results. The complete projected  interface operators for given GSO types of CFT1 and CFT2 can be written as: 
\beq\label{completeinterface1}
I_{1,2}^{\rm full} (\Lambda,\varphi, \eta_{\rm S}^i)=I_{1,2}^{\rm bos}(\Lambda,\varphi)\otimes
I_{1,2}^{\rm ferm}(\Lambda, \eta_{\rm S}^i)\ ,
\eeq
where the fermionic interface is charged if \ $\det\Lambda=\zeta  \eta_{\rm S}^1 \eta_{\rm S}^2$: 
\beqn\label{completeinterface2}
I_{1,2}^{{\rm ferm}, \, c \pm }(\Lambda, \eta_{\rm S}^i )= {1\over {2}}\left(
I_{1,2}^{\rm NS}(\Lambda_{\rm F})\pm I_{1,2}^{\rm R}(\Lambda_{\rm F})\right)  + (\eta\to -\eta) \ , 
\eeqn
 or  neutral if   $\det\Lambda= -\zeta  \eta_{\rm S}^1 \eta_{\rm S}^2$:
\beqn\label{completeinterface3}
I_{1,2}^{{\rm ferm}, \, n }(\Lambda, \eta_{\rm S}^i )= 
{1\over\sqrt{2}}\,   I_{1,2}^{\rm NS}(\Lambda_{\rm F})+ (\eta\to -\eta) \ . 
\eeqn
  From these normalizations, and taking into account that  the NS ground state contributes equally for the two values of $\eta$, 
 one finds the following relations for the $g$-factors of the projected interfaces:
$g = g_{\rm bos}$ in the charged case, 
and $g = \sqrt{2}\, g_{\rm bos}$ in the neutral one.  
 
 \smallskip
 
 For applications to type-II superstring theory separate GSO projections for 
 left-  and   right-moving fermions have to be imposed.  This introduces additional twisted sectors -- mixed NS-R and R-NS sectors of CFT1 and CFT2. 
   Following the same logic as above, only  interfaces which commute with the action of $(-1)^{F}$  acquire intertwiners for these mixed sectors; all other interfaces
   map the NS-R and R-NS states of CFT2  to zero. 
  
Interfaces  commuting with $(-1)^{F}$  cannot mix the left and right worldsheet fermions, 
     \ie the fermion-gluing
   matrix $\Lambda_{\rm F}$ and by supersymmetry also the gluing matrix $\Lambda$  for the bosonic currents,  \cf \eqref{lambdaprime},   
have to be elements of  $O(1)\times O(1)$. Hence, such interfaces are topological. 
   In  \cite{Satoh:2011an} it was argued that in the Green-Schwarz formulation  space-time supersymmetric interfaces
   are either topological or totally reflecting interfaces.  In the NSR formulation, on the other hand, the topological property
   follows  from the 
       requirement that the interfaces do not project out the mixed NS-R and R-NS sectors, which correspond to  space-time fermions.

As alluded to above, the GSO projection of the fermionic part of the theory is nothing but the Ising model, for which the conformal 
defect lines have been known.  
Let us briefly comment on relation of the interfaces $I^{\rm ferm}$ 
to these known defects.
The simplest of those are the topological ones, which can be constructed using the tools described in 
\cite{Petkova:2000ip}. Here, the modular invariant for the theory on either side of the defect is diagonal, and the defects carry the same labels $a$ as primary fields (in our case $a$ runs over the representations corresponding to the weights $h=0,1/2, 1/16$). The defects $I_a$ act on a bulk field in the representation $(b,\tilde{b})$ by multiplication by the quantum dimensions\footnote{$S$ denotes the modular $S$-matrix.}
\begin{equation}
f_{a,b}= \frac{S_{ab}}{S_{0b}} \ .
\end{equation}
Being topological, these defects act naturally on other interfaces via fusion. In particular, the defect labelled by $0$ is the identity defect, whereas the one labelled by $1/2$ acts as the identity in the NS sector, but inverts the  Ramond charge. Finally, $I_{1/16}$ does not couple to Ramond ground states and hence maps charged interfaces to uncharged ones. 

To translate to our language, we first pick $\zeta=1$ to ensure equal modular invariants on either side of the interface, and set $\eta_{\rm S}^1=\eta_{\rm S}^2$.
The fermionic interfaces $I^{\rm ferm}$ are topological if and only if the $O(1,1)$-matrix $\Lambda$ is diagonal, \ie $\Lambda=\pm {\bf 1}$ or $\Lambda=\pm{\rm diag}(1,-1)$, where the first case corresponds to charged and the second to uncharged interfaces. 
One can then identify
\begin{eqnarray*}
I_0&=&I^{{\rm ferm}, c+}(\Lambda={\bf 1})\\
I_{1/2}&=& I^{{\rm ferm},c-}(\Lambda={\bf 1})\\
I_{1/16}&=& I^{{\rm ferm}, n}(\Lambda={\rm diag}(1,-1))\,.
\end{eqnarray*}
General conformal defect lines in the Ising model have been constructed in 
\cite{Oshikawa:1996dj,Quella2}, where the tensor product of two Ising models was identified with a $\ZZ_2$ orbifold of a free boson compactified on a circle of radius 1. Via the folding trick, defects of the Ising model were constructed as boundary conditions for a single free boson on this orbifold. The latter come in two families, Dirichlet and Neumann boundary conditions. Both families are parametrized by a circle valued parameter, the position of the Dirichlet brane and the Wilson line parameter on the Neumann brane, respectively.

In our formalism, the fermionic interfaces in the GSO projected purely fermionic theory are parametrized by $\Lambda\in PO(1,1)=O(1,1)/\{\pm 1\}$. This group has two one-dimensional components, distinguished by the sign of $\det(\Lambda)$, \cf \eq{218}. The interfaces with $\det(\Lambda)=1$ are charged, and correspond to the Dirichlet boundary conditions of \cite{Oshikawa:1996dj,Quella2}. The interfaces with $\det(\Lambda)=-1$ on the other hand 
are neutral and correspond to the Neumann boundary conditions. 
Inclusion of purely reflective interfaces compactifies the two components of $PO(1,1)$ to circles parametrized by the angle variables $2\vartheta$ from \eq{218}, which corresponds to position and Wilson line parameters of the Dirichlet and Neumann boundary states, respectively.

    %%%%%%%%%%%%%%%%%%%%%%%%%%%%%%%%%%%%%
%%%%%%%%%%%%%%%%%%%%%%%%%%%%%%%%%%%%%

\section{Fusion and the defect monoid} \label{secfusion}

We now turn to the computation of fusion of the supersymmetric  
  interfaces constructed in the previous section. 
 The fusion of $\wh u(1)^2$ preserving bosonic interfaces between circle theories has already been 
calculated in \cite{arXiv:0712.0076}.
Because of a divergent Casimir energy 
this operation is in general singular, and requires regularization and renormalization. Only 
when one of the interfaces is topological, meaning that it commutes with both  left and right Virasoro algebras, fusion is
finite. 
 In this section we extend the analysis of  \cite{arXiv:0712.0076}  to the supersymmetric case. 
As anticipated  in \cite{bbdo},    ${\cal N} = 1$ supersymmetry  renders  the fusion  
of these free-field interfaces non-singular, 
 because the divergent Casimir energies of bosons and fermions cancel out.\footnote{In interacting SCFTs, or for more
general boundary conditions,   the interface self-energy is not the only potential counterterm. In principle, logarithmic divergences
 are allowed by  ${\cal N} = 1$ supersymmetry and cannot in general be excluded.}

 \subsection{Classical versus quantum}

\begin{figure}[h!]
\begin{center}
\includegraphics[height=9.5cm]{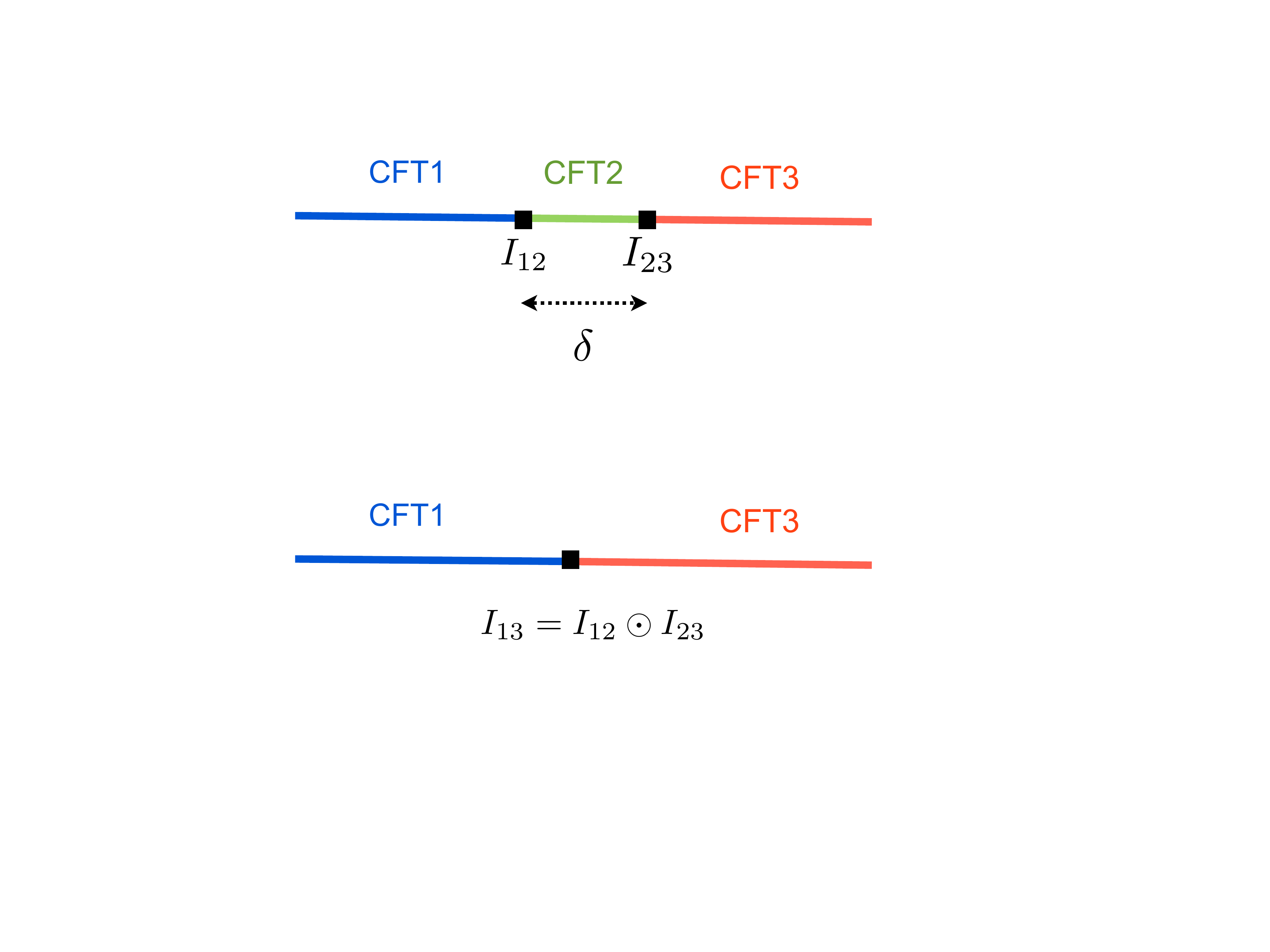}
\vskip -2 cm
\caption{\small The fusion of two interfaces corresponds to taking the size, $\delta$,  of the middle region to zero.
Only the $\tau$ axis is drawn in the figure. The $\sigma$ coordinate parametrizes either a circular space, or a  periodic Euclidean time.}\label{fusion}
\end{center}
\end{figure}

Consider 
three conformal field theories  (CFT3, CFT2 and CFT1)    on the cylinder  separated, 
  at  $\tau= 0$ and $\tau = \delta$,  by interfaces $I_{23}$  and $I_{12}$.  Fusion amounts
  to shrinking the middle region to zero size $\delta\to 0$, 
  so that   CFT1 and CFT3 are separated by a new local interface
  which we denote $ I_{12}\odot I_{23}$. This is shown schematically in figure \ref{fusion}. 

On the level of classical gluing conditions    
  fusion amounts to multiplication of  $O(1,1\vert  \mathbb{R})$  matrices. 
Indeed, let  $\Lambda$ and $\Lambda^\prime$ be the gluing matrices for the left and right
 $\widehat u(1)$ currents imposed by the interfaces    $I_{23}$  and $I_{12}$, so that 
\beq
 \vect{ a_n^1}{  -\tilde  a_{-n}^1} \, = \, \Lambda^\prime \vect{ a_n^2}{  -\tilde  a_{-n}^2} \Bigl\vert_{\tau =\delta}\  \  \ {\rm and}\ \ \
\vect{ a_n^2}{  -\tilde  a_{-n}^2} \, = \, \Lambda \vect{ a_n^3}{  -\tilde  a_{-n}^3} \Bigl\vert_{\tau =0}\ .  
\eeq
Taking $\delta\to 0$ leads, by continuity,  to the gluing condition
\beq
 \vect{ a_n^1}{  -\tilde  a_{-n}^1} \, = \, \Lambda^\prime\Lambda  \vect{ a_n^3}{  -\tilde  a_{-n}^3} \Bigl\vert_{\tau =\delta = 0}\ . 
\eeq
Likewise for the fermions, fusion leads to the gluing condition
 \beq\label{gluingconditionFcomp}
 \vect{ \psi^1_r }{ -  i \eta_{\rm S}^1\, \tilde  \psi^1_{-r} }   \, = \, \eta^\prime \eta\,  \Lambda^\prime\Lambda
   \vect{ \psi^3_r }{ -  i \eta_{\rm S}^3\, \tilde  \psi^3_{-r} }\Bigl\vert_{\tau =\delta =  0} \ , 
\eeq
provided  the two interfaces preserve the same supersymmetry in the middle region, so that
the factors of $\eta_S^2$ cancel out, \cf  equations \eqref{lambdaprime} and \eqref{gluingconditionFnn}. 
 In the sequel we will always assume this to be the case.

 \vskip 1mm

In the quantum theory, fusion is defined by  the composition of  interface operators, which, as alluded to above, 
requires regularization. One defines
  \beq\label{fusiondefn}
     I_{12}\odot I_{23}\, :=   \,   {\rm lim}_{\delta\to 0} \, {\cal R}_\delta [  I_{12}\, e^{-\delta H}  I_{23} ] \, ,  
     \eeq
     where $H\equiv L_0 + \tilde L_0$ is the Hamiltonian of CFT2.  (We drop the  -${c\over 12}$ term which commutes with the interface operators and therefore does not contribute to our analysis.)
      ${\cal R}_\delta$  denotes the renormalization procedure which, by the usual arguments of quantum field theory,
 can be achieved by  local counterterms. 
For the superconformal interfaces we study here, fusion turns out to 
 be finite without  renormalization, so that the symbol ${\cal R}_\delta$ can be omitted. 

\vskip 1mm

 Although the gluing conditions still compose according to
 multiplication in $O(1,1|\RR)$,   fusion of the quantum interfaces is much more subtle. 
Firstly,  as we have seen in Section \ref{secBos}, 
 the quantization of the $u(1)$ charges  restricts the gluing matrices to lie in dense subsets of $O(1,1|\RR)$
which are  isomorphic to the rational subgroup $O(1,1|\mathbb{Q})$. Furthermore, in order to respect
charge quantization the interface operators have to project to sublattices of 
the charge lattice, while the remaining sectors are projected out. 
If this sublattice is a proper sublattice, the respective interface
is not invertible. As a result,  the classical $O(1,1|\RR)$ group is replaced in the quantum theory by a semi-group. Moreover, quantum interfaces can be superposed, \ie the associated operators are added. In particular the superposition of interfaces with different values of the classically irrelevant moduli $\varphi$ can give rise to non-trivial effects. 
      
       For all these reasons the algebraic structure of quantum interfaces is richer and more interesting
 than that of their classical counterparts. This will be discussed in the rest of this paper.

%%%%%%%%%%%%%%%%%%%%%%%%%%%%%%%%%
%%%%%%%%%%%%%%%%%%%%%%%%%%%%%%%%%

 \subsection{Intertwiners for non-zero modes}
\label{section3.4}

 We will perform the fusion \eqref{fusiondefn} of the superconformal 
  interfaces by  separately composing  the bosonic and fermionic interface operators.
   According to
\eq{bosinterface} and \eq{ferminterfacemodes},  these latter can be written as tensor products of maps on the different frequency 
sectors of the (free) CFTs:
\beq
I_{1,2}=   \prod_{n>0} I_{1,2}^n  \,  I_{1,2}^0\equiv I_{1,2}^>I_{1,2}^0\,.
\eeq
As derived in Section \ref{sec3/2} the $I_{1,2}^n$ for $n>0$ can be expressed as exponentials of quadratic expressions 
of the bosonic,  respectively fermionic,  modes, \cf \eq{opsnonzeromodesb} and \eq{opsnonzeromodesf}.  
We recall that operators of CFT1 act on the zero-mode part 
from the left while  the operators of CFT2 act from the right.
\smallskip 
 
In order to obtain \eq{fusiondefn}, we first calculate $I_{1,2}^n e^{-\delta H} I_{2,3}^n$ for the tensor factors.
The bosonic expressions can be evaluated along the lines of 
 \cite{arXiv:0712.0076}. 
Pushing the Hamiltonian in the product
  $ I_{1,2}^{\, n, {\rm bos}} e^{-\delta H}  I_{2,3}^{\, n, {\rm bos}}$ to the left
 multiplies   the oscillators $a^2_n$ and $\tilde a^2_n$ in $I_{1,2}^{\, n, {\rm bos}}$ by a factor $e^{-\delta n}$.  
  Furthermore, the oscillators of CFT1 and CFT3 commute with every other operator   in this calculation,  and can be
 treated  as c-numbers.  This leaves us with the ground state matrix element of  exponentials
 that are either  linear or quadratic in the oscillators of CFT2.  The identity
 \beq\label{idd1}
 {\rm exp}\,\left( {1\over n} v a_{n}\right) \, f(a_{-n}) \, =   \, f(a_{-n}+ v)\,  {\rm exp}\,\left( {1\over n} v a_{n}\right) \ , 
 \eeq
 valid for any analytic function $f$ and any  commuting operator $v$, allows us to push to the right in  the  matrix element all  
 linear exponentials.
 We can then rearrange the quadratic terms with the use of the identity\footnote{The manipulations in this subsection are valid
 if the currents, and their modes $a_n$ and $\tilde a_n$, are $d$-dimensional vectors, so that  $M^\prime$ and $M$ are matrices.} 
$$
 \bra 0| \  {\rm exp}\,\left( {1\over n} a_{n} M^\prime\,  \tilde a_n \right)\,  {\rm exp}\,\left( {1\over n} a_{-n} M \,  \tilde  a_{-n} \right) 
$$
\beq\label{idd2}
=  \bra 0| \ {\rm det} (1- M^\prime M^T)^{-1}\, {\rm exp}\left( {1\over n}  a_{n}  (1 - M^\prime M^T)^{-1} M^\prime\,  \tilde a_{n}\right) \ . 
\eeq
  Finally, pushing the ensuing quadratic  exponential through the   linear terms on its right, and doing
 some straightforward algebra,   leads   to the following  result for the product:\footnote{For the calculation we will indicate the dependence of the interfaces on the orthogonal matrices $\OOO=\OOO(\Lambda)$ instead of the $O(d,d)$-matrices $\Lambda$.}
  \beq\label{bosoniccomp}
  I_{1,2}^{\, n, {\rm bos}}(\OOO^\prime)  e^{-\delta H}  I_{2,3}^{\, n, {\rm bos}}(\OOO) =  {\rm det} (1-  e^{-2n\delta} \OOO_{11}  \OOO_{22}^\prime)^{-1}\, 
   I_{1,3}^{\, n, {\rm bos}}(\OOO^{\prime\prime}(e^{-\delta n})) \ , 
 \eeq
 with
 \beq\label{compmat}
{ \scriptsize
  \OOO^{\prime\prime}(x)=\mat{\OOO\p_{11}+x^2\OOO\p_{12}(1-x^2\OOO_{11}\OOO\p_{22})^{-1}\OOO_{11}\OOO\p_{21}}
  {x\OOO\p_{12}(1-x^2\OOO_{11}\OOO\p_{22})^{-1}\OOO_{12}}
 {x\OOO_{21}(1-x^2\OOO\p_{22}\OOO_{11})^{-1}\OOO\p_{21}}
 {\OOO_{22}+x^2\OOO_{21}(1-x^2\OOO\p_{22}\OOO_{11})^{-1}\OOO\p_{22}\OOO_{12}} \, .
 }
 \eeq
 Collecting all the positive-frequency contributions of the bosonic intertwiners to \eq{fusiondefn} we obtain
 \beq\label{boscomp2}
 I_{1,2}^{>,{\rm bos}}(\OOO\p)e^{-\delta H}I_{2,3}^{>,{\rm bos}}(\OOO)=
 \prod_{n>0}\det(1-e^{-2\delta n}\OOO_{11}\OOO\p_{22})^{-1}I_{1,3}^{n, \rm bos}(\OOO^{\prime\prime} (e^{-\delta n}))\,.
 \eeq
In the limit  $\delta\to 0$ the matrices $\OOO^{\prime\prime}(e^{-\delta n})$ converge to the orthogonal matrix associated via \eq{SintermsofO}
 to the product of the gluing conditions $\Lambda\p$ and $\Lambda$, 
 \beq
 \OOO^{\prime\prime}(e^{-\delta n})\stackrel{\delta\to 0}{\longrightarrow}\OOO(\Lambda\p\Lambda)\ . 
 \eeq
 The product of determinants,  on the other hand,  exhibits a singular behavior  in this limit 
 due to a divergent Casimir energy  \cite{arXiv:0712.0076}. 
\smallskip
 
  Repeating the calculation for the fermionic intertwiners yields
   \beq\label{fermcomp}
     I_{1,2}^{\, r, {\rm ferm}}(\OOO^\prime)  e^{-\delta H}  I_{2,3}^{\, r, {\rm ferm}}(\OOO) =  
     {\rm det} (1-  e^{-2r\delta} \OOO_{11}  \OOO_{22}^\prime) \, 
   I_{1,3}^{\, r, {\rm ferm}}({\OOO}^{\prime\prime}(e^{-\delta r})) \ , 
   \eeq
  which combines to 
  \beq\label{fermcomp2}
 I_{1,2}^{>,{\rm ferm}}(\OOO\p)e^{-\delta H}I_{2,3}^{>,{\rm ferm}}(\OOO)=
 \prod_{r>0}\det(1-e^{-2\delta r}\OOO_{11}\OOO\p_{22})I_{1,3}^{r, \rm ferm}(\OOO^{\prime\prime}(e^{-\delta r}))\ 
 \eeq
 for the positive-frequency contributions to the fusion \eq{fusiondefn}.
The useful fermionic identities, analogous  to \eqref{idd1} and \eqref{idd2}, are
    \beq\label{iddf1}
 {\rm exp}\,(   \chi \psi_{r}) \, f(\psi_{-r}) \, =   \, f(\psi_{-r}+ \chi)\,  {\rm exp}\,(  \chi \psi_{r})  
 \eeq
for $\chi$  an operator  anticommuting with the fermionic oscillators, 
and  
 $$
\bra 0| \  {\rm exp}\,(   \psi_{r} M^\prime\,  \tilde \psi_r )\,  {\rm exp}\,(\psi_{-r} M \,  \tilde  \psi_{-r} ) 
$$
\beq\label{iddf2}
=  \bra 0 | \ {\rm det} (1- M^\prime M^T) \, {\rm exp}\left( \psi_{r}  (1 - M^\prime M^T)^{-1} M^\prime\,  \tilde \psi_{r}\right) \ . 
\eeq
Note that the determinant factors in expression \eq{fermcomp2} appear with opposite exponent as the ones in the corresponding bosonic formula \eq{boscomp2}. 
\smallskip

  When composing two superconformal interfaces, one should   replace 
  the  matrices   $\OOO\p$  and $\OOO$ in the expression \eqref{fermcomp2}  
   by the fermion-gluing matrices $\OOO\p_{\rm F}$  and $\OOO_{\rm F}$.  Nevertheless, 
   the determinant  that enters  in the formulae for the bosons and fermions  is the same. 
  Indeed, let  $(\eta\p, \eta_{\rm S}^1, \eta_{\rm S}^2)$ be the signs associated
with $I_{12}$, and  $(\eta, \eta_{\rm S}^2, \eta_{\rm S}^3)$  those associated with $I_{23}$, \cf 
\eqref{lambdaprime}. Then from \eqref{SintermsofO} we find: 
   \beq\label{SintermsofOa}
\OOO_{\rm F} \equiv \OOO(\Lambda_{\rm F}) =\mat{\eta_{\rm S}^2\,  \Lambda_{12} \Lambda_{22}^{-1}}
 {\eta\, \Lambda_{11}- \eta\,  \Lambda_{12}\Lambda_{22}^{-1}\Lambda_{21}} {\eta \eta_{\rm S}^2\eta_{\rm S}^3\, \Lambda_{22}^{-1}} 
{-\eta_{\rm S}^3\, \Lambda_{22}^{-1} \Lambda_{21}} \, , 
\eeq 
 and 
  \beq\label{SintermsofOb}
\OOO\p_{\rm F} \equiv  \OOO(\Lambda\p_{\rm F}) =\mat{\eta_{\rm S}^1\,  \Lambda\p_{12} (\Lambda\p_{22})^{-1}}
 {\eta\p\, \Lambda\p_{11}- \eta\p\,  \Lambda\p_{12}(\Lambda\p_{22})^{-1}\Lambda\p_{21}} {\eta\p \eta_{\rm S}^1\eta_{\rm S}^2\, (\Lambda\p_{22})^{-1}} 
{-\eta_{\rm S}^2\, (\Lambda\p_{22})^{-1} \Lambda\p_{21}} \, . 
\eeq 
It follows 
 from these expressions that  $(\OOO_{\rm F})_{11} (\OOO\p_{\rm F})_{22} = \OOO_{11}\OOO\p_{22}$, \ie all the
  supersymmetry-related signs cancel in this particular combination. 
Crucial for this to happen  is the assumption that the interfaces preserve the same supersymmetry in the CFT2 region between them, \ie 
that the same sign $\eta_{\rm S}^2$ is chosen for both $I_{12}$ and $I_{23}$.  
 \smallskip  \smallskip

Let us finally put together all the positive-mode bosonic and fermionic intertwiners
\beq
I_{1,2}^{>} =I_{1,2}^{>,{\rm bos}}\otimes I_{1,2}^{>,{\rm ferm}}\,.
\eeq
In the Ramond sector, where the fermionic-mode  frequencies $r$ are integer, 
  the determinant factors in \eqref{fermcomp2} exactly cancel the ones from the bosonic intertwiners \eq{boscomp2}. 
  Thus, one can take the limit $\delta\to 0$ to 
obtain
\beq\label{Rintert}
I_{1,2}^{>}(\Lambda\p,\eta\p,\eta_S^1,\eta_S^2)\, I_{2,3}^{>}(\Lambda,\eta,\eta_S^2,\eta_S^3)=I_{1,3}^>
(\Lambda\p\Lambda,\eta\eta\p,\eta_S^1,\eta_S^3)\quad {\rm R\, \, sector}\,.
\eeq

In the NS sector, on the other hand,  the $r$ are half integers, and the 
determinant factor from the bosonic sector is not cancelled by the one from the fermionic sector. 
However, its singular behavior for $\delta\to 0$ does cancel. 
This can be seen with the help of  the Euler-Maclaurin formula,
which implies 
$$
{\rm lim}_{\delta\to 0}\, \sum_{n\geq 1}  F( e^{-2\delta n} )  =  {1\over \delta}  \int_0^\infty dx \,  F( e^{-2x} ) -  {1\over 2} F(1)   
+ {\delta \over 6} F^\prime(1) + O(\delta^2)\  , 
$$
\beq\label{euler}
{\rm lim}_{\delta\to 0}\, \sum_{n\geq 1}  F( e^{-2\delta n + \delta} )  =  {1\over \delta}  \int_0^\infty dx \,  F( e^{-2x} ) 
-  {\delta \over 12} F^\prime(1) + O(\delta^2)\  .  
\eeq
for any function $F$ vanishing analytically at the origin.
Substituting   $F(z) \equiv  {\ln\, \det} (1 - z \OOO_{11}  \OOO_{22}^\prime )$ one finds
\beqn \label{NSintert}
&&I_{1,2}^{>}(\Lambda\p,\eta\p,\eta_S^1,\eta_S^2)\,  I_{2,3}^{>}(\Lambda,\eta,\eta_S^2,\eta_S^3)=\\
&&\qquad\qquad\qquad\sqrt{\det(1-\OOO_{11}\OOO\p_{22})}\,  I_{1,3}^> \, 
(\Lambda\p\Lambda,\eta\p\eta,\eta_S^1,\eta_S^3)\quad {\rm NS\, \, sector}\,.\nonumber
\eeqn
The NS fermions precisely cancel the divergent  Casimir energy of  the  bosons.
The final answer for the composition of oscillator intertwiners in the NS sector
 is   identical to the renormalized one in the purely  bosonic model  \cite{arXiv:0712.0076}.

%%%%%%%%%%%%%%%%%%%%%%%%%%%%%%%%%%%%%%%

\subsection{Zero modes and the defect monoid}
\label{3.3}

It follows from \eqref{Rintert}  and \eqref{NSintert} that the composition of positive-frequency
parts of the interface operators is consistent with the one in the classical theory,  which is given by group multiplication.
 In other words,
if  $(\Lambda, \eta)$ and $(\Lambda\p, \eta\p)$ are the data  that determine the 
positive-frequency parts  $I^>_{2,3}$ and $I^>_{1,2}$, then the data in the positive-frequency
 part  of $I_{1,3} = I_{1,2}\odot I_{2,3}$ is $(\Lambda\p\Lambda, \eta\p\eta)$.\footnote{Without loss of generality,
 we will from now,  and till further notice,   set all the  signs $\eta_{\rm S}^i$ to $+1$, \ie we will assume that
 the unbroken supersymmetry is given by the same combination of left and right supercharges
 in all CFTs.} The only subtlety is the appearance of the determinant in the NS sector. As we will
 see,  this is precisely what is needed in order for the  $g$-factors to compose as they should.  
 \smallskip 
 
   Consider first the unprojected  theory,  where the interface operators are those given in \eqref{completeinterface}. 
 The identity  maps  \eqref{zeroNStrivial}  between NS-fermion ground states compose trivially,  
 \beq
I_{1,2}^{0,{\rm NS}} I_{2,3}^{0,{\rm NS}}=I_{1,3}^{0,{\rm NS}}\,.
\eeq
To complete the calculation  of \eqref{fusiondefn}  
 we therefore  only have to compose the bosonic ground state maps  \eqref{sec2last}. A simple calculation gives
 \footnotesize
 \beqn
  I_{1,2}^{0, {\rm bos}}I_{2,3}^{0, {\rm bos}} &=& {
\left( g(\Lambda\p) \sum_{   \hat\gamma\p \in\ZZ^{1,1}}
 e^{2\pi i\varphi\p(  \hat\gamma\p)} |\hat \Lambda\p  \hat\gamma\p\ket\bra  \hat\gamma\p| \, \Pi_{\hat\Lambda\p} \right) 
\left( g(\Lambda)  \sum_{   \hat\gamma\in\ZZ^{1,1}}
 e^{2\pi i\varphi(  \hat\gamma)} |\hat \Lambda  \hat\gamma\ket\bra  \hat\gamma| \, \Pi_{\hat\Lambda}\right)} \nonumber\\
\label{101a}
&=& g(\Lambda\p) g(\Lambda) \, \sum_{   \hat\gamma\in\ZZ^{1,1}}
e^{2\pi i[ \varphi\p (\hat\Lambda \hat\gamma) +  \varphi(  \hat\gamma)] } \, 
|\hat \Lambda\p\hat\Lambda   \hat\gamma\ket\bra  \hat\gamma| \, \Pi_{\hat\Lambda\p\hat\Lambda} \Pi_{\hat\Lambda}\ , 
\eeqn 
\normalsize
 where  $g(\Lambda)= \sqrt{{\rm ind}(\hat\Lambda) \vert \Lambda_{22}\vert}\ $ is the $g$-factor of the interface. 
 The result  looks like the ground state map for an interface with gluing matrix $ 
 \Lambda^\prime\Lambda$, except for two
  important  differences:  (i) in general $g(\Lambda\p) g(\Lambda) \not=  g( \Lambda^\prime\Lambda)$, and
 (ii)  there is an  extra projector,   $\Pi_{\hat\Lambda}$, in addition to the projector $\Pi_{\hat\Lambda\p\hat\Lambda} $. 
 
 \smallskip
 
   Concerning the  normalization, note   that   the product of
  $g$-factors should  be multiplied by the  determinant   from the composition of the positive-frequency parts, \cf equation 
   \eqref{NSintert}.  In the case at hand   from \eqref{218} and \eqref{219} we find
   $\OOO_{11} = {\rm tanh} \alpha$ and $\OOO_{22}^\prime =  \varepsilon\p {\rm tanh} \alpha^\prime$, so that  the
   product of the determinant and of the two  $g$-factors yields
 \beqn
&& \sqrt{\det(1-\OOO_{11}\OOO\p_{22})}\,  g(\Lambda\p) g(\Lambda)\\
&&\qquad = \sqrt{\vert k\p_1k\p_2k_1k_2\vert}
 \sqrt{(1 + \varepsilon\p {\rm tanh} \alpha\p {\rm tanh} \alpha)({\rm cosh} \alpha\p {\rm cosh} \alpha)}\nonumber\\
 &&\qquad=
  \sqrt{\vert k\p_1k\p_2k_1k_2\vert} \sqrt{{\rm cosh}(\alpha+ \varepsilon\p \alpha\p)}
  = 
  \sqrt{{\vert k\p_1k\p_2k_1k_2\vert\over{\rm ind}(\hat\Lambda\p\p)}}g(\Lambda\p\p)\,.\nonumber
  \eeqn
  In the last step, we used 
    that  ${\rm cosh}(\alpha+ \varepsilon\p \alpha\p) = \vert \Lambda^{\prime\prime}_{22}\vert$ 
 where $ \Lambda^{\prime\prime}  = \Lambda^\prime\Lambda$. Thus,  if  ${\rm ind}( \hat \Lambda^{\prime\prime} )$ 
 were equal to   $\vert k\p_1k\p_2k_1k_2\vert $,  we would precisely obtain  
 $g(\Lambda^{\prime\prime} )$, \ie the $g$-factor of an elementary interface with gluing matrix $\Lambda^{\prime\prime} $. 
 
 \smallskip
 
 In general, however,  ${\rm ind}( \hat \Lambda^{\prime\prime} ) \not=  \vert k\p_1k\p_2k_1k_2\vert $
so that the fusion of two simple interfaces is not a simple  interface, but rather the sum
 of several simple interfaces.\footnote{We adopt here the language of reference
 \cite{Frohlich:2006ch}, and call ``simple interfaces" those  that cannot be written as the sum of two other interfaces.}
 To see this   let for example
 $\Lambda\p = \Lambda^{-1}$,  so that  the composition of gluing matrices is the identity matrix, $\Lambda^{\prime\prime}= {\bf 1}$. 
 Let also CFT1 and CFT3 be the same conformal  theory, so that the interface $I_{1,2}$  is the ``would-be inverse"   of  the interface $I_{2,3}$. 
 Clearly,  in this case $k\p_1/k\p_2 = k_2/k_1$ since $\hat\Lambda\p$ is the inverse of $\hat\Lambda$. 
 For simplicity we set   $\varphi\p = \varphi = 0$.  The ground state map  \eqref{101a} multiplied by  the
 determinant from the positive-frequency modes then gives
 \beq
  \vert k_1 k_2\vert  \,   \Pi_{\hat\Lambda}   =    \sum_{N,M}  \sum_{n,m=0}^{k_1,k_2}  e^{2\pi i ({N n\over k_1} + { M m\over k_2})} \vert N, M\rangle\langle N, M\vert \ ,  
 \eeq
\ie the sum of $ \vert k_1 k_2\vert$  identity interfaces, with phase moduli arranged in a periodic array so as to 
implement the projection on the charge sublattice  $k_1\ZZ \oplus k_2\ZZ$. 
Only for $\vert k_1\vert = \vert k_2\vert = 1$, \ie if  $\hat\Lambda\in O(1,1\vert\ZZ)$, does fusion
yield the identity interface. For all other $\hat\Lambda\in O(1,1\vert\mathbb{Q})$
   the projector is non-trivial,  and the corresponding  interface operators cannot be inverted.

\smallskip
      
   The algebraic structure of  $\widehat u(1)^2$ preserving interfaces in the unprojected-fermion theory 
   is the same as   in the purely bosonic theory
 \cite{arXiv:0712.0076, Fuchs:2007tx}, modulo a $\ZZ_2$ that changes the sign of the fermion field.  
  To describe this algebraic structure,  we first note   that two interfaces  can only be added if they separate the same CFTs.  They can only be fused if the CFT to the right
 of the first interface is the same as the CFT to the left of the second interface.  
 These conditions  are  automatically obeyed if we restrict attention to interfaces between identical CFTs. We will call such interfaces  
  ``defect lines".\footnote{In the literature the term  ``defect" is used   interchangeably with the term ``interface".} 
 Two defects in the same CFT  can be always added and fused,  and fusion is distributive over addition. 
 If we also allowed subtraction,  these defects would form  a ring. 
 But subtraction is not a physical operation since  negative $g$-factors correspond to imaginary entropy.
   So the  set of defects is   a monoid  (or semi-group) with respect to both, addition and fusion. 
 
 \smallskip 

   The  monoid of conformal defect lines is independent of the continuous moduli of the underlying CFT. This can be seen by fusing 
   from both left and right with special invertible interfaces 
      (called ``deformed identities" in \cite{arXiv:0712.0076})
which  parallel transport  the CFT along the connected components of its moduli space \cite{Brunner:2010xm}.  
 In the case at hand, these are the interfaces with $\hat\Lambda={\bf 1}$, $\varphi=0$ and $\eta=1$ in the notation of Section \ref{sect22}. 

Any $\widehat u(1)^2$ preserving interface between circle theories can in this way be
converted to a $\widehat u(1)^2$ preserving defect line in any given circle theory. Since this latter
is irrelevant for  the algebraic structure of the defects, we do not  have to  indicate it explicitly. 
We therefore parametrize the simple defects   by 
$(\hat\Lambda , \varphi , \eta )$, where the gluing matrix $\hat\Lambda\in O(1,1|\QQ)$. 
 
 \smallskip

   The fusion of any two defects can always be written as  the sum of simple  defects. 
  The rule for two simple  defects  reads
\beq\label{comp105}
 (\hat\Lambda\p,\varphi\p, \eta\p)  \odot
 (\hat\Lambda , \varphi , \eta)
= \sum_{\varphi^{\prime\prime}}\,   (\hat\Lambda\p\hat\Lambda, \varphi^{\prime\prime} ,  \eta\p\eta)\,,
\eeq
where the sum runs over an array of $K$ linear forms on the sublattice  that is
projected out by  $\Pi_{\hat\Lambda\p\hat\Lambda}$. These forms have the following property: their exponentials are
independent functions which,  when  restricted to the (in general smaller)  sublattice projected out  by 
$\Pi_{\hat\Lambda\p\hat\Lambda}\Pi_{ \hat\Lambda}$ obey
\beq
 e^{2\pi i {\varphi^{\prime\prime}}( \hat\gamma) } = e^{2\pi i {\varphi\p}(\hat\Lambda \hat\gamma)+ \varphi( \hat\gamma) }\qquad {\rm when}\ \ \ 
\Pi_{\hat\Lambda\p\hat\Lambda}\Pi_{ \hat\Lambda} \vert \hat\gamma\rangle = \vert \hat\gamma\rangle\   . 
\eeq
If we parametrize the matrices as in \eqref{rationalmatrices}, with  $(k_1,k_2)$ and $(k\p_1,k\p_2)$ coprime integers, 
 then  the number $K$ of terms in the sum  is given by 
\beqn
 K=\left\{\begin{array}{cc} {\rm gcd}(k_1k\p_1,k_2k\p_2)\,,& {\rm det}\hat\Lambda = 1 \ , 
\\{\rm gcd}(k_1k\p_2,k_2k\p_1)\,,& {\rm det}\hat\Lambda =-1\ . \end{array}  \right.  
\eeqn
The above rules determine completely the $\widehat u(1)^2$ preserving defect monoid in the non-GSO projected theories.

\smallskip

 Consider  next   the GSO projected theories. Instead of
 the fermion sign $\eta$,  elementary interfaces
 are now characterized by their R charge: they can have charge $\pm$ or be neutral, \cf  
 expressions  \eqref{completeinterface1}, \eqref{completeinterface2}
 and \eqref{completeinterface3}. 
  As discussed in Section \ref{projsection}, an interface is  charged 
   if  $\zeta \det \hat\Lambda = \zeta  \det\Lambda=+1$ and it is neutral if
 $\zeta  \det\Lambda=-1$, where  $\zeta=\pm 1$ 
  distinguishes whether the GSO projections on both sides of the interface are taken to be the same ($+1$) or opposite ($-1$). 
  If we insist that the GSO projection on both sides be the same, \ie $\zeta=1$, 
then the choice of $\hat\Lambda$ and the R charge are correlated.

  \smallskip

 When fusing the projected interfaces,  
  one has to  compose separately the NS and the R components of the interface operators. In the NS sector, the calculation only differs
  from the one in the unprojected theories by an additional normalization factor
    ${1/ 2}$ for  the  charged interfaces and ${1/\sqrt{2}}$ for  the neutral ones. 
  For simplicity, we suppress  the dependence on phase moduli $\varphi$, which is the same as in \eqref{comp105}. 
 Fusion of the NS components can be described by the following rules:
    \beq
 (\hat\Lambda\p, {\rm charge } \,  \pm )\odot
 (\hat\Lambda ,{\rm neutral }) =  (\hat\Lambda\p ,{\rm neutral }) \odot   (\hat\Lambda , {\rm charge } \,  \pm )  
= K  (\hat\Lambda\p\hat\Lambda,{\rm neutral } )\,,  \nonumber
\eeq
    \beq
 (\hat\Lambda\p ,{\rm neutral }) \odot 
  (\hat\Lambda  ,{\rm neutral }) = K  \left[ (\hat\Lambda\p\hat\Lambda,{\rm charge } \, +) + 
  (\hat\Lambda\p\hat\Lambda,{\rm charge } \, - ) \right]  \ , \nonumber
 \eeq 
  \beq \label{comp107}
 (\hat\Lambda\p, {\rm charge  } \,  s\p)\odot
 (\hat\Lambda ,{\rm charge } \,  s)
= K  (\hat\Lambda\p\hat\Lambda,{\rm charge } \, ss\p)\, .  
\eeq
Here $K$ is the number of elementary  defects  with phase moduli in an appropriate array, as discussed for the unprojected theory above.

Note that only in the third  line do the R sectors actually contribute to the fusion product. 
The  neutral operators in the second line have of course no R-sector terms, consistently with the fact 
that on  the right-hand-side of the equation  one sums over  interfaces with opposite  R charge, 
 so that the R-sector operators  precisely cancel.

  To verify that  the R-sector  operators compose as in the third line of  \eqref{comp107},   
recall  the expression \eqref{R0int}  for the ground state maps, and the expression for the defect  $g$-factor  (which can be found
in  \eqref{bos0int}).  Combining  these two expressions one finds
\beq
g(\Lambda\p) I_{1,2}^{0,{\rm R}}(\Lambda\p)
\, g(\Lambda) I_{2,3}^{0,{\rm R}}(\Lambda)=\sqrt{\vert k_1 k_2 k\p_1 k\p_2\vert}\, 
\imath^{\rm R}_{1,3} \, S(\Lambda\p\Lambda)  \nonumber
\eeq 
 \beq
 =\, K\,  g(\Lambda\p\Lambda)\,  I_{1,3}^{0,{\rm R}}(\Lambda\p\Lambda)\ . 
 \eeq
 Recall furthermore that there is no determinant  from  the positive-frequency modes in the R sector,
 where the bosonic contribution exactly cancels the contribution of fermions. 
 Finally, $I_{1,2}^{{\rm R}}$  has a coefficient $1/2$ in the full expression \eqref{completeinterface2}
 for  the interface operator, and we must  sum  over the two possible values of $\eta$. 
  Putting all these facts  together one finds
  that  the R-sector  operators compose indeed as indicated in  the third line of  \eqref{comp107}. 
 \smallskip
 
   The $\widehat u(1)^2$ preserving defect algebra in the GSO projected $c=3/2$ theory 
   can be described more succinctly as follows:
    it is  the tensor product of the $\widehat u(1)^2$ preserving defect  algebra in the bosonic $c=1$ theory, tensored with the fusion algebra 
    of the Ising model. 
   The latter reads
 \beq\label{verlinde}
  \epsilon \times \epsilon = 1\ , \qquad  \epsilon \times \sigma = \sigma\ , \qquad \sigma\times\sigma = 1 + \epsilon\ . 
 \eeq
 Identifying $1$ and $\epsilon$ with the two charged interfaces, and $\sigma$ with the neutral interface, 
reproduces precisely the pattern  \eqref{comp107} in the fermion sector. 

\smallskip

This is not a coincidence. The  conformal defects of the Ising model,   analyzed in \cite{Oshikawa:1996dj,  
Petkova:2000ip, Frohlich:2004ef,  Quella2}, can be described in our language 
  by the data $(\Lambda ,  \alpha)_{\rm Ising}$,  where    $\alpha \in  \{1, \epsilon, \sigma \}$  labels the R charge in the way just
  described, 
$\Lambda$ and $-\Lambda$ correspond to  identical defects, 
and ${\rm det}\Lambda = +1$ for charged defects and $-1$ for the neutral ones. 
One may compute the fusion of these defects,    without associating them necessarily to the  bosonic field, 
by subtracting the divergent Casimir energies as in \cite{arXiv:0712.0076}. The result  is
\beq
(\Lambda\p ,  \alpha\p)_{\rm Ising}\odot (\Lambda ,  \alpha)_{\rm Ising} = (\Lambda\p\Lambda , \alpha\p \times \alpha )_{\rm Ising}\ ,  
\eeq
where $\alpha\p \times \alpha$ is given by \eqref{verlinde} and the sum  of Ising primaries  indicates in the above equation
 the sum of
the corresponding  interface operators. 

The defect $(\Lambda ,  \alpha)_{\rm Ising}$ is  topological if and only if  $\Lambda\in O(1)\times O(1)$. The topological defects of the Ising model 
are known to be
in one-to-one correspondence with  primary fields, and their fusion algebra is the Verlinde algebra \cite{Petkova:2000ip}. 
This provides a consistency check of the   more general analysis presented here.

 %%%%%%%%%%%%%%%%%%%%%%%%%%%%%%%%%%%%%%%%%%%%%
 %%%%%%%%%%%%%%%%%%%%%%%%%%%%%%%%%%%%%%%%%%%%%

\section{Topological interfaces  as quasi-symmetries}  \label{sec5}

    The defects described in the previous sections  are specified
    by the following data: the moduli of the bulk CFT, \ie a radius $R_1=R_2$,     
the gluing matrix $\hat\Lambda\in O(1,1|\QQ)$ of the integer charges,  and
    the phase moduli $\varphi$. Furthermore  the fermionic gluing conditions require some extra data:
    the sign $\eta=\pm 1$   in the unprojected theory, and in the GSO projected theory, the
    Ramond charge ($\pm$, or neutral), or equivalently an Ising primary ($1,\sigma,\epsilon$). 
    Again, we fix the preserved supersymmetry algebras by setting $\eta_{\rm S}^j=1$. The gluing matrix for the fermion fields 
is thus given by $ \eta \Lambda$, where $\Lambda=U_2\hat\Lambda U_2^{-1}$ is
the  gluing matrix for   bosonic currents.
    
These defects are superconformal and preserve a $\widehat u(1)^2$ current algebra. 
Generically, they are not topological. However, as explained in Section \ref{sect22}, for any such defect 
$I_{R_2,R_2}(\hat\Lambda,\varphi)$,
 there is a unique radius $R_1= f_{\hat\Lambda}(R_2)$ such that parallel transport yields a topological interface
 between the theories of radius $R_2$ and $R_1$. Explicitly
 $I_{R_1,R_2}(\hat\Lambda,\varphi)=D_{R_1,R_2}\odot I_{R_2,R_2}(\hat\Lambda,\varphi)$ where
 $D_{R_1,R_2}$ is the deformed identity interface that transports the theory from $R_2$ to $R_1= f_{\hat\Lambda}(R_2)$,
 \cf the previous subsection. 
  In fact, this was only explained for the bosonic components, but due to supersymmetry, it immediately carries over to the fermions as well.

Since $R_2$ is arbitrary, parallel transport indeed yields an isomorphism between
the fusion algebra of $\widehat u(1)^2$-preserving conformal defect lines in any given circle theory (they are all isomorphic),
and the fusion algebra of $\widehat u(1)^2$-preserving 
topological interfaces between circle theories. To be more precise,
for any radius $R_3$, and any gluing matrices $\hat\Lambda\p,\hat\Lambda$ there are radii $R_2=f_{\hat\Lambda}(R_3)$ 
and $R_1=f_{\hat\Lambda\p}(R_2)$ such that the interfaces $I_{R_1,R_2}(\hat\Lambda\p,\varphi\p)$ and $I_{R_2,R_3}(\hat\Lambda,\varphi)$ are topological and their fusion is given by parallel transport of the fusion of the respective conformal defects in the theory with radius $R_3$:
\beq
I_{R_1,R_2}(\hat\Lambda\p,\varphi\p)\odot
I_{R_2,R_3}(\hat\Lambda,\varphi)
=D_{R_1,R_3}\odot I_{R_3,R_3}(\hat\Lambda\p,\varphi\p)\odot
I_{R_3,R_3}(\hat\Lambda,\varphi)\,.
\eeq
  [We have suppressed  the fermion-interface labels for simplicity]. 

Thus, the monoids of $\widehat u(1)^2$-preserving conformal defects and topological interfaces in torus models are isomorphic. 
The isomorphism actually
 breaks  down if the requirement of $\widehat u(1)^2$-symmetry is dropped. 
 This would allow  for example the addition of defects with different gluing conditions $\hat\Lambda$ and $\hat\Lambda\p$, 
 but topological interfaces can only be added if the theories on both sides agree, \ie if $f_{\hat\Lambda}(R)=f_{\hat\Lambda\p}(R)$.
 \smallskip
 
  In the next subsection, we will explain how the topological interfaces on the string worldsheet are related to the  $O(1,1|\RR)$ symmetry of classical supergravity compactified on a circle.

 %%%%%%%%%%%%%%%%%%%%%%%% 
 
  \subsection{Action on perturbative string states}
  
    Consider first  the purely bosonic theory and let  $\hat\Lambda$ be the gluing matrix for the integer charges. 
    If the topological condition $R_1= f_{\hat\Lambda}(R_2)$ is satisfied, the gluing condition
    $\Lambda = U_1^{-1} \hat\Lambda U_2\in O(1)\times O(1)=\{{\rm diag}(\pm1, \pm1)\}$, which implies that left and right Virasoro algebras commute separately 
     with the interface operator. 
   
 In the following, we will restrict our attention to the case $\Lambda = {\bf 1}$. The other cases can be obtained from this one by T-duality transformations,
 which are  implemented by invertible topological interfaces with $\hat\Lambda \in O(1,1\vert \ZZ)$. Since T-duality is well understood 
  \cite{Giveon:1994fu}, we refrain from giving any more detail on these other cases here.
     
  \smallskip 
  
 From
the  expressions \eqref{gluingcondition} and \eqref{sec2last} we deduce
   that the topological-interface operator  maps states in CFT2 to states in CFT1 as follows:
    \beq\label{map111}
   (\prod_{\{n_i\}} a_{n_i}^\dag )  (\prod_{\{\tilde n_j\}} \tilde a_{\tilde n_j}^\dag )  \vert  \hat\gamma \rangle \  \to \
   e^{2\pi i\varphi ( \hat\gamma)}\, 
  \sqrt{ \vert k_1 k_2 \vert }\,   (\prod_{\{n_i\}} a_{n_i}^\dag )  (\prod_{\{\tilde n_j\}} \tilde a_{\tilde n_j}^\dag )  \vert \hat\Lambda \hat\gamma \rangle  
   \eeq
if $ \hat\gamma\in k_1\ZZ\otimes k_2\ZZ$, while all other states are mapped to zero. 
 Here we used that $\Lambda_{22}= 1$ for $\Lambda = {\bf 1}$.
 
  The physical charge vector $\gamma: = U \hat\gamma$ is preserved by the above
  map, 
 \beq
    U_1 \hat\Lambda\hat\gamma =   \Lambda (U_2 \hat\gamma)  = U_2 \hat\gamma  \ , 
 \eeq
  and hence, the masses 
        \beq\label{mass121} 
   {\cal M}_{\rm pert}^2 =  8  \gamma^T  \gamma  + \sum_i 2n_i + \sum_j \tilde 2n_j \ . 
    \eeq
    of perturbative string states are also preserved. 
   [Our convention is $\alpha^\prime = 1/2$]. This of course is an immediate consequence of the property of topological interfaces to commute with left and right Virasoro algebras combined with the fact that 
      masses of perturbative string states are proportional to $(L_0+ \tilde L_0)$.

\smallskip 

 In a nutshell,  topological interfaces transform  moduli and perturbative charges in the same way as the $O(d,d\vert \mathbb{R})$
 symmetry of the low-energy  action.  But they have the `integrity' to only transform charges
 if this is consistent with charge quantization.
 In fact,  the transformations preserves a  larger set of observables 
 than the masses, as we will now explain. 
\smallskip 
   
Namely, any local operator $V$ 
  with $u(1)$ charges $ \hat\gamma\in k_1\ZZ\otimes k_2\ZZ$ is just multiplied by the factor 
  $e^{2\pi i \varphi( \hat\gamma)}\, \sqrt{\vert k_1k_2\vert}$ under the action of the interface operator.
  Thus,   $N$-point  correlation functions 
   on the sphere  transform by  a common  multiplicative factor, 
   \beq\label{map114}
   \langle V_{1} V_{2} \cdots V_{N}  \rangle_{\rm sphere} \ \mapsto\ 
   \vert k_1 k_2 \vert ^{N/2} \,  \langle V_{1} V_{2} \cdots V_{N}  \rangle_{\rm sphere}\ . 
   \eeq
  Note that the phase factors drop out from the expression on the right due to the $u(1)$-charge conservation. 
  
  Translated to string theory,  
  \eqref{map114} implies that  the tree-level scattering amplitudes of  states with vertex operators $V_j$ are 
   invariant   provided one also transforms  the effective
      string coupling constant according to 
   \beq\label{116coupling}
  {\lambda_{c} \over \sqrt{2\pi R}}  =:    \lambda_{\rm eff} \  \mapsto  \  \lambda_{\rm eff} \, \sqrt{ \vert k_1 k_2 \vert }\ .   
   \eeq
Here, $\lambda_c$ is the closed-string coupling constant in 26 dimensions, and $\lambda_{\rm eff}$ the effective coupling after
compactification on a circle of radius $R$. This effective  coupling  can be defined 
as the common normalization of all vertex operators  \cite{Polchinski:1998rq}. 
  We stress that only a part of the tree-level S-matrix is preserved by the topological map,
  the part restricted   to  asymptotic states  for which the $O(1,1\vert \mathbb{Q})$  
   transformation   respects the charge quantization. All other string states are projected out. 
\vskip 1mm

    The rescaling  \eqref{116coupling} of the coupling  is surprising, because it depends on arithmetic properties of the $O(1,1\vert \mathbb{Q})$ gluing matrix. Since it amounts to a redefinition of the Planck scale, it is invisible classically, even if all stringy $\alpha\p$ corrections are included in the closed-string action. 
    Nonetheless, it is  crucial for  the proper  transformation of D-brane charges and masses.  
     
 \smallskip
 
   Before proceeding to the treatment of D-branes, let us comment on a relation of our discussion with the orbifold construction. Indeed, since in the case at hand the quotient of the circle radii $R_1/R_2=k_2/k_1$ is rational, the theory with radius $R_1$ can be obtained from the one with radius $R_2$ by orbifolding with respect to the shift symmetry
   \beq\label{orb116}
 \phi \to   \phi + 2\pi R_1\ . 
 \eeq
The orbifold group generated by this symmetry is of order $k_1k_2$. The operator $\Pi_{\hat\Lambda}$  for $\hat\Lambda =
  {\rm diag} (   {k_2/k_1},  {k_1/k_2} ) $   projects
on untwisted states of the   orbifold, while  all other states  
in CFT1 arise as twisted sectors. 
 
  This  viewpoint   demystifies the relations 
  \eqref{map114}. These relations express  the well-known fact that
  the parent and the orbifold theory share the same sphere amplitudes  in   the untwisted sector. 

Indeed, our construction  fits in nicely with 
     the general framework of topological interfaces in rational CFTs put forward by 
   Fr\"ohlich et al  \cite{Frohlich:2006ch}. These authors single out  
   two  classes of special topological interfaces in  RCFT:  (i)  
    the so-called ``group-like" interfaces, which describe automorphisms of CFTs, and which form a groupoid under fusion, and (ii)
     the broader class of ``duality interfaces", which have the property that fusion with their parity-transform  results in a sum 
   of group-like defects. It has been argued in \cite{Frohlich:2006ch} that duality interfaces exist between
   a  parent theory and any  of its  orbifold descendants, and that such an  interface is group-like only when the 
   orbifold is the same CFT as the parent theory.
   
  Although the arguments of \cite{Frohlich:2006ch} were made  in the context of RCFT, they 
 extend  to  the circle theories studied here.  
All the topological interfaces associated to gluing matrices  $\hat\Lambda\in O(1,1\vert \mathbb{Q})$ are duality interfaces, whereas the ones with $\hat\Lambda\in O(1,1\vert \mathbb{Z})$ are group-like.
   The parent and orbifold theories of   \cite{Frohlich:2006ch} are nothing but the
theories at radius $R_2$, respectively $R_1$. 
     
   \smallskip
 
   We may extend this  analysis  from the bosonic string theory  to
    the type-0 or the type-II superstring theories  in the following way. 
 We  first  note that  for a bosonic gluing matrix $\Lambda =  {\bf 1}$, the fermionic one is given by 
 $\Lambda_{\rm F}= \pm {\bf 1} $, \cf 
\eqref{lambdaprime}.  Thus, left (right) fermions of CFT1 are glued to left (right) fermions of CFT2, \ie 
   the fermionic interface is automatically topological as well.   
This of course is a  consequence of  supersymmetry.  
The mass  of the perturbative string states, which is still 
equal to  the square root of $2(L_0 + \tilde L_0)$,   is   therefore still  preserved by the interface map  as in the purely  bosonic case. 

What needs to be checked is that the uniform rescaling \eqref{map114} is also valid for  states
in the Ramond sector.  We focus on charged interfaces, since the neutral ones 
anyway project out all Ramond  states.  Making use of 
$\vert {\rm sin}(2 \vartheta) \vert = \vert \Lambda_{22} \vert = 1$, and the property
$S(\pm {\bf 1}) = 
(\begin{smallmatrix}  1 & 0 \\ 0 & \pm1 \end{smallmatrix})$ of the spinor representation, it follows  
from \eqref{R0int}
that indeed  all  vertex operators transform with the same normalization factor.  
\vskip 1mm

This argument  applies to the type-0 superstrings. 
The type-II  superstring theory  has separate GSO projections for the left- and right-moving  fermion numbers. 
To implement these projections, we have to tensor the $c=3/2$ interfaces with the identity interface for the nine remaining non-compact dimensions. 
 Because $\Lambda_{\rm F} =\pm {\bf 1}$, such topological interfaces  commute or anticommute  with  $(-1)^F$ and
$(-1)^{\tilde F}$. In the first case the interface  must be  resolved by the addition of new twisted contributions, 
which  are intertwiners for  the mixed R-NS sectors of the type-II theories. The construction proceeds along the lines described in Section
\ref{projsection}. 
 It is tedious but straightforward to check  that for the ensuing topological interfaces, equation 
  \eqref{map114} still holds 
  for states from all  four  sectors of the type-II superstring theory.

 %%%%%%%%%%%%%%%%%%%%%%%%%%%
    
 \subsection{Action on D-branes}   
   
    As alluded to above, interfaces not only transform perturbative string states, but also act on D-branes. This action is given by fusion with the respective boundary condition. 
   
   We consider (super)string theory  compactified on a  $d$-dimensional torus, and  take any
    D-brane wrapped entirely around some of the torus directions, so that it  looks like a point particle in the non-compact  spacetime. 
    The D-brane can be described by a  boundary state $\vert {\cal B} \keti$ of the  
      $c=3d/2$   SCFT  [we focus for definiteness on the type-0 supersymmetric case]. 
     The mass of this point particle is proportional to  the $g$-factor of the  boundary state 
    \cite{Harvey:1999gq} 
    \beq\label{Dmass}
      {\cal M}_{\cal B } =  \ 4 (\sqrt{\pi})^{7-d}\, M_{\rm Planck} \, g_{\cal B } \ ,  
    \eeq 
where  $M_{\rm Planck} $ is the Planck mass in the effective $(10-d)$ dimensional theory.  It is given by
(see for instance \cite{deAlwis:1996ez} and recall that $\alpha^\prime = 1/2$)
\beq
M_{\rm Planck}^{-2} =     8 \pi^7 \,   \lambda_{\rm eff} ^2 \  
\eeq
 with the effective coupling $\lambda_{\rm eff} = \lambda_c/\sqrt{V_d}$ defined as above, where  $V_d$ denotes  the volume of the torus.
 Modulo a numerical constant,  ${\cal M}_{\cal B } \sim g_{\cal B } / \lambda_{\rm eff}$. 
It follows from this relation that ${\cal M}_{\cal B }$ is preserved by the operation of the topological interfaces, as were
 the masses of perturbative string states. 
\smallskip

To understand this, let us  fuse the D-brane state $\vert {\cal B}\keti$ with 
a  charged topological interface.\footnote{Recall that charged interfaces are the ones
that  extend   the $O(d,d\vert\mathbb{Q})$ action to Ramond states, and which therefore act  non-trivially on the Ramond charge. Notice also that
boundary states
are special interfaces for which the CFT on one side is the trivial theory. Fusing an interface and  a boundary is therefore a special
case of interface fusion.} 
  Since the interface is topological, the $g$-factors of interface and D-brane multiply
   \beq\label{g119}
  g_{\cal B } \mapsto  g_{\rm top}\,  g_{\cal B } =  \sqrt{\vert k_1 k_2 \vert}  \, g_{\cal B }\ .  
 \eeq
  This follows from the fact that 
 topological defect lines can be deformed as long as they do not cross any operator insertion.    
We have used  that $g_{\rm top} = \sqrt{\vert k_1 k_2 \vert }$ for any topological interface, \cf \eqref{sec2last}
 with $\vert \Lambda_{22} \vert = 1$.  Combining  \eqref{116coupling} and \eqref{g119}  shows that the D-brane masses  are 
 invariant,  as claimed. 
 
 \vskip 1mm
 
    It is instructive to also look at the transformation of the D-brane  charges.  For a single compact dimension
    there are two types of Ramond charge,  proportional respectively to  the number of D0-branes and of wrapped D1-branes. 
    We may  arrange  them  in a 2-component vector, 
    \beq
          \hat \gamma_D :=  \left(  \begin{matrix}  N_{D0} \\  N_{D1}  \end{matrix} \right)
          \qquad {\rm or} \qquad 
          \gamma_D :=       {1 \over \lambda_{\rm eff} }  \left(  \begin{matrix}  {1\over \sqrt{2R}} & 0 \\ 0 & \sqrt{R} \end{matrix} \right) 
           \left(  \begin{matrix}  {N_{D0} }   \\   N_{D1}  \end{matrix} \right)\ , 
    \eeq
   where  following the same convention as in  the perturbative case we use a hat  to distinguish the vector of  {\emph { integer}}
  as opposed to  {\emph {physical}}  charges. The physical charges  are the couplings to  Ramond gauge fields that  are canonically normalized  (modulo
  an irrelevant numerical constant). 
    \smallskip
    
      Consider now the fusion with a charged topological interface of gluing matrix $ \hat\Lambda = ( \begin{smallmatrix} k_2/k_1 & 0  \\ 0 & k_1/k_2    \end{smallmatrix} )$
    where $(k_1, k_2)$ are positive relatively-prime integers.\footnote{The GSO projection requires both the $ \hat\Lambda$ and $ -\hat\Lambda$
    gluing conditions, so we can choose the $k_i$ to be positive without loss of generality. 
    As for the second branch of $O(1,1\vert\mathbb{Q})$ matrices, this  can be obtained by composition with $ \hat\Lambda = ( \begin{smallmatrix} 0 & 1 \\ 1 & 0 \end{smallmatrix} )$,
   \cf \eqref{rationalmatrices}. The corresponding topological interface implements the radius-inverting T-duality transformation.
   For the action of T-dualities on D-branes see for example \cite{Obers:1998fb}.}
      Physical charges 
    transform   with  the spinor representation $S(\Lambda)$ of the gluing matrix $\Lambda$ for the currents. Since $\Lambda = {\bf 1}$
    for the topological interfaces, physical charges change at most by a sign. The integer Ramond charges,  on the other hand, 
  transform up to a sign with the following matrix:  
   \beq
    \sqrt{\vert k_1 k_2 \vert}\, S(\hat \Lambda)\, =\, 
     \sqrt{\vert k_1 k_2 \vert}\,  \left(
      \begin{matrix}  \sqrt{k_2/ k_1} & 0 \\ 0 & \sqrt{k_1/ k_2} \end{matrix} 
     \right)  \,
     =\,  \left( \begin{matrix}  k_2 &  0 \\ 0 & k_1  \end{matrix} \right) \ . 
   \eeq
    The  square-root of the index in the left-hand-side is due to the transformation  
   \eqref{116coupling} of the effective string coupling. It is   crucial to  ensure  that
   the topological map respects the quantization of Ramond charges. 
  
\smallskip

   We close this section by emphasizing how  the transformation  of perturbative  states differs from the transformation of D-branes. 
   For $|k_1k_2|\neq 1$, the former is non-invertible because it only acts  on a sublattice of rank $\vert k_1k_2\vert $ 
   of the perturbative charge lattice.
   The latter on the other hand acts  as an endomorphism  of the Ramond charge lattice, mapping the 
   entire lattice to a sublattice of rank $\vert k_1k_2\vert $. Both of these transformations are invertible only for $\vert k_1k_2\vert =1$,\ie $\hat\Lambda\in O(1,1\vert \mathbb{Z})$

  \smallskip  
       The transformations of the integer charges are accompanied by a change  
       of the radius of the bulk CFT,  
       as well as by the rescaling \eqref{116coupling} of the effective string coupling constant. 
       The combined transformation leaves all the physical charges  invariant up to  signs.

 \vskip 1mm

%%%%%%%%%%%%%%%%%%%%%%%%%%%%%%%%%
%%%%%%%%%%%%%%%%%%%%%%%%%%%%%%%%%

\section{Generalization to   torus models}

 The results  of the previous sections  generalize in a mostly straightforward manner to 
${\mathcal N}=(1,1)$ superconformal sigma models whose target spaces are tori of arbitrary dimension $d\geq 1$. 

These ``toroidal models" factorize into bosonic CFTs describing $d$ free bosons compactified on a torus, and the theory of $d$ free Majorana fermions. 
They exhibit 
left and right $\widehat u(1)^d$ symmetries, coming from the bosonic part, and 
they are determined
by the choice of the lattice of charges of the associated $u(1)^d\oplus u(1)^d$ zero mode subalgebra
 (left and right momenta in string-theory language). These are even self-dual lattices $\Gamma\subset\RR^{d,d}$, which are parametrized
by the coset space
\beq  \label{cosets}
   {  O(d \vert \mathbb{R}) \times O(d\vert\mathbb{R})  \, \backslash\,   O(d, d \vert  \mathbb{R}) \, /  \, O(d,d\vert  \mathbb{Z})} \ ,    
\eeq
where  $O(d,d\vert  \mathbb{Z})$ is the group of discrete lattice automorphisms (the group of 
 ``T-dualities'' in string theory).  One standard choice of parametrization  is 
  \beq
\Gamma = \left\{    \left( \begin{matrix} {1\over 2} E^{-1}N   &  E^T (1+B) M \\
-{1\over 2} E^{-1}N   &  E^T (1- B) M \end{matrix} \right) = U  \left( \begin{matrix} N \\ M \end{matrix} \right) \Bigl\vert  N, M \in \ZZ^d
\right\}= U \ZZ^{d,d}\ , 
\label{changeto cano}
\eeq
where $G= E E^T$ is the metric of the target space torus and $B$ the antisymmetric Neveu-Schwarz field.
The matrix $U$ is the ``vielbein" introduced in equation \eqref{vielbeins}. 
  In our context, it
is convenient to work with the  covering space  of the coset \eqref{cosets}
on which the T-dualities and the $O(d \vert \mathbb{R})\times O(d \vert \mathbb{R})$ automorphisms
are implemented by invertible interfaces.

In this section we will first construct $\widehat{u}(1)^{2d}$-preserving  interfaces between such torus models, 
which also preserve a worldsheet supersymmetry,  and then determine their fusion. 

%%%%%%%%%%%%%%%%%%%%%%%%%%%%%%%%%%%%%%%%

\subsection{Superconformal interfaces preserving $\widehat u(1)^{2d}$}

As in the case of circle theories ($d=1$) discussed in Section \ref{sec3/2}, 
the requirement of superconformal and $\wh u(1)^{2d}$ symmetry forces the interfaces 
to factorize  into   interfaces for the bosonic and fermionic degrees of freedom.

\subsubsection*{Bosonic interfaces in torus models}

The construction of the bosonic  interfaces is a straightforward 
extension of the discussion in Section \ref{secbosonic}.
Since the energy momentum tensor is quadratic in the currents, 
the corresponding interface operators
$I_{1,2}:{\mathcal H}_2\rightarrow{\mathcal H}_1$ between the Hilbert spaces of the torus models have to satisfy 
commutation relations
\beq\label{defbosglue}
\vect{a_n^1}{-\wt{a}_{-n}^1}I_{1,2}=I_{1,2}\,\Lambda\vect{a_n^2}{-\wt{a}_{-n}^2}\,,\quad \Lambda\in O(d,d|\RR)
\eeq
for the modes of the left and right $\widehat u(1)^d$ currents, which now are considered to be $d$-dimensional vectors.

Analogously to $d=1$,  
these commutation relations can be realized by linear combinations of  intertwiners
\beq\label{interbos}
I_{1,2}^{{\rm bos},\gamma_2}=     \prod_{n>0} I_{1,2}^{n,{\rm bos}}|\Lambda\gamma_2\ket\bra\gamma_2|\,,
\eeq
where the exponentials
\beq\label{opsnonzeromodesbd}
I_{1,2}^{n,{\rm bos}}=
    {\rm exp}\hskip -1mm \left(  {1\over n} ( a^1_{-n} \OOO_{11}  \tilde a^1_{-n}  
    -  a^1_{-n}\OOO_{12} a^2_n -\tilde a^1_{-n} \OOO_{21}^t \tilde a^2_n +  a^2_{n} \OOO_{22}^t \tilde a^2_{n}) 
    \right) 
 \eeq
are composed with maps  on the ground states 
implementing the zero-mode gluing conditions.
In this expression, the modes 
of CFT1 and CFT2 act on the left respectively right of the maps $|\Lambda\gamma_2\ket\bra\gamma_2|$.
Furthermore, the matrix $\OOO$ is related to the gluing matrix  $\Lambda$ by
\beq\label{SintermsofOd}
\OOO=\OOO(\Lambda) =\mat{ \Lambda_{12} \Lambda_{22}^{-1}}
{\Lambda_{11}-\Lambda_{12}\Lambda_{22}^{-1}\Lambda_{21}} {\Lambda_{22}^{-1}} {-\Lambda_{22}^{-1} \Lambda_{21}} \,.
\eeq
This is an immediate generalization of the $d=1$ case, where now the $\Lambda_{ij}$ are
 $d\times d$ blocks of the $O(d,d)$ matrix $\Lambda$ in a basis  in which the invariant metric  is given by $\eta= {\rm diag}({\bf 1},-{\bf 1})$. 

It is easy to see that the matrix  $\OOO$ is orthogonal, \ie $\OOO(\Lambda)\in O(2d)$ whenever  $\Lambda\in O(d,d)$. 
The  inverse to relation \eqref{SintermsofOd}  is given by
\beq\label{OintermsofSd}
\Lambda(\OOO)  =\mat{\OOO_{12}-\OOO_{11}\OOO_{21}^{-1}\OOO_{22}}{\OOO_{11}\OOO_{21}^{-1}}{-\OOO_{21}^{-1}\OOO_{22}}{\OOO_{21}^{-1}}\,.
\eeq

Note that  intertwiners \eq{interbos} only exist for those charge vectors 
$\gamma_2\in\Gamma_2$ of CFT2, which under the gluing condition map to a charge vector of CFT1,  
in other words for all   $\gamma_2$ for which  $\gamma_1=\Lambda\gamma_2\in\Gamma_1$.
These  form a sublattice
\beq
\Gamma_{1,2}^{\Lambda}=\{\gamma\in\Gamma_2\,|\,\Lambda\gamma\in\Gamma_1\}=\Gamma_2\cap\Lambda^{-1}\Gamma_1
\eeq
of the charge lattice  of CFT2. Similarly as in the case $d=1$  
one needs   $\Gamma_{1,2}^\Lambda$ 
 to be a maximal-rank  sublattice of 
$\Gamma_2$,  in order to be able to solve Cardy's condition for the interface. 
Gluing conditions which satisfy this requirement, 
  ${\rm rank}(\Gamma_{1,2}^\Lambda)=2d$,  will be referred to as  {\it admissible}.

       In the folded picture,  the orthogonal matrix $\OOO$ determines the orientation and  worldvolume gauge fields
       of a D-brane in the toroidal  tensor-product theory  CFT1$\otimes$CFT2${}^*$.  Admissibility translates to the conditions that this D-brane is compact,  
       and its worldvolume gauge fields obey Dirac's quantization
       condition. 
    
The admissibility  condition is more transparent  when expressed as a condition on
  the gluing of the integer $u(1)^{2d}$ charges. 
 Namely, representing the lattices of physical-charge vectors $\Gamma_i=U_i\ZZ^{d,d}$ 
  with $U_i$ the generalized vielbein defined
 in \eqref{changeto cano}, it is easy to see that 
\beq
\Gamma_{1,2}^{\Lambda}  = U_2 \left( \ZZ^{d,d} \cap (U_2^{-1} \Lambda^{-1}U_1) \ZZ^{d,d} \right)   
 \eeq
 is a maximal-rank sublattice of  $\Gamma_2 = U_2 \ZZ^{d,d}$ if and only if 
 the matrix inside the nested brackets has  only rational entries.  This can be written equivalently as
  \beq
\hat\Lambda\  \eqdef \ U_1^{-1}\Lambda U_2\in O(d,d|\QQ)\, , 
\eeq 
where $\hat\Lambda^T \hat\eta \hat\Lambda = \hat\eta$ with  $\hat\eta= \left( \begin{smallmatrix} 
0&{\bf 1} \\ {\bf 1}&0 
\end{smallmatrix} \right)$. 
 
\vskip 1mm

For  admissible gluing conditions   one can construct
the following (simple)   interface operators
\beqn\label{bosintd}
 I_{1,2}^{\rm bos}=\prod_{n\geq 0} I_{1,2}^{n,{\rm bos}}\,, \qquad \;{\rm with} \ \ \  \ \ 
 I_{1,2}^{0,{\rm bos}}=g_{1,2}^{\Lambda}\sum_{\gamma\in\Gamma_{1,2}^{\Lambda}}
e^{2\pi i\varphi(\gamma)} |\Lambda\gamma\ket\bra\gamma|\,.
\eeqn
 Here $\varphi\in(\Gamma_{1,2}^\Lambda)^*$ is some linear form on the lattice of intertwiners,\footnote{In the folded picture
 it determines position and Wilson lines of the respective D-brane.} 
  and  the normalization constant (the $g$-factor) 
 \beq
 g_{1,2}^\Lambda=\sqrt{\|\pi_{\Lambda}(\Gamma_{1,2}^{\Lambda})\|}
 \eeq
 is determined by the volume $\|\pi_{\Lambda}(\Gamma_{1,2}^{\Lambda})\|$
 of the hybrid lattice
 \beq
 \pi_{\Lambda}(\Gamma_{1,2}^{\Lambda})=
 \left\{\vect{\pi(\gamma)}{\wt{\pi}(\Lambda(\gamma))}\,\Big|\,\gamma\in\Gamma_{1,2}^{\Lambda}\right\}\,.
 \eeq
 In this formula $\pi$ and $\wt\pi$ denote the projections on left and right charge vectors, respectively.
 The above volume is given by the product of the index 
  \beq\label{index}
 {\rm ind}(\Gamma_{1,2}^{\Lambda}\subset\Gamma_2)=|\Gamma_2/\Gamma_{1,2}^{\Lambda}|
 \eeq
of the lattice of intertwiners in the lattice of all the charges of CFT2, and the volume
of the hybrid projection of the full charge lattice $\Gamma_2$,  
 \beqn
 \|\pi_{\Lambda}(\Gamma_{2})\|&=&
\Big|\!\det\left(
 \mat{\bf 1}{0}{0}{0}+\mat{0}{0}{0}{\bf 1}\Lambda\right)\!\Big|\nonumber\\
&=&\big|\det(\Lambda_{22})\big|=\big|\det(\Lambda_{11})\big|\ . 
 \label{volf}
 \eeqn
  Hence, the $g$-factor can be written as
 \beq\label{geng}
 g_{1,2}^\Lambda=\sqrt{ |\Gamma_2/\Gamma_{1,2}^\Lambda|\,
|\!\det(\Lambda_{22})|}\,.
 \eeq
 It is important to note that while the volume factor \eq{volf} depends   on 
 the  matrix $\Lambda$, which  varies continuously with the moduli of the bulk CFTs, 
   the index factor \eq{index}   depends on arithmetic properties of the rational 
 matrix $\hat\Lambda$ which is the gluing matrix for  integer charge vectors.

  It is straightforward to check that the index is determined by $\hat\Lambda$ as follows:
  \beq\label{indexdefn}
   |\Gamma_2/\Gamma_{1,2}^\Lambda| = {\rm smallest} \ \ K\in \mathbb{N}\qquad
   {\rm such\ that } \ \ \   K \hat\Lambda \in GL(2d, \ZZ)\ .  
  \eeq 
Put differently, $K$ is the least common multiple of all  (irreducible) denominators of the matrix elements
 $\hat\Lambda_{ij}$.  For $d=1$,  with the parametrization of the gluing condition chosen in Section \ref{secbosonic}, one finds
 \beq
 |\Gamma_2/\Gamma_{1,2}^\Lambda|= \vert k_1k_2\vert \,,\quad
  |\det(\Lambda_{22})|=\cosh(\alpha)={1\over \vert \sin(2\vartheta)\vert }\ . 
  \eeq
The general expression \eq{geng} for the $g$-factor, valid  for arbitrary $d$,  specializes  as it should  to the 
expression \eq{d=1g}  which was obtained for  $d=1$. 
 
 \vskip 1mm
 
 We will refrain from showing here 
 that the operators \eq{bosintd} indeed satisfy Cardy's consistency condition. 
 This could be done, as in  the one-dimensional case,
  by computing the annulus partition functions in the folded theory, 
  and checking the multiplicities in the open-string channel.  However, the
   analysis of the fusion of these operators,  carried out in Section \ref{secfusiond} below, 
  will  provide a stronger consistency check than Cardy's condition.

 \vskip 1mm
 
The interfaces \eq{bosintd} are  {\emph {simple}} or  {\emph {elementary}} interfaces, meaning that  
 their vacuum is non-degenerate. 
 Non-elementary interfaces  consistent with the $\widehat u(1)^{2d}$ symmetry 
  can be obtained by summing  simple ones with the same gluing condition $\Lambda$.
  In this way, it is possible to obtain 
interfaces which only involve (maximal rank) sublattices  $L\subset\Gamma_{1,2}^{\Lambda}$ of all the possible intertwiners for a given gluing condition.  
  To project out all intertwiners not in $L$ one needs to  sum   
over $|\Gamma_{1,2}^{\Lambda}/L|$ simple  interfaces $I_{1,2}^{{\rm bos}}(\Lambda,\varphi_i)$
with phase moduli $\varphi_i$ arranged in  an appropriate periodic array. 
The resulting interface operators read
 \beq\label{Lint}
 I_{1,2}^{0,{\rm bos}}(\Lambda,\varphi,L)=g_{1,2}^\Lambda
 |\Gamma_{1,2}^\Lambda/L|
 \sum_{\gamma\in L}e^{2\pi i\varphi(\gamma)}|\Lambda\gamma\ket\bra\gamma|\,,
 \eeq
where now $\varphi$ is a linear form on $L$. Note that due to the summation, the normalization of the defect received a factor of 
${\rm ind}(L\subset\Gamma_{1,2}^\Lambda)=|\Gamma_{1,2}^\Lambda/L|$.

Non-elementary interfaces are important in the discussion of fusion of interfaces. Namely, as in the one-dimensional case, the composition of elementary interfaces with gluing conditions $\Lambda\p$ and $\Lambda$ yields an interface with gluing condition 
$\Lambda\p\Lambda$. But in general not all intertwiners for $\Lambda\p\Lambda$ can be obtained by composing intertwiners for $\Lambda\p$ and $\Lambda$. So, a composition of two elementary interfaces produces a non-elementary interface in general.

\vskip 1mm
 
 In   reference  \cite{arXiv:0712.0076} it was shown that,  for $d=1$,  the $g$-factor is minimized by
   topological interfaces, and that  furthermore $g=1$ only for  the group-like invertible
   defects that generate the CFT isomorphisms. The following generalizes these results to any $d$: 
 
 \vskip 3mm
 
\underline{Lemma}: All $\widehat u(1)^{2d}$ invariant  interfaces have $g_{12}^{\Lambda}  \geq \sqrt{|\Gamma_2/\Gamma_{1,2}^\Lambda|}\geq 1$. 
The first inequality is saturated
 by topological interfaces for which $\Lambda$ belongs to   $O(d)\times O(d)$ so that $\vert {\rm det}  \Lambda_{22}\vert = 1$.  Furthermore, 
 all  $\widehat u(1)^{2d}$ invariant interfaces with $g=1$ generate isomorphisms of  torus CFTs.
  
\vskip 3mm
\underline{Proof}: it follows from  $\Lambda\in O(d,d)$ that
\beq
\Lambda_{22} \Lambda_{22}^t = 1 + \Lambda_{21} \Lambda_{21}^t \, \Longrightarrow\,
({\rm det}  \Lambda_{22})^2  = {\rm det} (1 + \Lambda_{21} \Lambda_{21}^t) \, \geq\,  1\ ,
\eeq
with equality holding if and only if $\Lambda_{21} =  \Lambda_{12}  = 0$. This in turn implies that 
$\Lambda\in O(d)\times O(d)$. In this case the interface operator commutes with left and right Virasoro algebras separately, \ie it corresponds to a topological interface.
 This can  also be verified by considering the reflection coefficient, which is zero if and only if the interface is topological. Following \cite{Quella2} it can be calculated to be 
 \beq
 {\cal R} = 1 - \vert {\rm det}  \Lambda_{22}\vert^{-2}\ . 
 \eeq 
Clearly the absolute minimum  $g=1$  can only be attained by topological interfaces, for which furthermore $\Lambda:\Gamma_2\rightarrow\Gamma_1$ is a lattice isomorphism. Being in $O(d)\times O(d)$ it therefore realizes an isomorphism of CFTs.
This shows the second part of the lemma.

\vskip 1mm

Using the cover $O(d|\RR)\times O(d|\RR)\backslash O(d,d|\RR)$ of the moduli space \eq{cosets} to parametrize toroidal CFTs, the interfaces with $g=1$ can be parametrized by elements of the group $O(d,d\vert \ZZ) \ltimes u(1)^{2d}$,
where $u(1)^{2d}$ parametrizes the moduli $\varphi$ of the interfaces, \cf \eq{bosintd}. As will be shown in Section \ref{secfusiond}, these defects indeed fuse according to the group multiplication in $O(d,d\vert \ZZ) \ltimes u(1)^{2d}$. Furthermore, defects with $g>1$ are not invertible with respect to fusion.

%%%%%%%%%%%%%%%%%%%%%%%%%%%%%%%%%%%%

\subsubsection*{Fermionic interfaces in torus models}

Also the construction of the fermionic interfaces for general $d$ parallels the discussion for $d=1$ in Section \ref{secfermionic}.

The aim is to construct superconformal interfaces between toroidal CFTs with specified ${\cal N}=(1,1)$ structures. The latter are determined by a choice of
supercurrents, which we take to be the normal ordered products
\beq
G=\sum_{i=1}^d:j^i\psi^i:\,,\qquad
\tilde G=\sum_{i=1}^d:\tilde\jmath^i\tilde\psi^i:\,,
\eeq
where the sums are taken over an orthonormal basis of $\RR^d$. This can always be attained by $O(d)\times O(d)$-rotations of the bosonic currents or the fermionic fields.
The requirement of supersymmetry 
\beq
  ( G_r ^1- i \eta_{\rm S}^1\, \tilde G_{-r}^1 ) I_{1,2}   =   \eta  I_{1,2}    ( G_r^2 - i \eta_{\rm S}^2\, \tilde G_{-r}^2 )\ . 
\eeq
combined with commutation relations \eq{defbosglue} for the bosonic modes forces commutation relations
with the fermionic modes $\psi^i_r$, which are now regarded as $d$-component vectors:
 \beq\label{gluingconditionFnnd}
 \vect{ \psi^1_r }{ \ -  i  \, \tilde  \psi^1_{-r} } I_{12} = 
  I_{12}  \,   \Lambda_{{\rm F}}  \vect{ \psi^2_r }{ \ -  i \, \tilde  \psi^2_{-r} } 
  \,   
\eeq
where the $O(d,d)$ matrix $\Lambda_{{\rm F}}$ is related to the bosonic gluing matrix $\Lambda$ by 
\beq\label{lambdaprimed>1}
\Lambda_{{\rm F}} =  \eta \mat{\bf 1}{0}{0}{\eta_{\rm S}^1{\bf 1}} \Lambda \mat{\bf 1}{0}{0}{\eta_{\rm S}^2{\bf 1}} \ ,  
\eeq

\vskip 1mm
 
In complete analogy with the $d=1$ case, one can write the fermionic intertwining operators in the NS and R sectors as 
\beq\label{fintd>1}
I_{1,2}^{\rm NS}=\prod_{r\in\NN-{1\over 2}} I_{1,2}^{r,{\rm ferm}} I_{1,2}^{0,{\rm NS}}\,,\quad
I_{1,2}^{\rm R}=\prod_{r\in\NN} I_{1,2}^{r,{\rm ferm}} I_{1,2}^{0,{\rm R}}\,.
\eeq
Here the modes of CFT1 and CFT2 in the exponentials
\beq\label{opsnonzeromodesfd}
I_{1,2}^{r,{\rm ferm}}=
    {\rm exp}\hskip -1mm \left(  -i  \psi^1_{-r}\OOO^{\rm F}_{11}  \tilde \psi^1_{-r} 
   +  \psi^1_{-r} \OOO^{\rm F}_{12}\psi^2_r   
  - \tilde \psi^2_r \OOO^{\rm F}_{21}  \tilde \psi^1_{-r} 
    -  i  \tilde \psi^2_{r} \OOO^{\rm F}_{22}   \psi^2_{r}  
    \right)   \ 
\eeq
act on the left respectively right of the maps $I_{1,2}^{0,{\rm NS}}$ and $I_{1,2}^{0,{\rm R}}$ between the NS and R ground states of the theory. Since there is only a single ground state in the NS sector the ground state part of the interface reads
\beq
I_{1,2}^{0,{\rm NS}}=|0\ket^1_{\rm NS}\,{}_{\rm NS}^{\phantom{,,}2}\bra0|\,.
\eeq

To describe the map on the Ramond ground states, we recall that the fermionic zero modes
$\psi_0^i$ and $-i\wt\psi_0^i$ for each of the two theories  form the 
Clifford algebra of $\RR^{d,d}$ and transform under the fundamental representation of $O(d,d)$.
The induced representation on the Ramond ground states is the spinor representation $S$, \ie 
\beq
\vect{\psi_0}{-i\wt{\psi}_{0}}{\mathcal S}(\Lambda_{\rm F})={\mathcal S}(\Lambda_{\rm F})\Lambda_{\rm F}\vect{\psi_0}{-i\wt{\psi}_{0}}\,.
\eeq
Thus, if we denote by $\imath_{1,2}^{\rm R}$ the isomorphism between the Ramond ground states of CFT1 and CFT2, commuting with the action of the fermionic zero modes,
then  the map $\imath_{1,2}^{\rm R}\, S(\Lambda_{\rm F})$ implements the zero mode 
part of the commutation relations \eq{gluingconditionFnnd}.

The normalization is fixed by Cardy's condition  which requires 
\beq\label{trace155}
{\rm tr}_{{\rm R}_2}\left(\left(I_{1,2}^{0,{\rm R}}\right)^* I_{1,2}^{0,{\rm R}}\right)=2^d\ , 
\eeq
where the trace is over the Ramond ground states of CFT2. The factor of $2^d$ on the right-hand-side  is absorbed
by the transformation of  the  annulus  partition function (in the folded picture) between  the
closed-string and the  open-string channels.  It   generalizes to higher $d$
  the   factor $2^{1\over 2}$  in the formula \eq{RRd1}
for the Ramond boundary state, \cf   \eq{annulus1}.  
The  conjugation $(\cdot)^*$  in CFT amounts to  Hermitean-conjugation of  the spinor
matrix  $S(\Lambda_{\rm F})$.  This does not give  in general  the inverse matrix because the group $O(d,d)$
is not compact. Instead one finds
\beq
{S}(\Lambda_{\rm F})^\dag  = {S}\left( {\scriptsize \mat{\bf 1}{0}{0}{-{\bf 1}}}  \Lambda_{\rm F}^{-1} {\scriptsize \mat{\bf 1}{0}{0}{-{\bf 1}}}\right) = {S}(\Lambda_{\rm F}^T)
\eeq
Thus  the left-hand-side of  \eqref{trace155} is equal to the spinor trace
${\rm tr}{S}(\Lambda_{\rm F}^T\,  \Lambda_{\rm F})$.

To calculate this trace  we note that the square of the spinor representation 
is isomorphic to the sum of the exterior powers of the fundamental representations
of $O(d,d)$:
\beq
{\mathcal R}:=\Lambda^*\RR^{d,d}\cong {S}\otimes {S}\,.
\eeq
Moreover, for any $A \in O(d,d)$
\beq
{\rm tr}_{\mathcal R}\left({\mathcal S}(A^T A )\right)=\det(1+A^T A)
=2^{2d}|\det(A_{11})|^2=2^{2d}|\det(A_{22})|^2\,.
\eeq
Taking everything together, the properly normalized Ramond ground state contribution of the interface operator is given by
\beq\label{rgsop}
I_{1,2}^{0,{\rm R}}={1\over \sqrt{|\det(\Lambda_{22})|}}\, \imath_{1,2}^{\rm R}S(\Lambda_{\rm F})\,. 
\eeq
Here we have used the fact that the absolute values of the determinants of the 2-2 blocks of
bosonic gluing matrix $\Lambda$ and fermionic gluing matrix $\Lambda_F$ agree. 
The   normalization of $I_{1,2}^{0,{\rm R}}$ exactly cancels the part of the bosonic $g$-factor \eq{geng} which continuously depends on the 
gluing condition $\Lambda$. 

%%%%%%%%%%%%%%%%%%%%%%%%

\subsubsection*{Fermion-parity projections} 

 In the unprojected theory,   where there is only an NS sector,   the complete interface operators are given by tensor products\beq\label{completeinterfaced}
I_{1,2}^{\rm full} (\Lambda,\varphi,  \eta)=I_{1,2}^{\rm bos}(\Lambda,\varphi)\otimes
I_{1,2}^{\rm NS}(\Lambda_{\rm F})\  ,   
\eeq
of bosonic and fermionic interface operators \eq{bosintd} and \eq{fintd>1}. For ease of notation we suppress the dependence on  $\eta$ and $\eta_S^i$.

\smallskip 

 The  GSO-projection of these interfaces   
   works exactly as in  the one-dimensional case discussed in Section \ref{projsection}.
    It 
    amounts to taking the orbifold with respect to the $\ZZ_2\times\ZZ_2$ generated by the $(-1)^{F_i+\wt F_i}$. 
 The complete operators are  products of operators for   bosons and fermions,
\beq\label{completeinterface1d}
I_{1,2}^{\rm full} (\Lambda,\varphi, h)=I_{1,2}^{\rm bos}(\Lambda,\varphi)\otimes
I_{1,2}^{{\rm ferm}, h}(\Lambda_F )\ . 
\eeq
The label $h$ takes three values, which can be identified  with the primary fields of the Ising model ($1, \epsilon$
and $\sigma$). The first two values correspond to charged interfaces, which exist whenever  
${\rm det}\Lambda_F  =  \zeta$, while $h= \sigma$  corresponds to  (simple) neutral interfaces
which exist   if ${\rm det}\Lambda_F = - \zeta  $.  We recall from Section \ref{projsection} that $\zeta$ distinguishes whether CFT1 and CFT2 are of the same ($\zeta=1$) or of opposite ($\zeta=-1$) GSO type. 
 
 \smallskip
 The  two  charged  fermionic interfaces are given by
\beqn\label{completeinterface2d}
I_{1,2}^{{\rm ferm}, \, c \pm } = {1\over {2}}\left(
I_{1,2}^{\rm NS}(\Lambda_{\rm F})\pm I_{1,2}^{\rm R}(\Lambda_{\rm F})\right)  + (\eta \to -\eta ) \ , 
\eeqn
 while the neutral ones, which have no Ramond component,  read
\beqn\label{completeinterface3d}
I_{1,2}^{{\rm ferm}, \, n } = 
{1\over\sqrt{2}}\,   I_{1,2}^{\rm NS}(\Lambda_{\rm F})+ (\eta \to -\eta ) \ . 
\eeqn
Note that changing $\eta$ to $-\eta$ just multiplies $\Lambda_F$ with $-1$.

  The    $\eta = \pm 1$ terms in  the 
sum correspond to the  orbit  of the interface operator when acted upon by
the fermion parity operator $(-)^{F_1+\wt F_1}$. These orbits are  normalized with the standard  $1/\sqrt{2}$ factor.

   From the above  expressions, and taking into account that  the NS ground state contributes equally  
   to  the two terms of the   orbit,   one finds the following relations for the $g$ factors of the projected interfaces:
$g = g_{\rm bos}$ in the charged case, 
and $g = \sqrt{2}\, g_{\rm bos}$ in the neutral one.

%%%%%%%%%%%%%%%%%%%%%%%%%%%%%%%%%%%%%

\subsection{Fusion of interfaces} \label{secfusiond}
  
The fusion of  the $d\geq 1$   interfaces   can now be analyzed easily
using the same approach which was applied to the treatment of the $d=1$ case in Section \ref{secfusion}. Indeed, the calculations
 for the fusion of the positive-frequency contributions
carry over immediately:\footnote{Here we indicate the dependence of the interfaces on the orthogonal matrices $\OOO=\OOO(\Lambda)$.}
 \beq\label{boscomp211}
 I_{1,2}^{>,{\rm bos}}(\OOO\p)e^{-\delta H}I_{2,3}^{>,{\rm bos}}(\OOO)=
 \prod_{n>0}\det(1-e^{-2\delta n} \OOO_{11}\OOO\p_{22})^{-1}I_{1,3}^{n, \rm bos}(\OOO^{\prime\prime} (e^{-\delta n}))\,, \nonumber
 \eeq
   \beq\label{fermcomp211}
 I_{1,2}^{>,{\rm ferm}}(\OOO\p)e^{-\delta H}I_{2,3}^{>,{\rm ferm}}(\OOO)=
 \prod_{r>0}\det(1-e^{-2\delta r} \OOO_{11}\OOO\p_{22})I_{1,3}^{r, \rm ferm}(\OOO^{\prime\prime}(e^{-\delta r}))\ , 
 \eeq
where the matrix $\OOO^{\prime\prime}(x)$ depends on 
$\OOO$, $\OOO\p$ and $x$ as follows:
 \beq\label{compmat11}
{ \scriptsize
  \OOO^{\prime\prime}(x)=\mat{\OOO\p_{11}+x^2\OOO\p_{12}(1-x^2\OOO_{11}\OOO\p_{22})^{-1}\OOO_{11}\OOO\p_{21}}
  {x\OOO\p_{12}(1-x^2\OOO_{11}\OOO\p_{22})^{-1}\OOO_{12}}
 {x\OOO_{21}(1-x^2\OOO\p_{22}\OOO_{11})^{-1}\OOO\p_{21}}
 {\OOO_{22}+x^2\OOO_{21}(1-x^2\OOO\p_{22}\OOO_{11})^{-1}\OOO\p_{22}\OOO_{12}} \, .
 }
 \eeq
Just as in the $d=1$ case, 
the matrices $ \OOO^{\prime\prime}(e^{-\delta n})$ appearing in these formulae converges to $\OOO(\Lambda\p\Lambda)$ for $\delta\to 0$, 
but the determinant factors exhibit a singular behavior in the limit. 
\vskip 1mm

The  singular behavior cancels whenever  the two interfaces $I_{1,2}$ and  $I_{2,3}$
preserve the same supersymmetry in CFT2, \ie the two interfaces must have the same $\eta_S$ for the CFT in their middle. 
\smallskip

In this case  the determinant factors coming from bosons and fermions exactly cancel each 
other in the Ramond sector. In the
 NS sector, on the other hand,  the cancelation leaves a finite remainder, which can be computed 
 with the help of  the Euler-Maclaurin formula \eq{euler} as in the case $d=1$. 
 The result for the fusion of the combined positive-frequency parts  is 
\beqn\label{fusionposmodes}
&&I_{1,2}^{>}(\Lambda\p ,\eta\p )\  I_{2,3}^{>}(\Lambda ,\eta)=\\
\nonumber
&&\qquad\qquad I_{1,3}^>
(\Lambda\p\Lambda, \eta\p\eta )\times\left\{\begin{array}{cc}
\sqrt{\det(1- \OOO_{11}\OOO\p_{22})}\, &{\rm NS\,sector}\  , \\
1\, &{\rm R\,sector\ . }\end{array}\right.
\eeqn

Let us next discuss the fusion of the zero-mode contributions, which can be composed without a regulator.
 In the bosonic sector the result is
 \beq\label{bos0fus}
I_{1,2}^{0,{\rm bos}}(\Lambda\p)I_{2,3}^{0,{\rm bos}}(\Lambda ) 
 =
  {|\Gamma_{1,3}^{\Lambda\p\Lambda }/\Gamma_{1,2}^{\Lambda\p}\odot\Gamma_{2,3}^{\Lambda }|
\over\sqrt{\det(1- \OOO_{11}\OOO\p_{22})}}\, 
 g_{1,3}^{\Lambda\p\Lambda }\!\!\!\!\!\!
  \sum_{\gamma\in\Gamma_{1,2}^{\Lambda\p}\odot\Gamma_{2,3}^{\Lambda}}
\!\!\!\!\!\! e^{2\pi i (\varphi\p\Lambda +\varphi)(\gamma)}|\Lambda\p\Lambda \gamma\ket\bra\gamma|\,,
 \eeq
  where the lattice
  \beq
\Gamma_{1,2}^{\Lambda\p}\odot\Gamma_{2,3}^{\Lambda}
 \eqdef (\Lambda\p\Lambda)^{-1}\Gamma_1\cap \Lambda^{-1}\Gamma_2\cap\Gamma_3
 \eeq
  is the sublattice of those intertwiners 
 for the composed gluing condition $\Lambda\p\Lambda$ which can be obtained by 
 fusion of intertwiners of $\Lambda\p$ and $\Lambda$ respectively. 
   Note that if $\Lambda\p$ and $\Lambda$ are admissible gluing conditions, \ie the lattices of intertwiners for 
   both of them are of maximal rank $2d$, so is $\Lambda\p\Lambda$. Moreover this is also true for
$ \Gamma_{1,2}^{\Lambda\p}\odot\Gamma_{2,3}^{\Lambda}$,  
 which is a maximal-rank sublattice of index
 \beq
 {\rm ind}( \Gamma_{1,2}^{\Lambda\p}\odot\Gamma_{2,3}^{\Lambda}\subset
 \Gamma_{1,3}^{\Lambda\p\Lambda})=|\Gamma_{1,3}^{\Lambda\p\Lambda}/\Gamma_{1,2}^{\Lambda\p}\odot\Gamma_{2,3}^{\Lambda}|
 \eeq
 in $\Gamma_{1,3}^{\Lambda\p\Lambda}$. 
 Thus, setting aside for the moment  the overall normalization,  one sees that the zero-mode contributions
 to the bosonic intertwiners multiply to one with composed gluing conditions. In general however, the result is not an elementary intertwiner.
  Instead it consists of  
$|\Gamma_{1,3}^{\Lambda\p\Lambda}/\Gamma_{1,2}^{\Lambda\p}\odot\Gamma_{2,3}^{\Lambda}|$
 elementary summands with different phases so as to project 
on the sublattice  $\Gamma_{1,2}^{\Lambda\p}\odot\Gamma_{2,3}^{\Lambda}$ of charges,
\cf the discussion around \eq{Lint}. 
\smallskip

Let us now show that the normalization of the right-hand-side of \eq{bos0fus} is indeed correct. 
 To show this we need to establish the identity
\beq
\left(g_{1,2}^{\Lambda\p} g_{2,3}^{\Lambda}\over g_{1,3}^{\Lambda\p\Lambda}\right)
= {
|\Gamma_{1,3}^{\Lambda\p\Lambda}/ \Gamma_{1,2}^{\Lambda\p}\odot\Gamma_{2,3}^{\Lambda}|
 \over\sqrt{\det(1- \OOO_{11} \OOO\p_{22} )}}\,.
 \eeq
 Consider first the factor of the $g$-functions \eq{geng} which depends continuously on the gluing conditions.
 Using the relation  \eqref{SintermsofOd} between $\OOO$ and $\Lambda$ we find
  
 $$ (\Lambda\p\Lambda)_{22} = \Lambda\p_{21}\Lambda_{12} + \Lambda\p_{22}\Lambda_{22} =
 \Lambda\p_{22} ( 1 +   \Lambda^{\prime\ -1}_{22} \Lambda\p_{21}\Lambda_{12} \Lambda_{22}^{-1})\Lambda_{22}
 $$
 $$
 =  \Lambda\p_{22} ( 1 -  \OOO\p_{22} \OOO_{11})\Lambda_{22}\ , 
 $$
 so that taking the determinants yields
 \beq\label{detidentity}
\det(1- \OOO\p_{22} \OOO_{11}) = \det(1- \OOO_{11}\OOO\p_{22})= {\det((\Lambda\p\Lambda)_{22})\over\det(\Lambda\p_{22})\det(\Lambda_{22})}\,.
\eeq
 To complete the proof of  \eq{bos0fus} it remains to 
  be shown that 
 \beq\label{indexidentity}
   {K\p K \over K^{\prime\prime}} \equiv { |\Gamma_2/\Gamma_{1,2}^{\Lambda\p}|\,
 |\Gamma_3/\Gamma_{2,3}^{\Lambda}|
 \over
 |\Gamma_3/\Gamma_{1,3}^{\Lambda\p\Lambda}|}
 =
|\Gamma_{1,3}^{\Lambda\p\Lambda}/\Gamma_{1,2}^{\Lambda\p}\odot\Gamma_{2,3}^{\Lambda}|^2\, . 
\eeq
 This index identity is proved  in Appendix \ref{indid}.
 \smallskip 
 
  The composition of the zero-mode contribution of the interfaces in the fermionic sectors is simpler. In the NS sector it is actually  trivial 
  \beq\label{173NS}
 I_{1,2}^{0,{\rm NS}} I_{2,3}^{0,{\rm NS}}=I_{1,3}^{0,{\rm NS}}\,.
 \eeq
Thus putting together   \eqref{fusionposmodes}, \eqref{bos0fus} and \eqref{173NS} we find,  in the full  unprojected
theory, that   the composition of two simple defects with indices $K\p$ and $K$ gives    $\sqrt{ K\p K / K^{\prime\prime}} $
defects with index $K^{\prime\prime}$. The index identity \eq{indexidentity} proves that this number is integer,  as it should. 
 The  defects that arise in this way have their phase moduli
arranged in a periodic array, so as to implement a projection on a sublattice of the lattice of all intertwiners
that are compatible with the transformation $\Lambda\p\Lambda$.  

%%%%%%%%%%%%%%%%%%%%%%%%%%%%
\smallskip

The calculation in the  GSO-projected theory 
  goes through exactly as in the $d=1$ case discussed at the end of Section \ref{3.3}.
  One only needs to check the composition of ground state intertwiners \eqref{rgsop} in the Ramond sector, 
    \beq
 I_{1,2}^{0,{\rm R}}(\Lambda\p_{\rm F}) I_{2,3}^{0,{\rm R}}(\Lambda_{\rm F})=
 \sqrt{\det(1-\OOO_{11}\OOO\p_{22})}\;\;
 I_{1,3}^{0,{\rm R}}(\Lambda\p_{\rm F}\Lambda_{\rm F})\,,
 \eeq
where use was made here of    \eq{detidentity}. 
The final result for the fusion can be summarized as follows:
 the fermionic part of 
GSO-projected interfaces is  labelled by   $h = 1, \epsilon, \sigma$, corresponding
to the three primary fields of the Ising model. The fusion  of these fermionic parts
follows the same pattern as the Verlinde algebra of the Ising model.

 This is the only difference
with the  unprojected theory, 
where the fermionic part  is  labelled by the sign  $\eta=\pm1$.  
Let us,  for the rest of this section,  fix the fermionic parts  by choosing the identity labels  ($\eta = 1$, 
  or $h= 1$) and  concentrate on the algebra of the 
  bosonic parts,   which is  the same in  the GSO-projected and in the unprojected theory.   
 \smallskip

  We can  give a more economic description of  this algebra  by enlarging  the set of simple interfaces
  to include interfaces  $I^L_{1,2}(\Lambda,\varphi)$, where $L$ is any (maximal rank) sublattice of  $\Gamma_{1,2}^\Lambda$. 
 This latter  is the lattice of   intertwiners contributing to the simple interface with gluing matrix $\Lambda$. 
If $L = \Gamma_{1,2}^\Lambda$ the interface is simple, otherwise it is a sum of  $\vert \Gamma_{1,2}^\Lambda/L\vert$  simple  interfaces
whose phase moduli are  arranged so as to enforce the projection on  $L$. 
   In terms of this larger  set of basic interfaces, the fusion of two interfaces  takes the following  elegant form: 
  \beqn\label{totalfusion}
  I^{L\p}_{1,2}(\Lambda\p,\varphi\p)\odot
 I^{L}_{2,3}(\Lambda ,\varphi ) 
  =I_{1,3}^{\Lambda^{-1}  L\p \cap L }(\Lambda\p\Lambda ,\varphi\p\Lambda +\varphi  )\,. 
\eeqn
 
 As mentioned before, it is clear that an interface $I^L_{1,2}(\Lambda)$ 
is invertible if and only if $L=\Gamma_{1,2}^\Lambda=\Gamma_2$ is the full charge lattice. Parametrizing  $\Gamma_i=U_i\ZZ^{d,d}$, this can only be achieved for
$\hat\Lambda \equiv U_1^{-1}\Lambda U_2\in O(d,d|\ZZ)$. 
 A special class of such interfaces are the deformation interfaces for which $\hat\Lambda = {\bf 1}$, 
\beq
D_{1,2}=I_{1,2}^{\Gamma_2}(U_1U_2^{-1},0)\,. 
\eeq
These encode the effect of deformations of the underlying bulk CFTs \cite{Brunner:2010xm}. 
 One can use them on both sides  to transport  any interface to a defect line in some reference torus model CFT0, 
\beq\label{setofinterfaces}
I_{1,2}^L(\Lambda,\varphi)=D_{1,0}\odot I_{0,0}^{U_0U_2^{-1}L} (U_0\hat\Lambda U_0^{-1},\varphi U_2 U_0^{-1})\odot D_{0,2}\,. 
\eeq
 Since the deformation interfaces are invertible,  the  fusion of   two arbitrary   interfaces 
 can be  completely determined by the fusion of the corresponding defect lines in the reference  CFT, which in turn does not depend on the choice of CFT0.
 \smallskip

 We may therefore  drop the explicit dependence on  CFT0 and characterize a defect  by  the data  $(\hat\Lambda ,  \varphi , \hat L)$,
 where $\hat\Lambda\in O(d,d|\QQ)$,  $\varphi$ is a linear form on $\ZZ^{d,d}$, and $\hat L$ a 
 maximal-rank sublattice     of the intertwiner lattice
 $ \ZZ^{d,d}\cap\hat\Lambda^{-1}\ZZ^{d,d}$  for the  integer charges. 
The composition rule for defect lines in this representation can be easily read off from 
\eq{totalfusion}:
 \beq\label{Fusion222}
 (\hat\Lambda\p ,  \varphi\p , \hat L\p) \odot  (\hat\Lambda ,  \varphi , \hat L) = 
 (\hat\Lambda\p\hat\Lambda ,\,  \varphi\p\hat\Lambda + \varphi , \,  \hat L \cap \hat \Lambda^{-1} \hat L\p)\ . 
 \eeq
 We note that since $\hat\Lambda\in O(d,d|\QQ)$, its inverse is also a matrix with rational entries so that
 $\hat L \cap \hat \Lambda^{-1} \hat L\p$ has maximal rank.
  
  \smallskip 
Invertible defects are those for which $\hat\Lambda\in O(d,d|\ZZ)$ 
and $\hat L=\ZZ^{d,d}$. They fuse according to the group $O(d,d|\ZZ)\ltimes u(1)^{2d}$, where
 the $u(1)^{2d}$ is generated by the phases $\varphi$. The fusion monoid ${\mathcal D}$ 
 for the more general defects is then  described by the semi-group extension 
\beq\label{extension}
1\longrightarrow {\mathcal S}\longrightarrow{\mathcal D}\longrightarrow O(d,d|\QQ)\ltimes u(1)^{2d}\longrightarrow 1
\eeq
of the group $O(d,d|\QQ)\ltimes u(1)^{2d}$ of all admissible gluing conditions and all phases by the semi-group ${\cal S}$, whose elements are maximal rank sublattices of $\ZZ^{d,d}$, and which multiply by taking intersections.\footnote{${\mathcal S}$ consists of the defects $({\bf 1},0,\hat L)$.}

%%%%%%%%%%%%%%%%%%%%%%%%%%%%%%%

\subsection{Fusion with boundary conditions}

Finally, let us sketch how the interfaces defined above fuse with boundary conditions. Since most of the calculations are similar to the ones done before, we will just state the result. 

A general $\widehat u(1)^d$ preserving boundary condition in a $d$-dimensional torus model with charge lattice $\Gamma$ is determined by the following objects. First the left and right currents are glued together by means of an orthogonal matrix $\Omega \in O(d)$, such that the respective boundary state is annihilated by the combinations $\{a_n+\Omega \tilde
 a_{-n}\,|\,n\}$. Such a gluing condition can only be realized by a boundary condition, if the lattice of Ishibashi states
\beq
\Gamma^\Omega =\Gamma\cap \left\{\vect{-\Omega x}{x}\,|\,x\in\RR^d\right\}
\eeq
has   rank $d$. This  guarantees that the volume of the corresponding D-brane is finite,  and 
the worldvolume gauge fluxes quantized. 
\smallskip

Then, as in Section \ref{DNstates}, 
for every choice of $\eta_{\rm S}\in\{\pm 1\}$, and $\varphi\in(\Gamma^\Omega )^*$ one finds a supersymmetric and $\widehat u(1)^d$ invariant elementary boundary state $|\Omega ,\varphi, 
\eta_{\rm S}\keti$.  In the GSO-projected theory the sign of ${\rm det} \Omega $ and $\eta_{\rm S}$ determine whether the D-brane is charged or neutral, 
whereas in  the unprojected theory these two signs are independent. 
 Furthermore, by summing suitable combintations of
$|\Gamma^\Omega /L|$ elementary boundary states one can construct new boundary states which only couple to a maximal rank sublattice $L\subset\Gamma^\Omega $ 
of possible Ishibashi states. 
We denote the result by
$|\Omega ,\varphi, \eta_{\rm S}\keti^L$. 

\smallskip 

The fusion of interfaces with boundary states is easy to compute. If  they preserve the same supersymmetry, the fusion is non-singular  and the result reads\footnote{We suppressed the 
$\eta_{\rm S}$-dependence.}
\beqn\label{181fuse}
&&I_{1,2}^{L\p} (\Lambda,\varphi\p )\odot
|\Omega ,\varphi \keti^{L}_2=\\
&&\qquad|(\Lambda_{12}-\Lambda_{11}\Omega )(\Lambda_{22}-\Lambda_{21}\Omega )^{-1},(\varphi+\varphi\p)\Lambda^{-1} \keti_1^{\Lambda(L\p\cap L)}\,.\nonumber
\eeqn
It is interesting to note that 
  $O(d,d)$ acts by fractional linear transformations on the gluing conditions in $O(d)$. 
\smallskip  
  
Using the invertible deformation interfaces, we can transport the above result   to
any reference model CFT0 with charge lattice $\Gamma = U \ZZ^{d,d}$.   The  fusion of a defect line with a boundary  condition 
in CFT0 can then be described by 
the action of the defect line on the  rank $d$   sublattice $U^{-1}\Gamma^\Omega$ of those 
integer charges in  $\ZZ^{d,d}$ which couple to the
boundary state. 
This sublattice is defined to be the kernel of $ ( {\bf 1} ,  \Omega ) U$ in $\ZZ^{d,d}$.
Here,  $ ( {\bf 1} ,  \Omega ) $ is a rectangular $2d\times d$
matrix.  The rank of $\Gamma^\Omega$ equals  $d$, if and only if the $d\times d$ matrix
\beq
\hat \Omega  \eqdef  (U_{11} + \Omega  U_{12})^{-1} (U_{21} + \Omega  U_{22}) \,   \in \, GL(d, \mathbb{Q})\ , 
\eeq
\ie it is invertible and has only rational entries. 
It follows from \eqref{181fuse} that  the $O(d,d|\QQ)$   matrix  $\hat \Lambda$  acts by
fractional linear transformations on  $\hat \Omega $, and that the corresponding lattices
compose according to $\hat\Lambda (\hat L\p \cap \hat L)$.

%%%%%%%%%%%%%%%%%%%%%%%%%%%%%%%%%%%%
%%%%%%%%%%%%%%%%%%%%%%%%%%%%%%%%%%%%%
%%%%%%%%%%%%%%%%%%%%%%%%%%%%%%%%%%%%%%

\section{Fusion of interfaces and geometric integral transformations}

There is another useful formula for the Ramond ground state contribution of the interface operators, which one obtains by first considering the associated folded boundary conditions. 
As discussed explicitly in the one-dimensional case in Section \ref{secfermionic}, the Ramond
ground state contribution $|\OOO_{\rm F}\ket_{\rm R}$ of the boundary states can be obtained by rewriting the folded gluing conditions \eq{fermbdgluing2} for the zero modes in terms of
\beq
\gamma^j_\pm \eqdef  {1\over\sqrt{2}}\left( \psi_0^j \pm  i \tilde\psi_0^j\right)\ . 
 \eeq
 This yields
\beq\label{fermbdgluing5d}
\left[ \vect{\gamma^1_{+}}{\gamma^2_{+}}  +   {\cal F}
 \vect{\gamma^1_{-}}{\gamma^2_{-}} \right]  \vert \OOO_{\rm F} \rangle_{\rm R}   
=0\, ,
\eeq
where, ${\cal F}$ is the antisymmetric matrix defined by
\beq\label{FintermsofOd}
\OOO_{\rm F} =  ({\bf 1}  +  {\cal F})^{-1} ({\bf 1} - {\cal F}) 
 \ \Longleftrightarrow\  {\cal F}  =  ({\bf 1} - \OOO_{\rm F}) ({\bf 1}+\OOO_{\rm F})^{-1}\ . 
\eeq
In case $\OOO_{\rm F}$ has an eigenvalue $-1$, we take ${\cal F}$ to be restricted to the orthogonal 
complement of the respective eigenspace $E_{-1}(\OOO_{\rm F})$. Furthermore, we pick a normalized volume form 
$\omega_{\OOO_{\rm F}}$ on this eigenspace and insert into it the $2d$-vector $(\gamma_-^1,\gamma_-^2)$. Denoting the result by $\omega_{\OOO_{\rm F}}(\gamma_-^i)$
 the normalized solution of equations   \eqref{fermbdgluing5} can be written as
 \beq\label{newRRd}
 \vert \OOO_{\rm F}  \rangle_{\rm R}\,  =  \,  [{\rm det}(1- {\cal F}  )]^{-{1\over 2}} 
    \, {\rm exp}\left( - {1\over 2} {\cal F} _{jl}\,   \gamma_-^l \gamma_-^j \right) \omega_{\OOO_{\rm F}}(\gamma_-^i) \vert {\bf 1}  \rangle_{\rm R} \, ,
 \eeq 
 where $|{\bf 1}\ket_{\rm R}$ is the normalized pure spinor state, \ie the normalized state annihilated by all the $\gamma_+^i$. Multiplied by $2^{d\over 2}g_{\rm bos}$, this is the Ramond charge vector of the boundary state. In non-linear sigma models, Ramond charges of 
 boundary conditions have a geometric meaning as Chern characters of the associated D-branes (see \eg Section 1 of \cite{Brodzki:2006fi} for a brief summary of the geometric aspects of Ramond charges). 
 
 The D-branes we are considering here are supported on affine subtori which are orthogonal to the $-1$-eigenspace of $\OOO$. They are equipped with $U(1)$-bundles whose curvature can be represented by the constant $2$-form $F={\mathcal F}$. Identifying 
the $\gamma_-^i$ with constant one-forms on the target space torus, 
 we indeed find 
 \beq
 \sqrt{{\rm vol}_T}\, e^{-F}{\rm PD}({\cal W})= \sqrt{{\rm vol}_T}\, Q^{\rm R}({\cal W},F)
 \eeq
 for the Ramond charge vector  of the corresponding D-brane. In this formula ${\rm vol}_T$ denotes the volume of the target space torus, 
 and ${\rm PD}({\cal W})$ is the Poincar\'e dual of the D-brane world volume ${\cal W}$.

 The state \eq{newRRd} can now be easily unfolded using the behavior of the $\gamma_\pm$ under folding. Namely, using \eq{spell}, one finds that 
 \beq\label{gammafoldd}
 \gamma_{\pm}^2\mapsto\mp\gamma_{\pm}^2\,.
 \eeq
 Thus, $|\OOO_{\rm F}\ket_{\rm R}$ unfolds to
 \beq\label{fgsinterfaces}
 I_{1,2}^{0,{\rm R}}=
 \,  \left({{\rm det}(1- {\cal F}  )}\right)^{-{1\over 2}} 
    \, {\rm exp}\left( - {1\over 2} {\cal F} _{jl}\,   \gamma_-^l \gamma_-^j \right) \,\omega_{\OOO_{\rm F}}(\gamma_-) \,\vert {\bf 1}  \rangle^1_{\rm R} {}_{\rm R}^2\bra-{\bf 1}|\, ,
    \eeq
    where as always, the modes of CFT1 and CFT2 act on the left, respectively right of the map $\vert {\bf 1}  \rangle^1_{\rm R} {}_{\rm R}^2\bra-{\bf 1}|$ mapping the pure anti-spinor state of CFT2 to the pure spinor state of CFT1.\footnote{
    The pure anti-spinor state $|-{\bf 1}\ket_{\rm R}$ is the Ramond ground state anihilated by all the $\gamma_-$'s. That the state $|{\bf 1}\ket_{\rm R}$ folds to ${}_{\rm R}\bra -{\bf 1}|$ follows from the folding behavior \eq{gammafoldd} of the $\gamma_i$ and the fact that $\gamma_\pm^*=\gamma_\mp$.}

Of course, also this formula has a geometric meaning. Here the $\gamma_-^i$ are the constant one forms on the target space tori $T_i$ of CFTi. Thus, up to the map $|{\bf 1}\ket_{\rm R}^1{}_{\rm R}^2\bra{-{\bf 1}}|$, it is nothing but 
\beq
\sqrt{{\rm vol}_{T_1}\,{\rm vol}_{T_2}}Q^{\rm R}({\cal W},F)\,,
\eeq
the Ramond charge of the D-brane on $T_1\times T_2$ associated to the respective folded boundary state. However, while
$|{\bf 1}\ket_{\rm R}^1$ just corresponds to the $0$-form $1$ on $T_1$, $_{\rm R}^2\bra-{\bf 1}|$ maps a $k$-form $\nu$ on $T_2$ to  
\beq
{1\over{\rm vol}_{T_2}}\int_{T_2}\nu\,.
\eeq
Thus, including all normalizations, the interface operator restricted on the Ramond ground states can be viewed geometrically as the following operation on forms $\nu$ on $T_2$:
\beq
I_{1,2}^{\Omega^*}: \nu\longmapsto \sqrt{{\rm vol}_{T_1}\over{\rm vol}_{T_2}}
\int_{T_2}Q^{\rm R}({\cal W},F)\wedge\pi_2^*(\nu)\,,
\eeq
where $\pi_i:T_1\times T_2\rightarrow T_i$ are the projections on the factors. 

Up to the square root normalization which is a relic of a particular choice of identification of the ground states, this formula describes what happens to D-brane charges under geometric integral transformations (see \cite{Andreas:2004uf} for a discussion of these transformations). Any D-brane ${\cal W}$ on a product $X_1\times X_2$ defines such a transformation mapping D-branes ${\cal W}_2$ on $X_2$ to D-branes
\beq
{\cal W}_2\mapsto (\pi_1)_*({\cal W}\otimes \pi_2^*({\cal W}_2))
\eeq
on $X_1$. ${\cal W}$ is referred to as the kernel of this transformation. If such a transformation is invertible, it is often called Fourier-Mukai transform. 

Thus, the interfaces act on Ramond ground states in the same way as the corresponding
geometric integral transformations do on cohomology -- a point first alluded to in reference \cite{bbdo}. 
We believe that this is in fact true on the level of the full D-brane category, and that in particular interfaces fuse in the same way as the corresponding geometric integral transformations compose.

That T-dualities can be described by Fourier-Mukai transformations has been known for some time. More general geometric integral transformations on tori 
have been analyzed in \cite{Bruzzo}. Although we have not shown it in general, in all examples we have studied the fusion of interfaces indeed agrees with the composition of the associated geometric integral transformations.
 
 \vskip 1mm
 In conclusion, we have  two formulae for the action of the $\widehat u(1)^{2d}$ symmetric
interface operators  on R-ground states. One involves the  spin representation of $O(d,d)$  times the square root
of the interface index, while the second one is the action induced by geometric integral transformations. 
By definition, the latter has to be an endomorphism of the R-charge lattice.
 We don't know if the relation between these two,  geometric and algebraic,  formulae has appeared before
in the mathematics literature.

%%%%%%%%%%%%%%%%%%%%%%%%%%%%%%%%%%%%
%%%%%%%%%%%%%%%%%%%%%%%%%%%%%%%%%%%%%
%%%%%%%%%%%%%%%%%%%%%%%%%%%%%%%%%%%%%%

\section{Topological realization of  the defect monoid}\label{sec8}

There is an important special class of interfaces with the property that they commute with both left and
 right Virasoro algebras separately \cite{Petkova:2000ip, Bachas:2004sy}. 
This means that correlation functions do not change under the deformation of their positions as long as no other interfaces or field insertions are crossed. 
For this reason they are called ``topological". 
If they are invertible they realize honest isomorphisms of conformal field theories. 

By definition, the $\widehat u(1)^{2d}$ preserving interfaces we have constructed are topological if and only if their gluing condition
 $\Lambda\in O(d)\times O(d)$. 

We have seen that by means of parallel transport, fusion of conformal interfaces can  be understood in terms of fusion of conformal defects in a single torus model.
 The latter is given by \eq{Fusion222}. The corresponding  monoid 
 can be described as the semi-group extension \eq{extension} of $O(d,d|\QQ)\ltimes u(1)^{2d}$. 
In the following, we will explain how the topological interfaces ``inherit"  this  semi-group structure. 
 
 \vskip 1mm

 The deformation space of torus models is the space of even self-dual charge lattices $\Gamma\subset\RR^{d,d}$.
These are determined  by the geometric and B-field moduli   packaged in the
 symmetric $O(d,d)$ matrix  $M \equiv  2 U^T U$, \cf   \eqref{M3} and \eqref{vielbeins}.  
The lattice $\Gamma$ is the lattice of ``physical" charges.
In terms of the lattice of integer charges it
 is  given by $\Gamma = U \ZZ^{d,d}$. 
 Two torus models are of course identified  if they only differ by the choice of basis of   left and right $\wh u(1)^d$ currents. 
 Such a change of basis is implemented by the action of $O(d)\times O(d)$ on the vielbein $U$ from the left. This   leaves 
  $M$ invariant.  
 Thus the  matrices $M$  parametrize  the  (homogeneous)  coset 
   space ${\mathcal D}_d=O(d)\times O(d)\backslash O(d,d|\RR)$. 
\smallskip

 In fact, two charge lattices $U\ZZ^{d,d}$ and $U\hat\Lambda \ZZ^{d,d}$ are identical, whenever $\hat\Lambda \in O(d,d|\ZZ)$ is 
an automorphism of  $\ZZ^{d,d}$. 
Thus, the moduli space of torus models is given by 
${\mathcal D}_d/O(d,d|\ZZ)$. However, while the two charge lattices $U\ZZ^{d,d}$ 
and $U\hat\Lambda\ZZ^{d,d}$ agree, the automorphism $\hat\Lambda$ acts non-trivially on the charges. In particular
 taking the $O(d,d|\ZZ)$ orbifold of  ${\mathcal D}_d$ 
creates non-trivial monodromies on the bundle of CFT-Hilbert spaces over it.
 In general these monodromies are not symmetries of the CFTs in the sense that they do not separately commute with left and right Virasoro algebras.

 Since we are interested in describing interfaces between different torus models, which can be realized as operators between different fibers of the Hilbert space bundle, it is convenient to work with the deformation space ${\cal D}_d$  on which the Hilbert space bundle is trivial.  A choice of flat connection then allows to identify all the fibers by means of parallel transport. On the level of charges this is realized as a specific ``gauge choice" for the vielbein $U$, 
 for instance  the choice  dictated by  the Iwasawa
 decomposition of $O(d,d|\RR)$, see reference  \cite{Obers:1998fb}.

 The deformation interfaces $D_{y\p,y}$ between any two torus models $y$ and $y\p$ incorporate
  the parallel transport in the Hilbert space bundle, hence they have gluing condition $U(y\p)U(y)^{-1}$. 
\smallskip

 Consider now the set   \eqref{setofinterfaces} of all   interfaces obtained by parallel transport   
 of   a conformal defect. As was explained in Section  \ref{secfusiond}, a defect is uniquely  specified by the 
  data $(\hat\Lambda , \varphi , \hat L)$,  where $\hat\Lambda \in O(d,d\vert\mathbb{Q})$ is  the gluing matrix for integer-charge vectors.   
 If  $y$ and $y\p$ are the two torus CFTs   on the left respectively right
  of the interface, then the gluing condition for their physical  charges reads
 \beq
 \Lambda=U(y\p)\hat \Lambda U(y)^{-1}\,.
\eeq
It can be shown  that for   given $y$ and $\hat\Lambda$
 there exists a unique $y\p$ for which this interface is topological, \ie such that 
$ \Lambda \in O(d)\times O(d)$.  Indeed, suppose there were two  theories for which this was true, say $y\p$ and $y^{\prime\prime}$.  
Then   both $U(y\p)\hat \Lambda U(y)^{-1}$ and $U(y^{\prime\prime})\hat \Lambda U(y)^{-1}$ would be
elements of $O(d)\times O(d)$, and hence so would  $U(y^{\prime\prime}) U(y\p)^{-1}$. 
This contradicts our assumption that $U(y)$  was a good parametrization of  the coset space 
  ${\mathcal D}_d$, which proves the claim. 

\smallskip 
 
Therefore, any given conformal defect  $(\hat\Lambda , \varphi , \hat L)$  of a reference  torus model  
 gives rise to a  topological interface between any given torus model  $y$ and a  model $y\p = f(y, \hat\Lambda)$, with the latter
 uniquely fixed by  $y$ and  $\hat\Lambda$.  
   Clearly, the converse statement is  also  true:  every topological interface can be parallel transported
   by fusing with deformation interfaces on the left and right  to a unique
  defect in some reference CFT0.  
  Hence, for all torus models $y$, there exists  a bijection between conformal defect lines   and the topological interfaces starting in $y$. 

Being valid for all $y$, this bijection allows to pull back the fusion of arbitrary fusable topological interfaces 
to the one of conformal defect lines.
 Thus, the fusion of topological interfaces is a representation of the monoid of conformal defect lines
  in a fixed torus model.

\subsubsection*{Connection with effective supergravities } 

 The relation of topological interfaces with the $O(d,d\vert \mathbb{R})$ symmetries of the low-energy supergravity
 has been discussed in the introduction,  and for $d=1$ in Section \ref{sec5}. The 
 generalization 
 to higher $d$ is straightforward, so we will only sketch  it very briefly.

As alluded to in the introduction,   the  continuous $O(d,d)$ symmetry of the effective low-energy supergravity 
acts on  the closed-string moduli, while leaving  the physical charges
invariant  modulo  $O(d)\times O(d)$ rotations.   Since the Einstein-frame metric does not transform, 
also the  masses of black hole solutions do not change. This fits in nicely with the fact that topological interfaces,
which implement the  transformations on the string worldsheet as we have proposed in this work, 
leave  invariant the masses of both fundamental string states
and D-branes.

   The mass-squared of a fundamental string is proportional to 
 $L_0 + \tilde L_0$ which   by definition commutes with topological interfaces. 
 For D-branes the story is more subtle, but  still follows from general facts: 
  the D-brane mass-formula \eqref{Dmass}; the fact that fusion with a topological interfaces multiplies the g-factor of the boundary condition with the one of the topological interface; 
  the form 
  $g_{\rm top}  =  \vert{{\rm ind}(\hat\Lambda)}\vert^{1/2}$ of the relevant topological interfaces here; and, finally the rescaling of the string coupling with the interface index as in \eq{coupling1}.
  Putting all these facts together  shows that D-brane masses are
  invariant under the  transformations by the topological interfaces considered here. 
  
  Note that the  argument does not  use properties of the boundary state; it even holds
  for D-branes that break some, or all,  of the  $\widehat u(1)$ symmetries.  
  \smallskip
  
     As we argued for $d=1$ above,  also for higher $d$  the classical symmetry group $O(d,d\vert\mathbb{R})$  is replaced
     in the quantum theory by a semi-group, the extension \eq{extension} 
     of $O(d,d\vert\mathbb{Q})$ by the semi-group of maximal-rank sublattices of $\ZZ^{d,d}$.  
     This is necessary for the preservation of charge quantization.
     As explained, this algebraic structure is completely captured by the fusion algebra of conformal defect lines. Its action on boundary conditions defines a homomorphism of this semi-group into $[\mathbb{R}^+ \times {\rm Spin}(d, d)] \cap GL( 2^d\vert \ZZ )$. 
          
       We end by repeating once more that  non-invertible transformations in this semi-group
              are not exact symmetries of string theory, but should be thought of as  orbifold equivalences. They are
        symmetries  at leading order in the string coupling  and  all orders in $\alpha\p$.  Even though  restricted in scope,  such
        orbifold equivalences can have non-trivial consequences, see for example   \cite{Bershadsky:1998mb}. 
        It would be interesting to look for similar applications in the present context.

%%%%%%%%%%%%%%%%%%%%%%%%%%%%%%%%%%%%%
%%%%%%%%%%%%%%%%%%%%%%%%%%%%%%%%%%%%
%%%%%%%%%%%%%%%%%%%%%%%%%%%%%%%%%%%%%

\subsubsection*{\bf Acknowledgements} We  would like to  thank   Benjamin Assel,  Jaume  Gomis, Michael Green,  Axel Kleinschmidt,  Suresh Nampuri 
 and Pierre Vanhove for useful conversations.  C.~B.  aknowledges the hospitality of the Newton Institute, where part of this work
 was completed, and thanks in particular the organizers of  the program on
``The Mathematics and Applications of Branes in String and M-Theory". He  also thanks  the string-theory group at LMU for their warm
hospitality during an extended visit made possible by  the Humboldt foundation.  
D.~R. is supported by DOE-grant DEFG02-96ER40949. I.~B. is supported by a EURYI award of the European Science Foundation.

 %%%%%%%%%%%%%%%%%%%%%%%%%%%%%%%%%%%%%
%%%%%%%%%%%%%%%%%%%%%%%%%%%%%%%%%%%%%

 %\vfil\eject

\appendix

%%%%%%%%%%%%%%%%%%%%%%%%%%%%%%%%%%%%%
%%%%%%%%%%%%%%%%%%%%%%%%%%%%%%%%%%%%%

\section{Conventions}\label{conventions}

The  fields of the ${\mathcal N}=(1,1)$  SCFT are a free massless boson $\phi$,  and a   
 free massless Majorana fermion with (left, right)  components ($\psi$, $\tilde\psi$). 
The mode expansion of $\phi$  on the circle parametrized by $\sigma \in [0,2\pi]$  reads
 \begin{equation}\label{phi}
\phi  =  \hat \phi_0 + {\hat N\over 2R}\tau+   \hat MR\sigma + \sum_{n\not= 0}^\infty
{i\over 2 n } \, \left( a_n e^{-in(\tau+\sigma)} + \tilde a_n e^{-in (\tau - \sigma)}  \right)\ , 
\end{equation}
where $\hat N$,  $\hat M$ are the integer-valued
momentum and winding operators, and $R$ is the compactification radius. 
 The fermion mode expansions likewise read
  \begin{equation}\label{psi}
  (\psi  , \tilde \psi ) = \sum_{r} \  ( \psi_r e^{-ir(\tau+\sigma)} \ , \ 
  \tilde \psi_r e^{-ir(\tau-\sigma)} )  
\end{equation} 
with  $r$ integer  in the Ramond sector, and half-integer  in the Neveu-Schwarz sector. 
These modes obey  the reality conditions $a_n^\dag = a_{-n}$, $\psi_r^\dag = \psi_{-r}$ and  
likewise for the right movers. 
The canonical  commutation relations are
\begin{equation}
[a_n, a_m]  =  [\tilde a_n, \tilde a_m]  
= n \delta_{n+m, 0}\ \ \ {\rm and}
\ \ \ [\hat\phi_0 , {\hat N\over R} ] = i\ ,  \nonumber
\end{equation}
 \begin{equation}
\{ \psi_r , \psi_s \} =\{ \tilde\psi_r , \tilde\psi_s \} =  \delta_{r+s, 0}\ . 
\end{equation}

\vskip 1mm\noindent
The currents   generating the two $\widehat u(1)$  Kac-Moody algebras are  
 \begin{equation}
 \jmath =  2\, \partial_+\phi \equiv  \sum_{n\in \mathbb{Z}}\  \jmath_n\,  e^{in(\tau+\sigma)}\ ,  \qquad  
\tilde \jmath =   2\, \partial_-\phi \equiv  \sum_{n\in \mathbb{Z}}\  \tilde \jmath_n \, e^{in(\tau-\sigma)}\ . 
\end{equation}
Comparing with (\ref{phi}) gives 
\begin{equation}
 \jmath_0 = {\hat N\over 2R} + \hat M R \ \    \ \ {\rm and}   \qquad  \jmath_n =  a_n \ \  {\rm for}\ \  \ n\not=  0\ , \nonumber
\end{equation}
 with similar expressions for the right movers. 
In terms of these modes  the fermionic generators of the super-Virasoro algebras  read
\begin{equation}
G_r =   \sum_{n\in \mathbb{Z}} \   \jmath_n \psi_{r-n} \ , \qquad \tilde G_r =   \sum_{n\in \mathbb{Z}} \  \tilde \jmath_n \tilde \psi_{r-n} \ . 
\end{equation}

%%%%%%%%%%%%%%%%%%%%%%%%%%%%%%%%%%%%%
%%%%%%%%%%%%%%%%%%%%%%%%%%%%%%%%%%%%%

\section{Proof of the index identity}\label{indid}

In this appendix we prove the index  identity \eqref{indexidentity}. This identity is independent of the moduli of the CFTs. Writing the charge lattices $\Gamma_i=U_i\ZZ^{d,d}$, and replacing the gluing conditions $\Lambda\mapsto U_1\Lambda U_2^{-1}$, we can formulate it entirely with respect to $\Gamma\equiv\Gamma_0=\ZZ^{d,d}$. Setting $\Gamma^\Lambda\equiv\Gamma_{0,0}^\Lambda$, the identity can be written as
 \beq\label{indid0}
     \left\vert  { \Gamma  \over \Gamma^{\Lambda\p}} \right\vert   
    \left\vert  { \Gamma  \over \Gamma^{\Lambda}} \right\vert\ 
     =\   \left\vert { \Gamma^{\Lambda\p \Lambda}\over 
     \Gamma^{\Lambda\p}\odot\Gamma^{ \Lambda} } \right \vert^{\, 2}\, 
     \left\vert  { \Gamma  \over \Gamma^{\Lambda\p\Lambda}} \right\vert \ . 
     \eeq
Here, all gluing conditions are admissible, \ie $\Lambda,\Lambda\p\in O(d,d|\QQ)$. 
Note that in the canonical basis the $O(d,d|\QQ)$ matrices have rational entries 
\begin{equation} 
\Lambda =\left[\begin{array}{ccccc} 
 {p_{11}/  q_{11} }  & {p_{12}/  q_{12} }  &  \dots & \dots &  {p_{1\, 2d}/  q_{1\, 2d} } \\ 
\vdots & \vdots &  \vdots &  \vdots &  \vdots \\
 {p_{2d\, 1}/ q_{2d\, 1} }  & {p_{2d\, 2}/  q_{2d\, 2} }  &  \dots & \dots &  {p_{2d\, 2d}/  q_{2d\, 2d} }
\end{array} \right]
\end{equation}
where $(p_{ab}, q_{ab})$ are pairs of relatively prime integers, and  
$|\Gamma/\Gamma^\Lambda|={\rm lcm}(q_{ab})$.
Similar expressions can be written for $\Lambda^\prime$ and $\Lambda^{\prime\prime} = \Lambda\p\Lambda$. 
One implication of the index identity is then that 
\beq
{{\rm lcm}(q\p_{ab})\times {\rm lcm}(q_{ab})\over  {\rm lcm}(q^{\prime\prime}_{ab}) } \ =\  \tilde K^2\ , \quad \tilde K\in\NN\,,
\eeq
i.e. the left-hand-side is a perfect square. Its square root  is the index of 
the sublattice $ \Gamma^{\Lambda\p}\odot\Gamma^{\Lambda} $
in $ \Gamma^{\Lambda\p \Lambda} $. 

\vskip 2mm 

  To prove identity \eq{indid0}, we first rewrite it in the equivalent form
  \beq
  \vert\hskip -0.5mm\vert \Gamma^{\Lambda\p} \vert\hskip -0.5mm\vert\, \vert\hskip -0.5mm\vert \Gamma^{\Lambda}\vert\hskip -0.5mm\vert
  \, \vert\hskip -0.5mm\vert \Gamma^{\Lambda\p\Lambda}\vert\hskip -0.5mm\vert \ =\ 
  \vert\hskip -0.5mm\vert \Gamma^{\Lambda\p}\odot\Gamma^{\Lambda}
    \vert\hskip -0.5mm\vert^2\ , 
  \eeq
where  $\vert\hskip -0.5mm\vert L \vert\hskip -0.5mm\vert$ is the volume of a unit cell of the lattice $L$, 
and $\vert\hskip -0.5mm\vert \Gamma  \vert\hskip -0.5mm\vert = 1$. Next note that 
$\Gamma^{\Lambda\p}\odot\Gamma^{\Lambda} = \Gamma \cap \Lambda{}^{-1} \Gamma \cap (\Lambda\p\Lambda)^{-1} \Gamma$
is a sublattice of  both $\Gamma^{\Lambda}$ and  $\Lambda^{-1}\Gamma^{\Lambda\p}$, in addition to
 $\Gamma^{\Lambda\p\Lambda}$. We can thus write \eq{indid0}  as
\beq
  \left\vert  {   \Gamma^{\Lambda} \over  \Gamma^{\Lambda\p}\odot\Gamma^{\Lambda}} \right\vert   
    \left\vert  {\Lambda^{-1}\Gamma^{\Lambda\p} \over  \Gamma^{\Lambda\p}\odot\Gamma^{\Lambda} } \right\vert\  =\  
     \left\vert  { \Gamma  \over \Gamma^{\Lambda\p\Lambda}} \right\vert \ , 
\eeq
where we used the fact that $O(d,d)$ transformations are volume preserving. It will be actually easier
to establish this identity in its dual form,
\beq\label{appidentity}
  \left\vert  {   (\Gamma^{\Lambda\p}\odot\Gamma^{\Lambda})^*\over (\Gamma^{\Lambda} )^*} \right\vert   
    \left\vert  { (\Gamma^{\Lambda\p}\odot\Gamma^{\Lambda} )^* \over (\Lambda^{-1}\Gamma^{\Lambda\p})^*} \right\vert\  =\  
     \left\vert  { ( \Gamma^{\Lambda\p\Lambda})^*\over\Gamma} \right\vert \ . 
\eeq
Here we employed the facts that $\Gamma^* = \Gamma$ is self-dual,  and that $(\Lambda L)^* = \Lambda L^*$, which holds because $O(d,d)$ transformations preserve the inner product. 

\vskip 1mm

   To prove this last identity we will make repeated use two more facts:  
    $(L_1 \cap L_2)^* = L_1^*\cup L_2^*$ for any two (maximal-rank) lattices $L_1$ and $L_2$,
    and $${A\cup B\over A} = {B\over A\cap B}$$
    for any sets $A$ and $B$. 
 With the help of these identities we can express  the first factor of equation \eqref{appidentity},   as follows:
 \beq
  \left\vert  {  (\Gamma^{\Lambda\p}\odot\Gamma^{\Lambda})^{*}   \over   (\Gamma^{\Lambda})^{*}     } \right\vert   
 =  \left\vert  {
  \Gamma \cup \Lambda^{-1}\Gamma \cup (\Lambda\p\Lambda)^{-1}\Gamma \over
   \Gamma \cup \Lambda^{-1}\Gamma
  } \right\vert
 =  \left\vert  {
  (\Lambda\p\Lambda)^{-1}\Gamma  \over 
     (\Gamma \cup \Lambda^{-1}\Gamma)    \cap  (\Lambda\p\Lambda)^{-1}\Gamma 
 } \right\vert\ . 
  \eeq
Rearranging the last denominator, 
$$
 (\Gamma \cup \Lambda^{-1}\Gamma)    \cap  (\Lambda\p\Lambda)^{-1}\Gamma  =
(\Lambda\p \Lambda)^{-1} ( \Lambda\p ( \Gamma \cup  \Lambda \Gamma) \cap \Gamma ) 
= (\Lambda\p\Lambda)^{-1} (  \Lambda\p  \Gamma^{\Lambda^{-1}}\cup  \Gamma   )^{*} \ , 
$$
and using also $\vert\hskip -0.5mm\vert    L^* \vert\hskip -0.5mm\vert   = \vert\hskip -0.5mm\vert   L \vert\hskip -0.5mm\vert^{-1}$,    leads to 
\beq\label{rel1}
 \left\vert  {  (\Gamma^{\Lambda\p}\odot\Gamma^{\Lambda})^{*}   \over   (\Gamma^{\Lambda})^{*}     } \right\vert   
 =   \vert\hskip -0.5mm\vert   \Gamma \cup   \Lambda\p  \Gamma^{\Lambda^{-1}}
  \vert\hskip -0.5mm\vert^{-1}\ . 
\eeq
Using the same reasoning, one can  likewise establish  the relation 
\beq\label{rel2}
 \left\vert  {  (\Gamma^{\Lambda\p}\odot\Gamma^{\Lambda})^{\ *}   \over   \Lambda^{-1} \Gamma^{\Lambda\p}{}^{\ *}     } \right\vert   
 =   \vert\hskip -0.5mm\vert  \Gamma  \cup  \Lambda\p{}^{-1}   \Gamma^{\Lambda} 
  \vert\hskip -0.5mm\vert^{-1}\ . 
\eeq
Now, the product of the two relations \eq{rel1} and \eq{rel2} can be written as
\beq
 \left\vert  {  (\Gamma^{\Lambda\p}\odot\Gamma^{\Lambda})^{*}   \over   (\Gamma^{\Lambda})^{*}     } \right\vert   
 \left\vert  {  (\Gamma^{\Lambda\p}\odot\Gamma^{\Lambda})^{\ *}   \over   \Lambda^{-1} \Gamma^{\Lambda\p}{}^{\ *}     } \right\vert   
=  \left\vert
{   \Gamma  \cup  \Lambda^{-1}   \Gamma^{\Lambda\p}
\over  
(\Lambda\p\Lambda)^{-1} ( \Gamma \cup   \Lambda\p  \Gamma^{\Lambda^{-1}} )^*  
}  \right\vert   \ , 
 \eeq
where the right-hand side makes sense
since the denominator lattice, 
  $$(\Lambda\p\Lambda)^{-1} ( \Gamma \cup   \Lambda\p  \Gamma^{\Lambda^{-1}} )^* 
= (\Lambda\p \Lambda)^{-1} \Gamma\cap ( \Lambda^{-1} \Gamma \cup \Gamma) \ , 
$$
 is contained in the numerator  lattice, 
$$ \Gamma  \cup  \Lambda\p{}^{-1}   \Gamma^{\Lambda} = \Gamma \cup  ( \Lambda\p{}^{-1} \Gamma \cap
(\Lambda\Lambda\p)^{-1} \Gamma ) \ , 
$$
by virtue of the obvious relation $ A \cap (B\cup C) = (A\cap B )\cup (A\cap C)  \subseteq (A\cap B)\cup C$. 
Simplifying the quotient by eliminating the summand 
$(A\cap B )$ in the numerator yields the desired identity \eq{appidentity}
\beq
  \left\vert  {   (\Gamma^{\Lambda\p}\odot\Gamma^{\Lambda})^*\over (\Gamma^{\Lambda} )^*} \right\vert   
    \left\vert  { (\Gamma^{\Lambda\p}\odot\Gamma^{\Lambda} )^* \over (\Lambda^{-1}\Gamma^{\Lambda\p})^*} \right\vert\  =\  
    \left\vert{\Gamma\over\Gamma^{\Lambda\p\Lambda}}\right\vert
=    
     \left\vert  { ( \Gamma^{\Lambda\p\Lambda})^*\over\Gamma} \right\vert \ .
\eeq
This proves \eq{indid0}.

\vskip 2cm

%%%%%%%%%%%%%%%%%%%%%%%%%%%%%%%%%%%%%%%%%%%%%%%%%%%%%

\end{document}